\documentstyle[sprocl,epsfig,epsf]{article}

\input{psfig}
\def\Journal#1#2#3#4{{#1} {\bf #2}, #3 (#4)}

\def\NPB{{\em Nucl. Phys.} B}
\def\NPA{{\em Nucl. Phys.} A}
\def\PLB{{\em Phys. Lett.} B}
\def\PRL{\em Phys. Rev. Lett.}
\def\PRD{{\em Phys. Rev.} D}
\def\PRC{{\em Phys. Rev.} C}
\def\ZPC{{\em Z. Phys.} C}
\def\SJNP{{\em Sov. J. Nucl. Phys.}~}

\def\be{\begin{equation}}
\def\ee{\end{equation}}
\def\bea{\begin{eqnarray}}
\def\eea{\end{eqnarray}}

\begin{document}

\begin{flushright}
CERN-TH/2000-296\\
BARI-TH/2000-394\\
\end{flushright}

\vspace{1.5cm}

\title{QCD SUM RULES, A MODERN PERSPECTIVE}

\author{PIETRO COLANGELO}

\address{ Istituto Nazionale di Fisica Nucleare, Sezione di 
Bari, Italy}

\author{ALEXANDER KHODJAMIRIAN~\footnote{
{\it On  leave from Yerevan Physics Institute, 
375036 Yerevan, Armenia}} }

\address{Theory Division, CERN, CH-1211 Geneva 23, Switzerland  \\}


\maketitle\abstracts{ An introduction to the method of QCD
sum rules is given for those who want to learn how to use this method.
Furthermore, we discuss various applications of sum rules,
from the determination of quark masses to the calculation 
of hadronic form factors and structure functions.
Finally, we explain the idea of the light-cone sum rules 
and outline the recent development of this 
approach.}

\vspace{1cm}

\begin{center}
{\it to be published in the Boris Ioffe Festschrift\\
''At the Frontier of Particle Physics / Handbook of QCD'',\\  
edited by M. Shifman (World Scientific, Singapore, 2001)}
\\
\end{center}

\newpage

\tableofcontents
\newpage

\section{Introduction}

The method of QCD sum rules, 
developed more than twenty years ago by Shifman, Vainshtein and Zakharov 
(SVZ),\cite{SVZ79} has become a widely used working tool 
in hadron phenomenology.\footnote{ 
The mere fact  that the original SVZ work~\cite{SVZ79} has already 
got more than 2200 citations reflects 
the amount of papers employing  QCD sum rules.}
The advantages of this method are well known. 
Instead of a model-dependent treatment in terms 
of constituent quarks, hadrons are represented 
by their interpolating quark currents taken at large virtualities. 
The correlation function of these currents 
is introduced  and treated in the framework of the 
operator product expansion (OPE), where  
the short- and long-distance quark-gluon 
interactions are separated. The former are calculated using QCD perturbation theory,
whereas the latter are parametrized in terms of 
universal vacuum condensates 
or light-cone distribution amplitudes. 
The result of the QCD calculation
is then matched, via dispersion relation, to a sum 
over hadronic states.
The sum rule obtained in this way allows to calculate observable 
characteristics of the hadronic ground state. Inversely, 
the parameters of QCD such as quark masses
and vacuum condensate densities can be extracted from  sum rules
which have experimentally known hadronic parts. 
What is also very important, the interactions of quark-gluon currents with QCD vacuum fields critically depend on the quantum numbers
(spin-parity, flavor content) of these currents. Therefore, interpolating hadrons with currents, one is able to understand why the hadrons with 
different quantum numbers are not alike.\cite{NSVZ81}
Numerous properties of hadrons with various 
flavor content have been calculated by the sum rule method.
The results are encouraging and in most cases reveal  a 
remarkable agreement with the experimental data. 
Therefore, whenever one needs
to determine an unknown hadronic parameter, the QCD sum rule 
prediction is among the reliable ones. 

However, the accuracy of  
this method is limited, on one hand, by  the approximations in the 
OPE of the correlation functions and, 
on the other hand, by a very complicated and largely unknown structure 
of the hadronic dispersion integrals. The latter are usually approximated 
by employing the quark-hadron duality  approximation. Consequently, 
the applicability of sum rules and the uncertainties 
of their predictions must be carefully assessed case by case. 

QCD sum rules are discussed in many  
reviews~\cite{6auth}$^{\!- }$\cite{deRafael97} 
(the list can be further continued) emphasizing various aspects of 
the method. In this review we try to present and update 
the subject in a more concise,
i.e. close to encyclopedic style,  
explaining the basics, presenting the most interesting 
applications, overviewing the recent developments and 
assessing the current and future potential of sum rules. 
In several cases we compare the predictions of this method with the 
results of lattice QCD. Our purpose is to convince the reader 
that the analytical QCD sum rules are to a large extent complementary 
to the numerical lattice simulations of hadrons.    
It is  a very special occasion to write a review on QCD sum rules
in honor of B.L. Ioffe who contributed to this field   
with landmark results, considerably enlarging the spectrum of 
hadronic problems treated within this method. 

The content of the review is as follows. 
Section 2  is an elementary introduction written for  readers who have no 
experience in QCD sum rules and would like to grasp the basics 
of the method. In Section 3 we present an 
overview of the current status of many important applications
 and discuss the possible improvements and the
perspectives for new investigations.
In Section 4 we consider the light-cone version of QCD sum rules, 
explaining the idea, outlining the derivation and presenting the main applications. 
Section 5  contains a summary.

\section{Understanding SVZ sum rules}

\subsection {Correlation function of quark currents}
The QCD Lagrangian  
\be
{\cal L}_{QCD}= -\frac{1}{4}G^{a}_{\mu\nu}G^{a \mu \nu} +
\sum_{q} \bar{\psi_q}(i\not\!\! D - m_q)\psi_q\,,
\label{qcd}
\ee
where $G_{\mu\nu}^a$ is the gluon field-strength tensor and $\psi_q$ 
are the quark fields with different flavors: $q=u,d,s,c,..$,
is discussed in detail in many  chapters of this book.  
It is our common belief that this Lagrangian governs all 
properties of hadrons and hadronic processes. 
However, a direct use of Eq.~(\ref{qcd}) and of the corresponding 
Feynman rules is possible only within the limited  
framework of perturbation theory.   
At least some of the quarks or gluons in a hadronic 
process have to be highly virtual. This condition 
guarantees the smallness of the corresponding effective quark-gluon 
coupling  $\alpha_s=g_s^2/4\pi$  and, thereby, a  legitimacy of 
the perturbative expansion. Usually, high virtuality is 
achieved in a scattering of hadrons at 
large momentum transfer. However, even for these 
specially configured hard scattering processes, a perturbative 
calculation of quark-gluon Feynman diagrams is not sufficient,
because the quarks  participating 
in the hard scattering are confined inside hadrons. Hence,  
one has to combine the perturbative QCD result  
with  certain wave functions or momentum distributions
of quarks in hadrons. To calculate these characteristics, 
one needs to know  the QCD dynamics at distances of 
order of the hadron size: $R_{hadr} \sim 1/\Lambda_{QCD}$,
the scale at which perturbation theory in $\alpha_s$ 
is not applicable.  

\begin{figure}
\hspace{1.5cm}
\psfig{figure=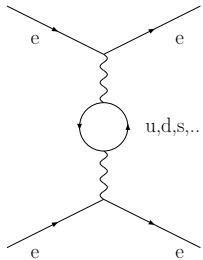,height=2.0in}
\caption{Quark-antiquark creation and annihilation
by the virtual photon in the electron-electron scattering.}
\label{fig:ee}
\end{figure}

In order to avoid long-distance problems, one could consider  
processes with {\em no} initial and final hadrons, and 
with all quarks propagating at short distances
(during short times).  Such configurations are 
not that hypothetical as it may seem. They are realized 
in nature, when  the quark-antiquark pair is  
produced  and absorbed by an external source, 
e.g. by a virtual photon  in the electron-electron elastic 
scattering, as shown in Fig.~\ref{fig:ee}. 
The propagation of the intermediate 
quark-antiquark pair in this process adds a very small 
(order of $\alpha_{em}=e^2/4\pi$ ) quantum correction to the 
cross section of $e^-e^-\to e^-e^-$.  
Nevertheless, taken separately, 
the amplitude of the  quark-pair creation and annihilation 
is an extremely useful object from the QCD point of view.
The formal expression for this amplitude can be written as
\be
\Pi_{\mu\nu}(q)= i\int d^4 x \; e^{iq\cdot x} 
\langle 0 \!\mid T\{j_{\mu}(x)j_{\nu}(0)\}\mid \! 0\rangle
= (q_\mu q_\nu-q^2g_{\mu \nu})\Pi(q^2)\,, 
\label{pimunu}
\ee
where $q$ is the four-momentum of the virtual photon with $q^2 \!<\! 0$,  
 $j_\mu= \bar{\psi}\gamma_\mu \psi$ is 
the colorless quark current with a given flavor $\psi=u,d,s,c,.. $   
(for simplicity, we have omitted the electromagnetic coupling from 
this definition). In the amplitude (\ref{pimunu}),
the initial and final states contain no hadrons and are therefore 
identified with the vacuum state of QCD. 
The Lorentz-structure of the r.h.s.$\!$ of Eq.~(\ref{pimunu}) 
is dictated by the conservation of  the electromagnetic current: 
$\partial_\mu j^\mu=0$, so that the single invariant amplitude 
$\Pi(q^2)$ encodes all dynamical effects.

The amplitude $\Pi_{\mu\nu}$ represents an important 
example of a two-point {\em correlation function} (correlator)
of quark currents.
If the four-momentum squared transfered to the quarks is 
large, $Q^2\equiv -q^2 \gg \Lambda_{QCD}^2$, 
this correlation function turns into the genuine short-distance object 
we are looking for. This is simply because 
the integral in  Eq.~(\ref{pimunu}) is dominated by small spatial 
distances and time intervals:
\be
|\vec{x}| \sim x_0 \sim 1/\sqrt{Q ^2} \ll R_{hadr}\, . 
\label{deltax}
\ee 
This condition can be directly inferred from a 
general analysis of Eq.~(\ref{pimunu}) in the case of massless quarks, 
an approximation justified 
for the light $u,d,s$ quarks. After contraction of the Lorentz 
indices, the vacuum average in Eq.~(\ref{pimunu}) can only 
depend on the space-time interval $x^2=x_0^2-\vec{x}~^2$:
\be 
\langle 0 \mid T\{j_{\mu}(x)j^{\mu}(0)\}\mid 0\rangle 
= \int d\tau \; e^{i\tau x^2} f(\tau)~,
\label{repr}
\ee
where the Fourier transform of this functional dependence 
is introduced. Using the representation (\ref{repr}) and 
shifting the variable $x$, one obtains from Eq.~(\ref{pimunu}): 
\be
3q^2 \Pi(q^2) 
= -i\int d \tau \int d^4 x \; e^{i\tau x^2}e^{iQ^2/4\tau}f(\tau)~.
\label{deriv}
\ee
The integrand on the r.h.s. of this equation is suppressed 
if at least one of the exponential functions rapidly oscillates. 
Therefore, dominant contributions to $\Pi(q^2)$ stem 
from the regions where both $ \tau \sim 1/x^2$ and $ \tau \sim Q^2 $. 
To fulfill these two conditions simultaneously, one has to demand
\be
x^2\sim 1/Q^2\,, 
\label{lc}
\ee
so that, at $Q^2 \to \infty$ the quarks 
propagate near the light-cone, $x^2\sim 0$ . 
This condition is necessary but not yet sufficient
for the short-distance dominance. To demonstrate the latter 
it is convenient to choose the reference frame $q_0=0$, so that 
$\vec{q}\,^2=Q^2$,  and the exponent in Eq.~(\ref{pimunu})
is simply equal to $\exp(-i\vec{q}\cdot \vec{x})$. 
Again, to avoid a fast oscillating integrand one 
needs 
\be
|\vec{x}| \sim 1/\sqrt{Q^2}\,,
\label{short}
\ee
which, together with Eq.~(\ref{lc}), yields Eq.~(\ref{deltax}).   
Hence, at large $Q^2$ the quarks  
in the diagram in Fig.~\ref{fig:ee} propagate predominantly at short distances 
and during short time intervals.
Due to the asymptotic freedom of QCD, the quark-gluon
interactions are then suppressed. Therefore, as a first approximation, 
one  may calculate the correlation function 
(\ref{pimunu}) representing  virtual quarks  
by the free-quark propagators inferred directly 
from the Lagrangian (\ref{qcd}). In the case of heavy quark
currents ($\psi=c,b$), the situation is even
simpler because the quark mass $m_{c,b} \gg \Lambda_{QCD}$ 
introduces an intrinsic large energy scale. One has an asymptotically 
free quark-antiquark fluctuation already at small 
$q^2 \ll 4~m_{c,b}^2$. In this case, the characteristic  distances 
in the correlation function are determined by the 
inverse heavy quark mass $ |\vec{x}|\sim x_0 \sim 1/(2~m_{c,b})$.

\subsection{Summing up hadrons: the unitarity relation}

Before turning to the actual calculation of
$\Pi_{\mu\nu}$, let us discuss 
how this object is related to physically observed hadrons.
Note that the invariant 
amplitude $\Pi(q^2)$ 
is an analytic function of $q^2$ defined at both 
negative (spacelike) and positive (timelike) values of $q^2$  
and, formally, even at complex values of this variable.
At positive $q^2$ the underlying electromagnetic process 
is the  cross-channel of the electron-electron scattering, 
i.e. the annihilation $e^+e^- \to e^+e^-$ with the total
c.m. energy $E_{e^+}+E_{e^-}=\sqrt{q^2}$.
If $q^2$ is shifted from large negative to positive values,
the average distance between the points $0$ and $x$ in 
the quark amplitude (\ref{pimunu}) grows. The long-distance quark-gluon interactions 
become important and, eventually, the quarks form hadrons.
In particular, a quark-antiquark 
pair created by the current $j_\mu$  
with the spin-parity $J^P=1^-$  materializes as a neutral vector meson. 
For the currents $\bar{u}\gamma_\mu u$ and $\bar{d}\gamma_\mu d$, 
one has to respect the isospin symmetry,
because the mass difference $m_d-m_u \sim O(\mbox{MeV})$ 
is much smaller than the QCD scale $\Lambda_{QCD}$. 
The ground-state vector mesons 
with the isospin $I=1$ and $I=0$ are $\rho$ and $\omega$  
with the quark content 
$(\bar{u}u-\bar{d}d)/\sqrt{2}$ and $(\bar{u}u+\bar{d}d)/\sqrt{2}$,
respectively. The lightest vector meson created by the 
current $\bar{s}\gamma_\mu s$ is $\phi$. For the 
heavy quark currents $\bar{c}\gamma_\mu c$ and $\bar{b}\gamma_\mu b$, 
the ground states are $J/\psi$ and $\Upsilon$, 
respectively. Physically, vector mesons are observed in a form 
of resonances in $e^+e^-$ annihilation at energies 
$\sqrt{{q^2}}=m_V$, ($ V=\rho,\omega,\phi, J/\psi, \Upsilon, ...$). 
Not only the ground-state, but also 
excited vector mesons and a continuum of 
two- and  many-body hadron states with the  quantum numbers of $V$ 
contribute to  $\Pi_{\mu\nu}$.

A rigorous way to quantify a very complicated hadronic content of  
$\Pi_{\mu\nu}$ at $q^2 >0$ is provided by the {\em unitarity relation}
obtained by inserting a complete set of intermediate hadronic states
in Eq.~(\ref{pimunu}):
\be
2\mbox{Im}\,\Pi_{\mu\nu}(q) = 
\sum _n  \langle 0 \!\mid  j_{\mu} \mid\! n \rangle   
\langle n \! \mid j_{\nu} \mid \! 0 \rangle ~d\tau_n
(2\pi)^4\delta^{(4)}(q-p_n)\,, 
\label{unitar}
\ee
where the summation goes over all possible 
hadronic states $\mid \! \!n \rangle$ created by the quark current
$j_\mu$ including sums over polarizations, and $d\tau_n$ denotes
the integration over  the phase space volume of these states.    

The one-particle, vector meson  
contribution to the hadronic sum (\ref{unitar}) is 
\be
\frac1{\pi}\mbox{Im}~\Pi^V_{\mu\nu}(q^2)= 
(q_\mu q_\nu - m_V^2g_{\mu\nu})f_V^2 \delta(q^2-m_V^2)\,,
\label{imV}
\ee
where the total decay width of $V$ is neglected for 
simplicity, and the decay constant $f_V$ is defined by 
the matrix element of the current $j_\mu$ 
between the vacuum and the vector-meson states:
\be 
\langle V(q) \! \mid  j_{\mu} \mid \! 0 \rangle =  
f_V m_V\epsilon^{(V)*}_\mu,
\label{fV}
\ee 
$\epsilon^{(V)}_\mu$ being the polarization vector
of $V$ ($\epsilon^V \cdot q =0$).   
Note that $f_V$ is a typical hadronic parameter 
determined by the long-distance dynamics. 
 
The contributions of continuum hadronic states 
to the unitarity relation 
(\ref{unitar}) are more involved. Each individual state 
$|n\rangle$ yields a  continuous imaginary part at $q^2>m_n^2$,
$m_n$ being  the sum of hadron masses in this state.
Moreover, the corresponding hadronic matrix elements
for multiparticle states    
$\langle n \! \mid  j_{\mu} \mid \! 0 \rangle$
are not just constants but depend on $q^2$.

For convenience, we single out the ground-state vector-meson contribution 
on the r.h.s. of the unitarity relation (\ref{unitar}) and 
introduce a compact notation for the rest of 
contributions including excited vector mesons and 
continuum states:
\be
\frac{1}{\pi}\mbox{Im}~\Pi(q^2)= f_V^2 \delta(q^2-m_V^2)+
\rho^h(q^2)\theta(q^2-s_0^h)~,  
\label{higher}
\ee
where $s_0^h$ is the threshold of the lowest continuum state. 
Notice that in the light quark channels this 
threshold, set by two- and three-pion states, is lower than 
$m_V$. In the heavy quark channels the pattern
is different. There are several heavy quarkonium resonances
below the threshold of the heavy flavored 
meson pair production and, therefore, Eq.~(\ref{higher}) has
to be slightly modified: $\mbox{Im}~\Pi^V(q^2)\to \sum_V\mbox{Im}~\Pi^V(q^2)$
including the sum over all below-threshold quarkonium states.

\subsection{Deriving the dispersion relation} 

\begin{figure}
\hspace{8cm}
\psfig{figure=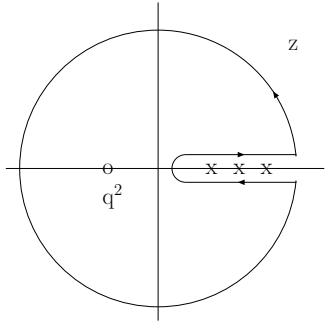,height=2.0in}
\caption{The contour in the plane of the complex variable $q^2=z$.
The open point indicates the $q^2<0$ reference point of the QCD 
calculation. Positions of hadronic thresholds
at $q^2 >0$ are indicated by crosses.}
\label{fig:contour}
\end{figure}

From the above discussion  we learned that 
the correlation function (\ref{pimunu}) 
is an object of {\em dual} nature. At large negative $q^2$   
it represents a short-distance 
quark-antiquark fluctuation 
and can be treated in perturbative QCD, whereas at positive $q^2$ it has a 
decomposition in terms of hadronic
observables. The next step is to derive 
a dispersion relation linking $\Pi(q^2)$    
at an  arbitrary point $q^2<0$  to the hadronic sum (\ref{unitar}). 
For that, one employs the Cauchy formula for the analytic function 
$\Pi(q^2)$, choosing the contour shown in Fig.~\ref{fig:contour}:
\bea
\Pi(q^2) &=& \frac {1}{2\pi i} \oint\limits_C dz \frac{\Pi(z)}{z-q^2}
= \frac {1}{2\pi i} \oint\limits_{|z|=R} dz \frac{\Pi(z)}{z-q^2} 
\nonumber
\\
&&
+ \frac {1}{2\pi i}\int\limits_0^R dz \frac{
\Pi(z+i\epsilon)- \Pi(z-i\epsilon)}{z-q^2}
\,.
\label{Cauchy}
\eea
The radius $R$ of the circle in this contour can be 
put to infinity. This simplifies the r.h.s. of Eq.~(\ref{Cauchy}) 
considerably, because, if the correlation function vanishes 
sufficiently fast at $|q^2|\sim R \to \infty$, 
(if $\mbox{lim}_{|q^2|\to \infty}\Pi(q^2) \sim 1/|q^2|^\epsilon$, 
with any $\epsilon >0$) the integral over the circle tends 
to zero. Below we shall discuss a necessary modification 
of Eq.~(\ref{Cauchy}) if $\Pi(q^2) $ does not vanish. 
The second integral on the r.h.s. of Eq.~(\ref{Cauchy})
can be replaced by an integral over the imaginary part of $\Pi(q^2)$.
One makes use of the fact that $\Pi(q^2)$ is real 
at $q^2 < t_{min}=\mbox{min}\{m_V^2,s_0^h \}$. Hence, 
according to the Schwartz reflection principle: 
$ \Pi(q^2+i\epsilon) -\Pi(q^2-i\epsilon)= 2i~\mbox{Im}\Pi(q^2)$
at $q^2 > t_{min}$.  After this replacement, we obtain the
{\em dispersion relation}:
\be
\Pi(q^2) = \frac{1}{\pi}\int\limits_{t_{min}}^\infty \!ds\frac{\mbox{Im}~\Pi(s)}{s-q^2-i\epsilon}\,.
\label{disp1}
\ee
The infinitesimal $-i\epsilon$ will not be shown explicitly hereafter.
As we shall see in the next subsection,
the correlation function (\ref{pimunu}) 
is ultraviolet divergent.
Consequently, the imaginary part 
$\mbox{Im}\,\Pi(s)$ does not vanish at $s \to \infty$
and the dispersion integral (\ref{disp1})  diverges. 
A standard way to cure this problem is to subtract from 
$\Pi(q^2)$ first few terms of its Taylor expansion at 
$q^2=0$. For the correlation function (\ref{pimunu}) 
one subtraction is sufficient:
\be
\overline{\Pi}(q^2)= \Pi(q^2)-\Pi(0).   
\ee
The dispersion relation (\ref{disp1}) is modified 
in the following way:
\be
\overline{\Pi}(q^2) = 
\frac{q^2}{\pi}\int\limits_{t_{min}}^\infty \!ds\frac{\mbox{Im}\,\Pi(s)}{s(s-q^2)}\,.
\label{disp2}
\ee
Using the hadronic representation (\ref{higher}), one finally obtains
\be 
\Pi(q^2)= \frac{q^2f_V^2}{m_V^2(m_V^2-q^2)} + 
q^2\int\limits_{s_0^h}^{\infty}ds \frac{\rho^h(s)}{s(s-q^2)} +\Pi(0)\,.
\label{dispSR}
\ee
Notice that in our case $\Pi(0)=0$, due to 
the gauge invariance of the electromagnetic interaction. Nevertheless,
we  retain $\Pi(0)$ in Eq.~(\ref{dispSR}), having
in mind a generic case, where 
a subtraction constant or a finite polynomial in $q^2$ 
appear in the resulting dispersion relation.

The dispersion relations similar to Eq.~(\ref{dispSR}) 
are central objects of our review. With the correlation
functions  calculated in QCD in a certain approximation,
these relations establish {\em sum rules}, i.e. nontrivial
constraints on the sums over hadronic parameters.

\subsection{Applying the Borel transformation}

The sum rules in the form (\ref{dispSR})  are not   
yet very useful, e.g. for estimating  the parameters of the 
lowest-lying hadronic state. They are in general plagued
by the presence of unknown subtraction terms. 
More importantly,  little is known about the spectral function
$\rho^h(s)$ of
excited and continuum states.  The situation can be substantially 
improved~\cite{SVZ79}  
if one applies to both sides of Eq.~(\ref{dispSR}) the 
Borel transformation 
\be
\Pi(M^2) \equiv 
{\cal B}_{M^2}\Pi(q^2)=\lim_{\stackrel{-q^2,n \to \infty}{-q^2/n=M^2}}
\frac{(-q^2)^{(n+1)}}{n!}\left( \frac{d}{dq^2}\right)^n \Pi(q^2)~.
\label{Borel1}
\ee
Two important examples are :
\be
{\cal B}_{M^2}(q^{2})^k=0, ~{\cal B}_{M^2}\left(\frac1{(m^2-q^2)^k}\right)=
\frac1{(k-1)!}\frac{\exp(-m^2/M^2)}{M^{2(k-1)}}\,,
\label{Borel2}
\ee
at $k> 0$. Transformations of more complicated functions  
can be found in the literature.\cite{NSVZtechn,NardeRaf}
Applying Eqs.~(\ref{Borel1}) and (\ref{Borel2}) to 
Eq.~(\ref{dispSR}), a more convenient form of the sum rule
is obtained:
\be
\Pi(M^2)= f_V^2e^{-m_V^2/M^2}+
\int_{s^h_0}^\infty ds ~\rho^h(s)e^{-s/M^2}.
\label{BorelSR}
\ee
The Borel transformation removes subtraction terms in the dispersion
relation and exponentially suppresses the
contributions from excited resonances and continuum states
heavier than $V$.
In the case of the heavy quark-antiquark 
currents, instead of the Borel transformation, it is more useful 
to apply a simpler procedure of $n$-times 
differentiation of Eq.~(\ref{dispSR}) at
$q^2=q^2_0\leq0$ :
\be
M_n(q_0^2)\equiv \frac{1}{n!}\frac{d^n}{dq^{2n}}\Pi(q^2)|_{q^2=q^2_0}=
\frac{f_V^2}{(m_V^2 -q_0^2)^{n+1}}+ \int_{s_0^h}^{\infty} \!ds
\frac{\rho^h(s)}{(s-q_0^2)^{n+1}}\,.
\label{moments}
\ee
One gains a power suppression 
of heavier states and again removes the subtraction terms.

\subsection{Calculating the correlation function in QCD: the perturbative
part}
\begin{figure}
\vskip 0.5cm
\hspace{2cm}
\psfig{figure=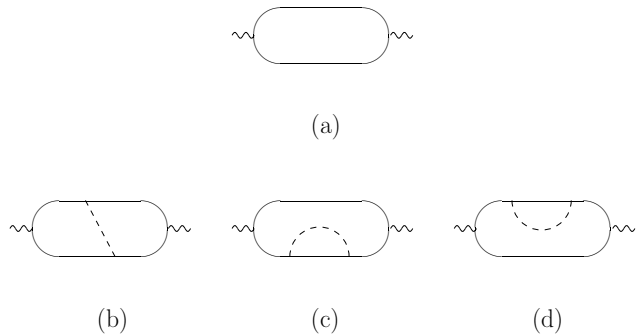,height=2.0in}
\caption{ Diagrams determining the correlation function 
(\ref{pimunu}): the free-quark loop (a), the perturbative 
QCD corrections (b,c,d). Solid lines denote quarks, dashed lines
gluons, wavy line external currents.}
\label{fig:loop}
\end{figure}

\begin{figure}
\hspace*{1.5cm}
\psfig{figure=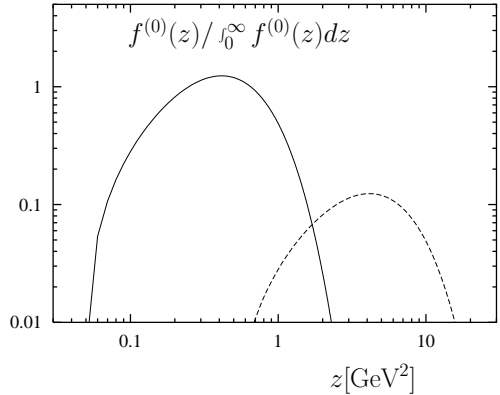,height=2.5in}
\caption{The distribution of the quark virtuality in the 
loop diagram after the Borel transformation at 
$M^2=1~\mbox{GeV}^2$ (solid line) and $M^2=10~\mbox{GeV}^2$ (dashed line).}
\label{fig:zdistr}
\end{figure}

We now turn to the next important step in the sum rule
derivation and describe how the QCD calculation of the correlation
function (\ref{pimunu}) is done. 
As  explained in subsection 2.1, at very large 
$Q^2=-q^2$ the function $\Pi_{\mu\nu}$
can be approximated by the free-quark  loop diagram shown in  
Fig.~\ref{fig:loop}a. 
In a generic case of a single quark 
flavor with the mass $m$, one has to contract all quark fields
in (\ref{pimunu}) considering the quark propagators 
in the free-quark approximation: 
\be
S_0^{ij}(x,y)=-i \langle 0 |T\{\psi^i(x)\overline{\psi}^j(y)\}|0\rangle =
\delta^{ij}\int \! \frac{d^4p}{(2\pi )^4}e^{-ip\cdot(x-y)}
\frac{\not\!p+m}{p^2-m^2}\,,
\label{prop0}
\ee
where the quark color indices are explicitly shown.
After integrating over $x$, shifting to 
$D\neq 4$ dimensions and taking traces,
we obtain
\be
q^2\Pi^{(0)}(q^2) = -\frac{12i}{(D-1)}\int\limits_0^1 \!\!dv\!\! 
\int \!\! \frac{d^Dp}{(2\pi)^D}
\frac{(2-D)(p^2-q^2v(1-v)) + Dm^2}{\left(p^2+q^2v(1-v)-m^2 \right)^2}.
\label{pioD}
\ee
One may now proceed, performing the momentum integration 
in $D$ dimensions. However, at  
this point we prefer to modify 
the standard procedure, and apply the Borel transformation 
before the integration in $p$ is done. The divergence  
disappears and the limit $D\to 4$ can safely be restored.
After the Wick rotation to the Euclidean space $p^2 \to -p^2 \equiv z$ 
and the angular integration in the 
four-dimensional integral, one obtains, in the massless quark case, 
\be
{\cal B}_{M^2}(q^2\Pi^{(0)}(q^2)) = \int\limits_0^\infty dz f^{(0)}(z)\,,
\label{distrib1}
\ee
where 
\be
f^{(0)}(z)= \frac{1}{2\pi^2} z\int\limits_0^1\! 
\!\frac{dv}{v(1-v)}\left( -1+ \frac{2z}{v(1-v)M^2}\right)
\exp\left \{ -\frac{z}{M^2v(1-v)} \right\}\,.
\label{distrib}
\ee
The numerical result for $f^{(0)}(z)$ is shown in 
Fig.~\ref{fig:zdistr}.
We see that the average $\langle z \rangle$
characterizing the quark virtuality in the loop diagram,
is of the order of $M^2$, and that the region of small
$z$, e.g., $z\leq 0.1~M^2$  is strongly suppressed. Thus, at 
sufficiently large $M^2$, the quarks are predominantly far off-shell.  
In the case of heavy $c,b$ quarks it is more useful to 
differentiate Eq.~(\ref{pioD}) $n$ times 
over $q^2$ at $q^2 =0$, revealing
that the dominant contribution stems 
from $|p^2|\sim m_{c,b}^2/n$. Thus, an average high virtuality of heavy quarks is guaranteed, 
if $n$ is not very large. 

The most convenient final expression of $\Pi^{(0)}(q^2)$ 
is in the form of the dispersion integral:
\be
\Pi^{(0)}(q^2)= \frac{q^2}{\pi} 
\!\int ds \frac{\mbox{Im}\,\Pi^{(0)}(s)}{s(s-q^2)}
\label{disp0}
\ee
with the imaginary part 
\be
\mbox{Im}\,\Pi^{(0)}(s)= \frac1{8\pi}v(3-v^2)
\theta(s-4m^2)\,,
\label{imloop0}
\ee
where $v=\sqrt{1-4m^2/s}$.
One should be 
careful in interpreting the imaginary parts of quark-loop diagrams. 
In QCD quarks are confined (the full quark propagators 
have no poles). 
Hence, the imaginary part  (\ref{imloop0}) is a purely mathematical object. 
The free-quark approximation $\Pi^{(0)}(q^2)$ 
is especially simple in the light-quark case 
$m^2 \ll Q^2$, yielding 
\be
\Pi^{(0)}(q^2)\simeq \frac{q^2}{4\pi^2} 
\!\int\limits_{4m^2}^{\infty}\!\frac{ds}{s(s-q^2)}
\simeq -\frac{1}{4\pi^2}\ln\frac{Q^2}{4m^2} + 
O\left(\frac{m^2}{Q^2}\right)\,, 
\ee
where the $O(m^2/Q^2)$  correction is numerically 
important only in the $s$-quark case.

To improve the free-quark approximation, one has to calculate 
the $O(\alpha_s)$ perturbative correction corresponding 
to the diagrams in Figs.~\ref{fig:loop}b,c,d. Since we 
already convinced ourselves that the average quark virtualities 
in the loop are of $O(M^2)$, it is conceivable 
to use QCD perturbation theory for these diagrams  
taking the quark-gluon coupling 
$\alpha_s$ at the scale $M$. The calculation of two-loop diagrams is 
technically quite involved but, fortunately, the result 
can be directly taken from QED, employing the Schwinger  
interpolation formula~\cite{Schwinger} for the $O(\alpha_{em})$ 
radiative correction to the electron polarization operator.  
One has simply to replace $\alpha_{em} \to \alpha_s C_F$  ($C_F=4/3$).
After adding the $O(\alpha_s)$ correction, the perturbative part 
of the correlation function, $\Pi^{(pert)}(q^2)$, is given 
by the dispersion relation (\ref{disp0}) with the imaginary part
\bea
\mbox{Im}\,\Pi^{(pert)}(s)\! &=& \!\mbox{Im}\,\Pi^{(0)}(s)
\left\{1+\alpha_s C_F
\left[\frac{\pi}{2v}-\frac{v+3}4\left(\frac{\pi}{2}
-\frac3{4\pi} \right) \right]  \right\}.
\label{PiMcorr}
\eea
In the case of $u,d,s$ quark currents, the $O(\alpha_s)$ 
correction in (\ref{PiMcorr})
reduces to  $\alpha_s/\pi$ and is numerically small. Virtual
gluon exchanges are potentially important for the correlation functions 
of heavy quark-antiquark currents since the $\alpha_s/v$ term 
in (\ref{PiMcorr}) becomes anomalously large at $v \ll 1$, 
i.e., at $s$ close to the threshold $4m^2$.
The  $\alpha_s/v$ terms 
can be traced back to Coulomb-type interactions between
quarks and antiquarks. In the nonrelativistic approximation, it is  
possible~\cite{Voloshin79}  
to sum up all  $(\alpha_s/v)^n$ terms in the correlation 
function, taking into account not only the one-gluon exchange 
but the whole ladder of such exchanges. This summation 
is usually applied  
to the correlation functions of $\bar{b}\gamma_\mu b$  currents
related to $\Upsilon$ resonances.\footnote{At $O(\alpha_s)$ 
accuracy one should take care 
of a proper definition of the heavy quark mass which is
a scale-dependent parameter in perturbative QCD. In
Eq.~(\ref{PiMcorr}) the so called ``pole'' mass 
is used. We do not discuss this important issue here, and refer to the
chapters by Uraltsev and Chetyrkin in this book.}

A careful reader may have noticed that in all perturbative diagrams 
considered here the regions of small quark 
and gluon virtualities are automatically included, e.g., the
integration in (\ref{distrib}) spreads over small 
quark virtualities. The QCD perturbation theory is invalid in 
this region and the 
free propagators cannot be used. Nevertheless, one may still 
argue that no large numerical error is being introduced, as long as $M^2$ is kept large. For instance, in the integral 
(\ref{distrib1}) the region with $ z\leq \mu^2$ contributes 
with a suppression of $O(\mu^4/M^4)$.

\subsection{Vacuum condensates and operator product expansion} 

The fact that the perturbative part of 
$\Pi_{\mu\nu}$ 
has been reliably calculated  does not yet imply that {\em all} 
important contributions to the correlation function  
have been taken into account. The complete calculation~\cite{SVZ79} 
has to include the effects due to the fields of soft gluons 
and quarks populating the QCD vacuum. 
The problem of vacuum fields in QCD with its many 
interesting aspects is discussed elsewhere in this book. 
We only mention that vacuum fluctuations in QCD  
are due to the complicated 
nonlinear nature of the Lagrangian (\ref{qcd}). The ultimate solution 
of QCD equations of motion and the resulting complete picture 
of the vacuum fields are unknown. Various nonperturbative
approaches (instanton models, lattice simulation of QCD, etc.) 
indicate that these fields  fluctuate with  
typical long-distance scales $\Lambda_{vac}\sim \Lambda_{QCD}$. 
It is clear that the quark-antiquark pair created by the
external current at one point and absorbed  at another 
point  interacts with the vacuum  fields. This interaction 
is beyond QCD perturbation theory and has to be 
taken into account separately.

\begin{figure}
\hspace{1cm}
\psfig{figure=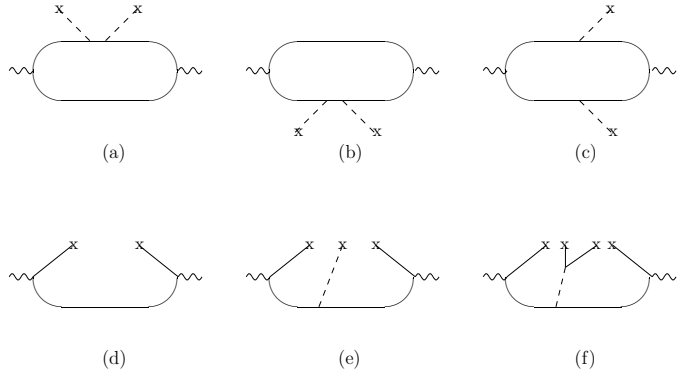,height=2.2in}
\caption{ Diagrams corresponding to the 
gluon (a,b,c), quark (d), quark-gluon (e) and 
four-quark (f) condensate contributions to  
the correlation function (\ref{pimunu}).}
\label{fig:cond}
\end{figure}

A practical way to calculate the vacuum-field 
contributions to the correlation functions relies on the following
qualitative arguments. At large $Q^2 \gg \Lambda_{QCD}^2$, 
the  average distance between the points 
of the quark-antiquark emission and absorption is essentially smaller than the 
characteristic scale of the  
vacuum fluctuations. Therefore, propagating in the 
QCD vacuum, the quark-antiquark pair 
acts as a short-distance probe of long-distance fields 
and perceives static, averaged characteristics of these fields. 
At the same time, the emission of quarks and antiquarks does not 
significantly disturb the vacuum state. Hence, in the 
first approximation,  quarks with large momenta 
$\sim \sqrt{Q^2}$ scatter over external static  fields composed 
of soft vacuum gluons and quarks. The corresponding diagrams 
are shown in Fig.~\ref{fig:cond}. In the case of  
light quarks, there are several important effects: 
the vacuum gluons are emitted and absorbed by virtual quarks 
(Figs.~\ref{fig:cond}a,b,c), the quarks and antiquarks are 
interchanged with their vacuum counterparts (Fig.~\ref{fig:cond}d) 
and, finally, a combined quark-gluon interaction takes place 
(Figs.~\ref{fig:cond}e,f). For the heavy 
quarks only the interactions with 
the vacuum gluons are important. 
 
A quantitative framework which follows this picture 
and incorporates both short- and long-distance
contributions was developed~\cite{SVZ79} 
in a form  of  a  generalized Wilson OPE. 
To apply this method to the 
correlation function (\ref{pimunu}), one has to expand the product
of two currents in a series of local operators:
\bea
i\int d^4x \;
e^{iq \cdot x }~T\{\bar{\psi}(x)\gamma_\mu \psi(x),
\bar{\psi}(0)\gamma_\nu \psi(0) \}
\nonumber
\\
= (q_\mu q_\nu -q^2g_{\mu\nu})\sum_d C_d(q^2) O_d\,,
\label{exp2}
\eea
so that 
\be
\Pi(q^2) = \sum_d C_d(q^2)\langle 0 \!\mid O_d \mid \!0 \rangle~.
\label{ope112}
\ee  
In this expansion, the operators are ordered according to their
dimension $d$. The lowest-dimension operator with $d=0$ is the unit
operator associated with the perturbative contribution: 
$C_0(q^2)=\Pi^{pert}(q^2)$, 
$\langle  0\! \mid\! O_0\! \mid \!0 \rangle \equiv 1$. 
The QCD vacuum fields 
are represented in (\ref{ope112}) in a form of the so 
called {\em vacuum condensates}, 
the vacuum expectation values of the $d\neq 0$ operators,
composed of quark and gluon fields, $\bar{\psi}$, $ \psi$   
and $G^a_{\mu\nu}$. 
The contributions of high-dimensional condensates corresponding  to the diagrams with multiple 
insertions of vacuum gluons and quarks, are suppressed by large 
powers of $\Lambda_{vac}^2/Q^2$.  Therefore, even at intermediate 
$Q^2\sim 1 $ GeV$^2$, the expansion (\ref{ope112}) can safely be truncated after 
a few terms. 

One may interpret the OPE (\ref{ope112}) in the following 
way. The product of quark currents acts as a quasi-local ``antenna''
having a small size $O(1/ \sqrt{Q^2} )$
and probing the static vacuum fields. This interaction 
depends  on the properties of the ``antenna'', 
i.e. on the  quantum numbers and flavor content of the quark 
currents. In Eq.~(\ref{ope112}), this dependence is accumulated
in the  
Wilson coefficients $C_d(q^2)$  receiving dominant contributions 
from the regions of short distances (large momenta). 
In addition, the current-vacuum interaction  is determined 
by the long-distance dynamics represented in Eq.~(\ref{ope112})
by the universal condensates $\langle 0 \!\mid O_d \mid \!0 \rangle$ with $d\neq 0$, 
which are independent of the properties of the quark currents.  
The separation of distances is a 
key point in the OPE (\ref{ope112}). 
Representing $\Pi(q^2)$ in this form, one  introduces a 
certain scale $\mu$ which separates the regions of 
short and long distances. The interactions at momenta $p^2>\mu^2$ are 
included in the coefficients $C_d(q^2)$, while the effects at $p^2<\mu^2$
are absorbed into the vacuum condensates. The scale $\mu$ should 
be large enough in order to justify the calculation
of $C_d$ in QCD perturbation theory. In practice,
using the standard methods of the Feynman-diagram technique,
an explicit separation of distances is impossible in 
the quark-loop diagrams. One is forced to take into account 
both the soft parts of perturbative diagrams and the 
long-distance condensate effects simultaneously. This 
yields  a certain amount of double counting, which is,
fortunately, in many cases numerically insignificant, because 
the condensate contributions turn out to be much larger than the 
soft ``tails''of perturbative diagrams. Moreover, if one does not go beyond 
the two-loop diagrams in Figs.~\ref{fig:loop}b-\ref{fig:loop}d 
and uses the leading-order Wilson coefficients $C_d$ for $d\neq 0$,
it is possible to rearrange the OPE including the soft 
($p^2 <\mu^2$) contributions of the perturbative diagrams 
in the definition of the condensates.   
This ``practical version'' of OPE is 
discussed in more detail in a recent review.\cite{Shifman98}

The list of the operators with low dimension entering Eq.~(\ref{exp2}) starts with 
\be
O_3 =\bar{\psi}\psi 
\ee
and
\be
O_4=G_{\mu\nu}^a G^{a\mu\nu}, 
\ee
whose vacuum averages are known as 
the quark and gluon condensates, respectively. It is important 
that in QCD there are 
no colorless operators with lower dimensions, $d=1,2$.

The operators with $d=5,6$ are 
\be
O_5=\bar \psi\sigma_{\mu\nu} \frac{\lambda ^a}2 
G^{a\mu\nu}\psi\,,
\ee
\be
O_6^{\psi}=(\bar{\psi}\Gamma_r \psi)( \bar{\psi}\Gamma_s \psi)~,
\label{oper2}
\ee
where $\Gamma_{r,s}$ denote various combination of Lorentz and color matrices,
and
\be
O_6^{G}=f_{abc}G_{\mu\nu}^a G^{b \, \nu}_ \sigma G^{c\sigma\mu}\,. 
\ee
The vacuum averages of the above operators are, correspondingly 
the quark-gluon, four-quark and three-gluon condensates.
The condensates with $d>6$  usually play
a minor role in the most QCD sum rule applications
and we will not consider them in detail. 

One should mention that, in addition to rather mild
effects of the quark scattering over the vacuum fields,
there exist specific vacuum fluctuations at short 
distances $\sim 1/\sqrt{Q^2}$, which absorb the whole momentum 
of the external quark current.
These effects, known as  ``direct instantons,''~\footnote{ They can 
be modeled  employing the QCD instanton solution.
Instanton-induced effects are discussed in the chapter by 
Shuryak in this book.}
violate the condensate 
expansion. For the 
vector currents considered  
here, short-distance nonperturbative effects  may appear~\cite{SVZ79} 
only at very high dimensions ($d>10$) and do not 
play any role in the truncated OPE. However, 
in the correlation functions of
pseudoscalar ($ J^P=0^-$) and scalar ($J^P=0^+$)  
quark and/or gluon currents, 
direct instantons are enhanced 
and important already at intermediate $Q^2$.

To proceed with the derivation of the QCD answer for $\Pi_{\mu\nu}$, 
on has to evaluate the Wilson coefficients of the condensate terms 
in (\ref{ope112}). For illustration, we demonstrate 
how the simplest contribution of the quark condensate 
is calculated. 
The relevant diagram is shown in Fig. 3d.
One factorizes out all contributions to $\Pi_{\mu\nu}$ 
containing one antiquark and one quark field, the remaining
quark fields being contracted in the free-quark propagators:
\bea
\Pi_{\mu\nu}^{(\bar{\psi}\psi)}(q)=i\int \! d^4x e^{iq \cdot x} 
\langle 0 \mid 
\{ \bar{\psi}^i(x)\gamma_\mu S^{ij}(x,0) \gamma_\nu \psi^{j}(0)
\nonumber
\\
+ 
\bar{\psi}^j(0)\gamma_\nu S^{ji}(0,x) \gamma_\mu \psi^{i}(x)
\}\mid 0 \rangle\,. 
\label{pimunu4}
\eea
In the above, the fields $\psi$ and $\bar{\psi}$ 
have to be treated as external vacuum fields with  
negligible momenta as compared with the momenta of the freely
propagating off-shell quarks. In other words, it is
possible to expand $\bar{\psi}(x)$ and $\psi(x)$
around $x=0$ :
\bea
\psi(x) = \psi(0) + x^\rho\stackrel{\rightarrow}{D}_\rho\psi(0)+ ..\,,
\nonumber 
\\
\bar{\psi}(x) = \bar{\psi}(0) + \bar{\psi}(0) \stackrel{\leftarrow}{D}_\rho
x^\rho+ ..~,
\label{expans}
\eea
where $D_\rho$ is the covariant
derivative, and the higher orders in this expansion are only relevant for
the operators with $d\geq 5$.
Substituting (\ref{expans}) in (\ref{pimunu4}), one encounters 
the following vacuum matrix elements:
\be
\langle 0 \mid \bar{\psi}^i_\alpha\psi^{j}_\beta \mid 0 \rangle= 
A\delta^{ij}\delta_{\alpha\beta}\,, 
\label{AA}
\ee
\be
\langle 0 \mid 
\bar{\psi}^i_\alpha\stackrel{\rightarrow}{D}_\rho\psi^{j}_\beta
\mid 0 \rangle =B\delta^{ij}(\gamma_\rho)_{\beta\alpha}\,, ~~~
\langle 0 \mid \bar{\psi}^i_\alpha\stackrel{\leftarrow}{D}_\rho
\psi^{j}_\beta(x)
\mid 0 \rangle =\overline{B}\delta^{ij}(\gamma_\rho)_{\beta\alpha}\,, 
\label{BB}
\ee
where $\alpha,\beta$ are the bispinor indices, and  the r.h.s. represent
the most general decompositions in color and Dirac matrices, obeying
 color and spin conservation.
The constants $A$,$B$ and $\overline{B}$ are easily obtained by multiplying
both sides of Eqs. (\ref{AA}) and (\ref{BB}) by
$\delta^{ij}\delta_{\alpha\beta}$ and 
$\delta^{ij}(\gamma^\rho)_{\alpha\beta}$, respectively,
and taking traces. The result is :
\be
A= \frac1{12}\langle 0 \mid \bar{\psi} \psi \mid 0 \rangle\,,
\ee
\bea
B= \frac{1}{48} \langle 0 \mid \bar{\psi} 
\not\!\stackrel{\rightarrow}{D} \psi \mid 0 \rangle
= -\frac{im}{48}\langle 0 \mid \bar{\psi} \psi \mid 0 \rangle
\,,~~\overline{B}=-B.
\eea
In the last two relations the Dirac equation
for the quark field $\not\!\stackrel{\rightarrow}{D}\psi(x) =-im\psi(x)$
was applied.    
Substituting the expansion (\ref{expans}) in (\ref{pimunu4}) and 
using the expressions (\ref{AA}) and (\ref{BB}) for the matrix elements, 
one obtains, after the integration over $x$:
\be
\Pi_{\mu\nu}^{(\bar{\psi}\psi)}(q)=
(q_\mu q_\nu -q^2g_{\mu\nu})
\frac{2m}{q^4} \langle 0 \mid \bar{\psi}\psi\mid 0\rangle\,,   
\ee
yielding the Wilson coefficient
\be
C_3(q^2)= \frac{2m}{q^4}\,.   
\label{qcondensate}
\ee
The proportionality of the above expression 
to the quark mass is expected 
on general grounds. The quark condensate
violates chiral symmetry and its contribution 
should vanish in the chiral limit $m=0$.

The derivation of higher dimensional terms of the 
OPE (\ref{ope112}) is more involved. A very useful 
tool, simplifying the calculational procedure, 
is the Fock-Schwinger gauge for 
the gluon field: 
\be
(x-x_0)_\mu A^{a\mu}(x)=0 \;\;.
\label{gauge}
\ee
In this gauge, the gluon 4-potential $A^a_\mu$ is directly 
expressed in terms
of the gluon field-strength tensor $G_{\mu\nu}^a$. 
This, and many other aspects of the calculational  technique  
are explained in the review~\cite{NSVZtechn}
serving as a very useful handbook for QCD sum rule practitioners
(see also Ref. 15). 

The final result for the OPE, 
with  all Wilson coefficients up to $d=6$ taken 
into account, reads: 
\bea
\Pi(q^2)\!&=&\! -\frac{1}{4\pi^2}
\left(1+\frac{\alpha_s}{\pi} \right) \ln \frac{-q^2}{4m^2}
+ \frac{2m \langle \bar{\psi}\psi\rangle }{q^4}+
\frac{\alpha_s \langle  G_{\mu\nu}^a G^{a\mu\nu} \rangle}{12\pi q^4}
\nonumber
\\
\!&&\!
+\frac{m^3}{3q^8}\langle g_s \bar \psi\sigma_{\mu\nu} 
\frac{\lambda ^a}2 G^{a\mu\nu}\psi\rangle
+ \frac{2\pi\alpha_s}{q^6}\Big[
\langle (\bar{\psi}\gamma_\mu\gamma_5\frac{\lambda^a}2\psi) 
(\bar{\psi}\gamma^\mu\gamma_5\frac{\lambda^a}2\psi) \rangle
\nonumber
\\
\!&&\!
+ \frac{2}9\langle (\bar{\psi}\gamma_\mu\frac{\lambda^a}2\psi) 
(\bar{\psi}\gamma^\mu\frac{\lambda^a}2\psi)\rangle\Big]\,,
\label{expanlight}
\eea
where the shorthand notation 
$\langle O \rangle \equiv \langle 0 \!\mid O \!\mid \!0\rangle$
is introduced. The above expression is 
valid for the light quarks $\psi=u,d,s$. In this case the quark-gluon
condensate contribution (diagram in Fig.~\ref{fig:cond}e) 
is suppressed by an extra factor $m^2/Q^2$. 
Note that the three-gluon condensate term vanishes 
for the correlators of massless quarks.\cite{DubSmil} 
The accuracy of the OPE for $\Pi(q^2)$  is not limited
by the above expression. Currently, 
$\Pi^{(pert)}$ is known up to $O(\alpha_s^3)$,\cite{kataev} 
and the $O(\alpha_s)$ corrections to many of the coefficients $C_n$ 
at $d\neq 0$ are also available.
If the quark is heavy ($\psi =c,b$), 
the quark condensate terms are 
suppressed~\footnote{Heavy quarks do not develop 
their own vacuum condensates, being far off-shell at the momentum scale 
$\Lambda_{vac}$, and the interactions of virtual heavy quarks 
with the light-quark condensates appear in higher order in 
$\alpha_s$.} and one has, to $d=4$ accuracy:\cite{SVZ79}  
\bea
\Pi(q^2)&=& 
\frac{q^2}{\pi} 
\!\int\limits_{4m^2}^{\infty} \! ds 
\frac{\mbox{Im}\,\Pi^{(pert)}(s)}{s(s-q^2)} 
+\frac{\langle \alpha_s G_{\mu\nu}^a G^{a\mu\nu} \rangle}{48\pi q^4}
f(a)\,, 
\label{expanheavy}
\eea
where $a=1-4m^2/q^2$,
$$
f(a)=\frac{3(a+1)(a-1)^2}{2a^{5/2}}\ln\frac{\sqrt{a}+1}{\sqrt{a}-1}
-\frac{3a^2-2a+3}{a^2}\,,
$$
and $\mbox{Im}\,\Pi^{(pert)}(s)$ is given in (\ref{PiMcorr}).
Note that at $q^2=0$ the gluon condensate term has 
a suppression factor $1/(4m_{c,b}^2)^2$ with respect 
to the perturbative part. 
The Wilson coefficients of $d=6,8$ terms in the OPE for the heavy quark correlator, 
corresponding to diagrams with three and four vacuum gluons
(e.g. the three-gluon condensate 
contribution) have also been calculated.\cite{Rad3G}

\subsection{What do we know about the vacuum condensates ?} 

The vacuum condensates introduced in the OPE 
are purely nonperturbative parameters, hence, their 
numerical values (the condensate densities) cannot be directly 
calculated and have to be determined by other methods, to be
briefly summarized now.

The quark condensate plays a special role  
being responsible for the observed spontaneous breaking of the chiral
symmetry in QCD (discussed in the chapters by Leutwyler,
by Diakonov and Petrov, and by Meissner in this book). 
For this reason, the value of the 
quark condensate density was known long before it was used 
in QCD sum rules:
\be
\langle \bar \psi \psi   \rangle = 
-\frac{f_\pi^2m_\pi^2}{2(m_u +m_d)}
\simeq-(240 \pm 10 ~\mbox{MeV})^3\,,
\label{condens}
\ee 
for $\psi=u,d$ and, in SU(3)-flavor approximation, also for
$\psi=s$. The above value corresponds to the normalization scale 
$\mu=1$ GeV.\footnote{ The condensate density is logarithmically dependent 
on the normalization scale if the underlying operator, in this case
$\bar{\psi}\psi$, has a nonvanishing anomalous dimension.
A natural choice is the scale $\mu$ separating the long and 
short distances in OPE.}

From first principles, very little is known about other condensates. 
Some attempts exist to calculate them on the lattice 
or in the models of the instanton vacuum.
At present, it is still more safe, as it was done in the 
original work,\cite{SVZ79} to determine the condensate  densities
empirically, by fitting certain QCD sum rules to experimental data. 
Being universal, the condensates extracted from one sum rule
can be used in many others. 

The gluon condensate density  was originally derived from 
the sum rule (\ref{moments}) 
for the correlation function of $j_\mu=\bar{c}\gamma_\mu c$
currents. The hadronic spectral
density was saturated by the 
experimentally known masses $m_V$ and decay constants $f_V$ 
of the charmonium levels  $ V=J/\psi,\psi',...$, employing 
quark-hadron duality (explained in the 
next subsection) for the heavier states.
Substituting the correlation function
(\ref{expanheavy}) in the l.h.s. of (\ref{moments}), and fitting it
to the r.h.s. at not very large $n$ (in order to safely 
neglect $d\geq 6$ terms) the 
gluon condensate density, 
\be
\langle \frac{\alpha_s}{\pi}
G^a_{\mu\nu}G^{a\mu\nu} \rangle = (0.012 ~\mbox{GeV}^4)\pm 30\%\,,
\label{glue}
\ee
has been obtained.\cite{SVZ79} 
Note that the above quantity is scale-independent.
The estimate (\ref{glue}) has survived after many years, although
claims urging to revise it appear from time to time in 
the literature. In fact, an independent
check of Eq.~(\ref{glue}) has been carried out~\cite{NSVZUse} 
by considering SVZ sum rules for two different correlators
in the pion channel, one of 
them sensitive to the gluon condensate and the 
other one to the quark condensate.

The quark-gluon condensate is  ``invisible'' in the 
sum rules for vector mesons and has to be extracted 
from the correlation functions for baryon currents 
(they are considered in Sec.~3). The conventional parametrization is 
\be
\langle g_s \bar \psi \sigma_{\mu\nu}
\frac{\lambda^a}{2}G^{a\mu\nu}\psi \rangle =
m_0^2 \langle  \bar \psi \psi  \rangle\,,
\label{qqbarG2}
\ee
where the numerical value of $m_0^2$ has been estimated long ago~\cite{IoffeBelyaev} 
and is still accepted:
\be
m_0^2(\mbox{1 GeV}) = 0.8 \pm 0.2 ~\mbox{GeV}^2\,.
\label{m02}
\ee

Four-quark condensates with different 
combinations of $\Gamma_r$ matrices (one of the 
corresponding diagrams is shown in Fig.~\ref{fig:cond}f) 
can be treated in the  
factorization approximation~\cite{SVZ79}  which relies 
on the dominance of the intermediate vacuum state
and allows to reduce each separate four-quark condensate
to the square of the quark condensate. The general factorization
formula reads:
\be
\langle
 \bar{\psi}\Gamma_rq \bar{\psi}\Gamma_s \psi\rangle=
\frac{1}{(12)^2} \{ (Tr\Gamma_r)(Tr\Gamma_s)-Tr(\Gamma_r\Gamma_s) \}
\langle  \bar{\psi}\psi\rangle^2~.
\label{vac4q2}
\ee
Applying it to the four-quark condensate terms 
in (\ref{expanlight}) one obtains, instead of the two terms
in the square bracket, a more compact expression
$
\frac{112}{81}\langle \bar \psi \psi  \rangle^2~. 
$
The product of the four-quark condensate and $\alpha_s$ 
in (\ref{expanlight}) has a negligible scale-dependence.

The three-gluon condensate contribution is not important for 
most of the interesting sum rules. An order of magnitude 
estimate of its density based on the instanton model is~\cite{SVZ79}
\be
\langle g_s^3 f_{abc} G_{\mu\nu}^a G^{b \, \nu}_ \sigma G^{c\sigma\mu}
\rangle \simeq 0.045 \;\; \mbox{GeV}^6~.
\ee
In the context of a recent development, let us mention the estimates 
of $d=7$ condensates obtained in the model of the instanton 
vacuum~\cite{MPolyakov} and using a 
factorization ansatz.\cite{Oganesian} 

One should admit that  
the accuracy of the condensate densities 
obtained from sum rules and/or invoking factorization 
is not very high and there is still room for 
a considerable improvement. 
Because of that, one cannot fully benefit from the 
improved Wilson coefficients
available  from  perturbative calculations. 
On the other hand, as we already noticed, perturbative 
loop diagrams include regions of soft momenta. Therefore, 
any update of condensates 
has to be combined with a more delicate procedure identifying and 
separating these regions in the perturbative 
coefficients, in spirit of the ``practical version'' of OPE. 
Regardless to their particular relevance 
for SVZ sum rules, the vacuum condensates are important 
nonperturbative characteristics of QCD in general, thus deserving
dedicated studies.

\subsection{Use of quark-hadron duality} 

Let us continue  our derivation.
Performing the Borel transformation of
the Eq.~(\ref{expanlight}), the QCD answer for $\Pi(M^2)$  
in the sum rule (\ref{BorelSR}) can now be obtained. 
The result reads:
\bea
f_V^2e^{-m_V^2/M^2} + 
\int\limits_{s^h_0}^\infty\!ds ~\rho^h(s)e^{-s/M^2}
=
\frac{1}{4\pi^2}\left(1+\frac{\alpha_s(M)}{\pi}\right)
\int\limits_{0}^{\infty}\! ds~e^{-s/M^2} 
\nonumber
\\
+ \frac{2m\langle\bar{\psi}\psi\rangle}{M^2}  
+\frac{\langle \frac{\alpha_s}{\pi} G^a_{\mu\nu}G^{a\mu\nu} 
\rangle}{12M^2} 
-\frac{112 \pi}{81} \frac{\alpha_s\langle \bar{\psi} \psi \rangle^2 }{M^4}\,.
\label{sample}
\eea
In the above,  the four-quark condensates are factorized and 
$\alpha_s$ is taken at the scale $M$.\footnote{To do it more precisely, 
the Borel transformation has to be applied to the 
logarithmic  dependence of $\alpha_s$ on the scale $q^2$.}

In addition, it is possible to estimate the integral over the 
excited and continuum states in Eq.~(\ref{sample}) using
the following arguments. In the deep spacelike region 
$q^2 \to -\infty$, where 
all power-suppressed condensate contributions can safely be neglected,   
the limit $\Pi(q^2) \to \Pi^{(pert)} (q^2)$  
is valid yielding an approximate equation of the corresponding 
dispersion integrals: 
\be
q^2 \!\int\limits_{t_{min}}^\infty\!ds \frac{\mbox{Im} \Pi(s)}{s(s-q^2)}\simeq 
q^2 \!\int\limits_{4m^2}^\infty\!ds \frac{\mbox{Im} \Pi^{(pert)}(s)}{s(s-q^2)}\,,
\label{displimit}
\ee
(again at $q^2 \to-\infty$). In order to satisfy the above 
relation, known as the {\em global quark-hadron duality},\footnote{For a more 
detailed
discussion of quark-hadron duality see the chapter by Shifman in this book.}
the integrands on both sides of it should have the same asymptotics:
\be
\mbox{Im}\,\Pi(s) \to \mbox{Im}\, \Pi^{(pert)}(s)
~~~\mbox{at}~s \to +\infty \,,
\label{locald}
\ee
where, in our case, $\mbox{Im}~\Pi(s)$ is given by Eq.~(\ref{higher})
and $\mbox{Im}~\Pi^{(pert)}(s)$ by Eq.~(\ref{PiMcorr}).
One could still allow Im~$\Pi (s)$ to oscillate 
around the perturbative QCD limit.
Eq.~(\ref{locald}) is an example of the {\em local} 
quark-hadron duality.
From (\ref{displimit}) and (\ref{locald}) 
it is postulated that at sufficiently large $Q^2=-q^2$ 
the following approximation is valid: 
\be
q^2\int\limits_{s_0^h}^{\infty} ds \frac{\rho^h(s)}{s(s-q^2)}\simeq 
\frac1{\pi}q^2\int\limits_{s_0}^{\infty} 
ds\frac{\mbox{Im}\, \Pi^{(pert)}(s)}{s(s-q^2)}\,,
\label{dual1}
\ee
where $s_0$ is an effective threshold parameter 
which does not necessarily coincide with $s_0^h$.
After the Borel transformation one obtains:
\be
\int\limits_{s_0^h}^{\infty} ds \,\rho^h(s)e^{-s/M^2}\simeq 
\frac1{\pi}\int\limits_{s_0}^{\infty} 
ds\,\mbox{Im}\, \Pi^{(pert)}(s)e^{-s/M^2}\,.
\label{Boreldual}
\ee
The relation (\ref{dual1}) and its Borel transformed version 
(\ref{Boreldual}) represent the  
quark-hadron duality approximation used in 
SVZ sum rules to replace the integrals over excited and continuum states.  
The threshold parameter $s_0$  which, in general, has to be fitted, 
is expected to be close to the mass 
squared of the first excited state of $V$. 

The assumption (\ref{dual1}) is certainly weaker than  
local duality, because it involves
integrals and, in particular, it is insensitive to oscillations 
of $\rho^h(s)$ around $\mbox{Im}~\Pi^{(pert)}(s)$. Moreover, at 
very large $Q^2$, for the positive definite 
correlation functions, such as  (\ref{pimunu}),
the validity of (\ref{dual1})  is simply 
a mathematical consequence of the global duality.
On the other hand, employing the duality approximation
(\ref{dual1}) in sum rules at intermediate $Q^2$ (or,
equivalently, Eq.~(\ref{Boreldual}) 
at intermediate $M^2$) one also relies
on the validity of the local duality 
approximation (\ref{dual1}),
starting from a certain finite $s_0$. The latter assumption is not 
a strict consequence of the asymptotic freedom.
It is therefore fair to call Eq.~(\ref{dual1}) ``semilocal'' 
duality. In this situation,   
the Borel transformation performed in the sum rule 
(\ref{sample}) is of a great importance, 
because, due to the exponential suppression of the
integral on the l.h.s., the sensitivity 
to the duality approximation (\ref{Boreldual}) 
of this integral is not high. 

It is also important to recall that quark-hadron duality 
was confirmed in the channels accessible in $e^+e^-$ annihilation and 
in $\tau$ lepton decays, where the  
spectral density $\rho^h(s)$ of excited and continuum states was 
measured and the  hadronic dispersion integrals were compared 
with their perturbative QCD counterparts. 
For example, using the experimental data 
on $J/\psi,\psi',..\to l^+l^-$ decay widths, 
the  quark-hadron duality for the charmonium channel 
was shown to hold to good accuracy.\cite{GeshkenMarinov} 

Using  the duality approximation (\ref{Boreldual}) 
in  Eq.~(\ref{sample}), one can simply subtract the 
integral over $\rho^h(s)$ from the perturbative part on the r.h.s. 
This is the last step in the derivation procedure outlined 
and explained in the previous subsections. Our goal is achieved: the SVZ sum rule 
for the parameters of the ground-state hadron $V$ can now be written 
explicitly. For definiteness, let as choose 
the case $V=\rho$.  
It corresponds to the correlation function 
(\ref{pimunu}), where the $I=1$ combination of 
$u$ and $d$ quark currents is taken:
\be
j_\mu^{(\rho)}=\frac12(\bar{u}\gamma_{\mu}u-\bar{d}\gamma_{\mu}{d})\,
\label{currrho}
\ee 
with the decay constant defined as in Eq.~(\ref{fV}): 
\be
\langle \rho^0(p) \mid j_\nu^{\rho} \mid 0 \rangle = 
\frac{f_\rho}{\sqrt{2}}m_\rho \epsilon^{(\rho)*}_\nu\,.  
\label{rhoconst}
\ee
The resulting sum rule reads:
\bea
f_\rho^2= M^2 e^{m_{\rho}^2/M^2}
\Big[\frac1{4\pi^2}\left( 1-e^{-s_0^\rho/M^2} \right) 
\left( 1+\frac{\alpha_s(M)}{\pi}\right) 
\nonumber
\\
+ \frac{(m_u+m_d)\langle\bar{\psi}\psi\rangle}{M^4}  
+\frac{1}{12}\frac{\langle \frac{\alpha_s}{\pi} G^a_{\mu\nu}G^{a\mu\nu}
\rangle}{M^4} 
-\frac{112\pi}{81} 
\frac{\alpha_s\langle \bar{\psi} \psi \rangle ^2 }{M^6}\Big]\,,
\label{SVZrho}
\eea
$s_0^\rho$ being the duality threshold for the $\rho$ meson 
channel.

\subsection{SVZ sum rules at work}

Historically, Eq.~(\ref{SVZrho}) was 
one of the first successful applications of the method. 
Recently, this sum rule was reanalyzed~\cite{Shifman98} 
highlighting many interesting theoretical details.   
Here we will follow a more pragmatic procedure, trying to reveal 
the predictive power and the actual accuracy of the method.

We start with discussing the choice of the Borel parameter. 
The sum rule (\ref{SVZrho}) is not
applicable at too small $M^2$ because the missing 
terms with higher-dimensional condensates and, potentially, 
also the short-distance 
nonperturbative effects, all proportional to large powers of $1/M^2$,
may become too important to be neglected. Usually, the low limit 
on $M^2$ is adopted by demanding that in the truncated OPE 
the condensate term with the highest dimension 
remains a small fraction  of the sum of all terms. 
This limit keeps the convergence of 
the condensate expansion under control and guarantees that  
one does not introduce a large error neglecting the higher-dimensional terms.
At too large $M^2$ the quark-hadron duality approximation cannot be trusted.
Therefore, one also has to choose an upper 
limit on $M^2$, so that the exponentially suppressed 
contribution of the states above $s_0^\rho$ 
remains a small part of the total dispersion integral. 
The value of $s_0^\rho$  is not completely arbitrary,  
being  correlated with the onset of 
excited states in the channel of the current $j^\rho_\mu$.
According to the experimental data, the resonance activity
related to the first excited states  
in the $\rho$ channel shows up at
$s \sim 1.5 \div 2.0 $ GeV$^2$, hence $s_0$ in this vicinity
can be expected. We have checked that in the sum rule 
(\ref{SVZrho}), in the range
\be 
0.5 < M^2 < 1.2~\mbox{GeV}^2 \;\;\;, 
\label{borelint}
\ee
the $d=6$ four-quark contribution is less than 10\% and,
simultaneously, the $s>s_0^\rho$ part of the dispersion 
integral is less than 30 \% of the total r.h.s..
One should emphasize that in certain sum rules the {\em Borel window} 
similar to Eq.~(\ref{borelint}) simply
does not exist, that is, the lower 
limit of $M^2$ overshoots the upper one. 
The channels where the sum rules fail are usually plagued 
by ``direct instantons'', and therefore the physical reason
of the failure is understandable. 

After the range of $M^2$ is determined, we can 
fit the decay constant $f_\rho$ and the threshold parameter $s_0^\rho$  
from Eq.~(\ref{SVZrho}) by demanding the maximal stability of 
$f_\rho$ within this range.
The mass $m_\rho$ is fixed by its experimental value.\footnote{The SVZ method  has enough predictive 
potential to yield also the $\rho$ meson mass $m_\rho$
by combining the sum rule (\ref{SVZrho}) with the 
same sum rule differentiated over $1/M^2$. We do not 
discuss this procedure here, concentrating on the  determination of
hadronic matrix elements such as $f_\rho$.}
%
\begin{figure}
\hspace{0.7cm}
\psfig{figure=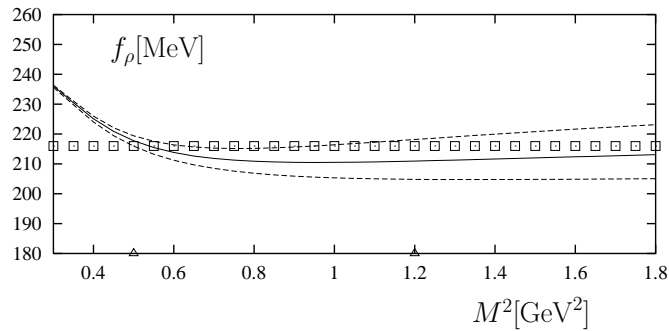,height=2.5in}
\vspace{-1cm}
\caption{ $\rho$ meson decay constant $f_\rho$
calculated from the SVZ sum rule (\ref{SVZrho}) 
as a function of the Borel parameter $M^2$, fixing the threshold at 
$s_0^\rho$=1.7 GeV$^2$ (solid line). For comparison, the 
results at $s_0^\rho$=1.5 GeV$^2$ and  $s_0$=2.0 GeV$^2$
are shown by the lower and upper dashed lines, respectively. 
The central experimental value is indicated by boxes, 
the triangles on the $M^2$ axis mark the allowed range
of the Borel parameter.}
\label{fig:frho}
\end{figure}
The resulting numerical values of $f_\rho$ are shown in Fig.~\ref{fig:frho}   
as a function of $M^2$, at the central values of the condensates
(\ref{condens}),(\ref{glue}) and at $\alpha_s(1 \mbox{GeV})=0.5$. 
Within the interval (\ref{borelint}), the maximally stable
$f_\rho$  corresponds to $s_0=1.7$ GeV$^2$.  

In order to assess  the predictive power
of the sum rule (\ref{SVZrho}), it is necessary to 
estimate the theoretical uncertainties. They
originate from the following sources:

(a) dependence on the Borel parameter.
The sum rule is an approximate relation, 
therefore it is not surprising that the final numerical result 
for the constant parameter $f_\rho$ changes with $M^2$. 
As can be seen in Fig.~\ref{fig:frho}, the variation of 
$f_\rho$  within the Borel window is small, 
\be
f_\rho^{Borel}= 210 \div 217 \;\;\mbox{MeV}, 
\label{fBorel}
\ee
and one can claim the success of the sum rule (\ref{SVZrho}). 
Indeed, a large instability with respect to the variation
of $M^2$  would indicate absence of 
important condensate contributions or may cast doubt over 
the reliability of the duality approximation.
Apparently, all values of $f_\rho$ in the interval (\ref{fBorel})
can be equally trusted as sum rule predictions, therefore, the variation within the Borel window has to be included in the total theoretical uncertainty. 

(b) inaccurate knowledge of the condensate densities. Varying them within 
errors indicated in Eqs.~(\ref{condens}) and (\ref{glue}) we obtain
a small, $\pm 1\%$ change of the prediction (\ref{fBorel}).

(c) neglect of the $d\geq 6$ terms in OPE. We assume 
that the neglected condensate terms are 
altogether not larger than the $d=6$ contribution.
This yields less than $\pm 5 \%$ uncertainty in $f_\rho$.   

(d) limited accuracy of perturbative contributions. Varying 
the scale of $\alpha_s$ and adding  the known higher-order 
corrections to the perturbative part one again observes 
a small, about $\pm 3\%$, variation of $f_\rho$. Absence of  $O(\alpha_s)$ 
corrections to $d\neq$0 Wilson coefficients has
a negligible impact on the result. 

Adding the individual uncertainties (a)$-$(d) linearly, 
which is a rather conservative attitude, 
we get about $\pm 10\%$  in total and obtain the following interval
for the SVZ sum rule prediction:
\be
f_\rho = 213 \pm 20~\mbox{MeV}\,, 
\label{finalfrho}
\ee
where the form of writing is just for convenience 
and there is no preferable central value.
The uncertainties (c),(d) may   
somewhat decrease after future theoretical work. 
Note that the variation of $s_0^\rho$ should not be counted
as an independent uncertainty, since this parameter is fitted from the sum 
rule together with $f_\rho$. 
On the other hand, it is helpful to have several alternatives 
for the quark-hadron duality ansatz, including the radial 
excited or continuum states and increasing the effective threshold. 
In the $\rho$ channel one can try different models, e.g., $\rho+\rho'+$continuum with higher $s_0^\rho$, including 
also the finite widths of these resonances. The results for $f_\rho$ 
are practically very close to what one obtains with 
the simplest SVZ construction: $\rho$ + continuum.
Our estimate (\ref{finalfrho}) is in a good agreement with 
the experimental number 
\be
f_\rho^{exp} = 216 \pm 5~\mbox{MeV} 
\ee
obtained from the measured leptonic width~\cite{pdg00} 
$\Gamma(\rho^0 \to e^+e^-)= 6.77 \pm 0.32$ keV. 
\begin{figure}
\hspace{0.7cm}
\psfig{figure=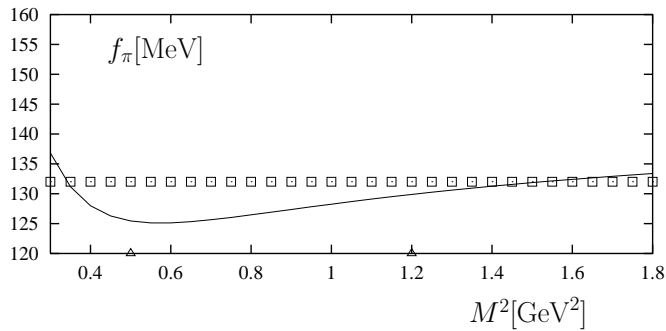,height=2.5in}
\vspace{-1cm}
\caption{ The decay constant $f_\pi$
calculated from the SVZ sum rule at 
$s_0$=0.7 $GeV^2$ (solid line). 
For comparison, the experimental value
is indicated with boxes, the triangles at 
the $M^2$ axis mark the allowed range
of the Borel parameter.}
\label{fig:fpi}
\end{figure}

Another  illustration of the method is provided by the SVZ sum rule
for the pion decay constant $f_\pi$, defined as 
\be
\langle \pi(p)|j_\mu^{(\pi)}|0\rangle=
-ip_\mu f_\pi \,,
\label{pionconst}
\ee
where 
\be
j_\mu^{(\pi)}= \bar{u}\gamma_\mu\gamma_5 d
\label{eq:axialcurrent}
\ee
is the axial-vector current interpolating the pion.
The sum rule~\cite{SVZ79} obtained from the correlation
function of two $j_\mu^{(\pi)}$ currents
looks very similar to Eq.~(\ref{SVZrho}):
\bea
f_\pi^2= M^2 
\Big[\frac1{4\pi^2}\left( 1-e^{-s_0^\pi/M^2} \right) 
\left( 1+\frac{\alpha_s(M)}{\pi}\right) 
\nonumber
\\
+\frac{1}{12}\frac{\langle \frac{\alpha_s}{\pi} G^a_{\mu\nu}G^{a\mu\nu}
\rangle}{M^4} 
+\frac{176\pi}{81} 
\frac{\alpha_s\langle \bar{\psi} \psi \rangle ^2 }{M^6}\Big]\,,
\label{SVZpi}
\eea
but, importantly, differs in the sign of the four-quark condensate term.
This difference between the nonperturbative interactions of the currents 
$j_\mu^{(\rho)}$ and $j_\mu^{(\pi)}$ with the QCD 
vacuum, to a large extent explains why the hadrons in the 
vector and  axial channels are not alike, i.e. why there is a single $\rho$ 
resonance in the first channel and a combination of pion with $a_1$ meson in 
the second.\cite{SVZ79} 
The sum rule prediction with the estimated uncertainty,
\be
f_\pi= 127 \pm 15~\mbox{MeV}\,,
\ee
is in excellent agreement with the experimental 
number $f_\pi=132 $ MeV. 
Many other  ``classical'' examples of the SVZ method  
can be found in the original papers~\cite{SVZ79} or 
in the reviews.\cite{RRY85} 

There are two important messages from the above analysis.
The first one is that  QCD sum rules are (and will remain) approximate and their accuracy cannot be improved beyond certain limits.
The second message is more encouraging. Within  the SVZ method
one is able to estimate the 
theoretical uncertainty of the predicted hadronic parameter. 
Similar  estimates are impossible,
e.g., in many quark models of hadrons,
where the inputs are nonuniversal and have no direct relation to  
QCD. 


\section{Applications and development of the method} 

In the past two decades, QCD sum rules  have been applied to
many problems of hadron physics, with an accuracy, 
in several cases, substantially improved with respect 
to the original analyses. An incomplete list of applications includes:
\begin{itemize}
\item determination of the light ($u$, $d$, $s$) and 
heavy ($c$, $b$) quark masses; 
\item masses and decay constants of light and heavy mesons and baryons; 
\item form factors of mesons and baryons ;
\item valence quark distributions and spin
structure functions of the nucleon;
structure functions of the photon, $\rho$ meson and  pion;
\item  hadronic matrix elements relevant for the description
of $K^0-\overline{K^0}$, $B_d-\overline{B}_d$, 
$B_s-\overline{B}_s$ mixing;
\item  strong couplings and magnetic moments of mesons and baryons; 
\item
calculation of the parameters of effective theories, such as
chiral perturbation theory ($\chi$PT), heavy quark effective theory (HQET),
nonrelativistic QCD (NRQCD);
\item
spectroscopy and properties of non $q \bar q$ hadrons (gluonia, hybrids);
\item 
hadrons in nuclear matter, properties of hadronic matter at 
high temperature and  density.
\end{itemize}

It is hardly possible to include in one review an exhaustive detailed presentation 
of all important results obtained in twenty years of investigations. 
For the interested reader, a lot of important information, not covered here, 
is accumulated in the papers listed in the bibliography. 
In this Section we focus only on  a part of the topics listed above, 
with an  emphasis  on the results which seem to be particularly important for
current and future experimental studies.\footnote{ For an earlier survey, see 
the Appendix on QCD sum rules 
in the ``The BaBar physics book''.\cite{babarbook} Some of the results 
collected there are updated here.} Our intention is also to discuss the possibilities to
improve these results and to outline the problems that still
need to be solved.

Before starting  the presentation, let us again emphasize that 
QCD sum rules are approximate relations. The predictions  
obtained from their analyses are characterized by uncertainties which are
estimated by varying the input parameters 
($\alpha_s$, condensates, Borel parameters, etc.)
in the allowed ranges. 
For a convenient comparison with the results obtained 
by other methods, we will 
quote the sum rule predictions  as: central value $\pm$~uncertainty. 
The latter cannot be considered as a statistical error.
It just indicates  the range of variation of the result
from the (central value $-$ uncertainty) to the (central value + uncertainty). 

\subsection{Light quark masses: $m_u, \; m_d, \; m_s$}
A precise determination of the quark masses 
is a task of paramount importance 
for the Standard Model and its extensions.
Chiral perturbation theory allows to determine the ratios of
the light quark masses:
\begin{equation}
{ m_s \over  m_d} = 18.9 \pm 0.8 \,, \;\; 
{ m_u \over m_d } = 0.553 \pm 0.043 \,, \;\;
{m_s\over {1\over 2} (m_u+m_d)}=24.4 \pm 1.5 \,,
\label{cratios}  
\end{equation}
obtained at the next-to-leading order from
the measured masses of the pseudoscalar mesons.\cite{leut96}
The determination of the individual quark masses has attracted a lot of
attention in the QCD sum rule community. The 
correlators of the divergence of light quark ($u$, $d$) 
axial-vector  currents have been studied
to calculate $m_u+m_d$.  In addition, 
correlators of the divergence of vector and axial-vector
strange currents have been 
analyzed to determine $m_s$, using  experimental
information on $K \pi$ and $K \pi \pi$ resonances.  

Let us present in more detail this set of results. For
further use, it is important to mention that the quark masses
depend on the normalization scale $\mu$ through the 
renormalization group equation:
\begin{equation}
\mu {d \over d \mu} m_q(\mu)=-\gamma(\alpha_s)m_q(\mu) .
\label{rge}
\end{equation}
The scale dependence  can be expressed as:
\begin{equation}
m_q(\mu)=\hat m_q R(\mu) , 
\label{mqmu}
\end{equation}
where  $\hat m_q$ is a  renormalization invariant parameter and
\be
R(\mu)=\Bigg({\beta_0 \over 2} {\alpha_s(\mu) \over \pi} 
\Bigg)^{2 \gamma_0 /\beta_0} \Bigg\{1+ \Bigg(2 {\gamma_1 \over \beta_0} -
{\beta_1 \gamma_0 \over \beta_0^2} \Bigg){\alpha_s(\mu) \over \pi} +
{\cal O}(\alpha_s^2) \Bigg\}\hskip 1pt;
\label{r}
\ee
$\gamma_i$ and $\beta_i$ are
the first coefficients of the mass anomalous dimension $\gamma$
and of the QCD $\beta$ function, respectively;
in the $\overline{MS}$ scheme, for three colors and $N_f$ flavors,
they read:
\begin{equation}
\gamma_0=2 \hskip 3 pt,\hskip 3 pt
\gamma_1={101 \over 12}-{5 \over 18 } N_f \hskip 3 pt,
\label{eq:gamman}
\end{equation}
and 
\begin{equation}
\beta_0=11-{2 \over 3 } N_f \hskip 3 pt,\hskip 3 pt
\beta_1=51-{19 \over 3 } N_f \hskip 3 pt.
\label{betan}
\end{equation}

In order to determine $m_u + m_d$
one can study the two-point correlation function
\begin{equation}
\Psi_5(q^2)= 
i \int d^4 x \; e^{i q \cdot x} \langle 0|T\{j_5(x) j_5^\dagger(0)\} |0 \rangle 
\,,
\label{cormu}
\end{equation}
$j_5$ being the divergence of the axial current (\ref{eq:axialcurrent}): 
$j_5=\partial_\mu ( \bar u \gamma^\mu \gamma_5 d)$. The correlator
(\ref{cormu}) is particularly sensitive to the $u$ and $d$ quark masses,
since $j_5=(m_u + m_d){\overline u} i \gamma_5 d $.

In the perturbative contribution to (\ref{cormu}) 
up to  4-loop corrections~\cite{kataev,chetyrkin97a} are available.
In the condensate expansion terms up to $d=6$ has been taken into account. 
The hadronic spectral function  is
expressed in terms of the pion pole and of 
higher resonances, and is constrained to satisfy the behavior predicted by 
chiral perturbation theory at the $(3\pi)$ threshold.

In Table~\ref{table:mq} we present a set of the results for
${\overline m_u} +{\overline m_d}$, in the 
$\overline {MS}$ scheme at  $\mu=1$ GeV
(see also Refs.~28--30). 
In the same Table we also include recent lattice QCD determinations, which
are typically provided for the combination 
$({\overline m_u}+{\overline m_d})/2$ 
at  $\mu=2$ GeV;  the corresponding values at  $\mu=1$ GeV 
are obtained using Eqs.~(\ref{r})-(\ref{betan}).
\begin{table}[h]
\begin{center}
\begin{tabular}{| c | c | c |}
\hline
$({\overline m_u}+{\overline m_d})(\mu=1$ GeV) (MeV) & Ref. & comments\\ \hline 
                &                            &             \\
$15.2 \pm 2.0 $ & DR87~\cite{dominguez87} &   LO       \\
$15.6 \pm 3.4$  & N89~\cite{N89}          &   ``       \\
                &                            &            \\
$12.8 \pm 2.5$  & P98~\cite{prades98}  &   ${\cal O}(\alpha_s^3)$  \\

                &                            &            \\
\hline \hline
                &                            &             \\
$9.5  \pm 1.6$  &G97~\cite{gough97}      & (quenched) lattice QCD \\
$12.6 \pm 1.3$  & BGLM00~\cite{BGLM00}    & ``                     \\
$11.6 \pm 1.1$  & A00~\cite{alikan00}    & ``                     \\
                &                            & \\
\hline
\end{tabular}
\caption{QCD sum rule results for 
${\overline m_u} + {\overline m_d}$ ($\mu=1$ GeV), compared to
the
recent lattice QCD determinations (renormalized to the same scale).
\label{table:mq}}
\end{center}
\end{table}
One observes that, although the uncertainty has not substantially 
improved in the years, the most recent sum rule determination of 
${\overline m_u} +{\overline m_d}$
predicts a value smaller than 
the previous ones. This is related to
the account of higher order QCD corrections and to the improvement
of the hadronic spectral function. 
Notice that, being combined with the chiral ratios in Eq.~(\ref{cratios}), the
results in Table~\ref{table:mq} allow one to
individually determine $m_{u}$, $m_{d}$ and $m_s$.
 
A similar analysis can be applied to determine the strange
quark mass. One considers the
two-point correlator of the divergences of the strangeness changing vector
current 
$j_S=\partial_\mu (\bar s \gamma^\mu u)$:
\begin{equation}
\Psi(q^2)= 
i \int d^4 x \; e^{i q \cdot x} \langle 0|T\{j_S(x) j_S^\dagger(0)\} |0 \rangle\,, 
\label{corrms}
\end{equation}
which is sensitive to $m_s$,  since 
$j_S=i (m_s-m_u)\bar s u$.

It is worth describing this  calculation in some detail.
In QCD, the second derivative 
 of the correlation function (\ref{corrms}),
$\Psi^{''}(q^2)= ( \partial^2/(\partial q^2)^2 ) \Psi(q^2)$,
 is obtained from the dispersion relation:
 \begin{equation}
\Psi^{''}(q^2) = {2 \over \pi} 
\int_0^\infty ds \;\; {{\rm Im} \; \Psi(s) \over (s - q^2)^3} \; \;.
\label{psisec}
\end{equation}
The hadronic spectral function 
$\displaystyle \rho(s)= {1 \over  \pi} {\rm Im} \; \Psi(s)$ can be obtained  
inserting a set of intermediate states with strangeness $|S|=1$ , $J^P=0^+$ and
$I={1 \over 2}$ in the correlator (\ref{corrms}), starting from the
two-particle states: $|K \pi\rangle$, $|K \eta\rangle$,
$|K \eta^\prime \rangle$, etc.
The  $|K \pi\rangle$ contribution  to $\rho(s)$ can be written as
\begin{equation}
\rho^{K\pi}(s) = {3 \over 32 \pi^2} {\sqrt{(s-s_+)(s-s_-)}\over s}
|d(s)|^2 \theta(s-s_+) ,
\end{equation}
where $s_\pm=(M_K \pm M_\pi)^2$. The function $d(s)$ is 
the scalar $K \to \pi $ form factor; in the 
low-s region  it admits
 a linear expansion, with the parameters fixed by the one-loop
chiral perturbation theory. Above the $K \pi$ threshold, the function  $d(s)$, 
hence $\rho^{K\pi}(s)$, can be reconstructed 
assuming the dominance of the scalar, $|S|=1$, 
$K^*_0(1430)$ and $K^*_0(1950)$  
resonances.\cite{jamin95,chetyrkin97} 
This procedure, however,  only partially
uses the experimental information on the  scalar $K \pi$
system, where a  sizeable  non-resonant  component is 
observed.\cite{LASS} 
Another possibility for reconstructing $\rho^{K\pi}(s)$
 consists in using the Omn\'es formula to account for 
information on the scalar $K\pi$ system
from the measured $I={1\over 2}$ $K\pi$ scattering phase shift. 
With this procedure, the
normalization condition of $\rho^{K\pi}(s)$
at the  $K \pi$ threshold is automatically satisfied.\cite{colangelo97}

Due to the positivity of the spectral function 
$\rho(s)$, the procedures described above yield either
a lower bound on $m_s$, if  only
the low hadronic states are taken into account, or a 
determination of $m_s$, if
the rest of contributions is  approximated by duality.
Concerning the perturbative 
contribution to the sum rule,  
early calculations included 
${\cal O}(\alpha_s^2)$  corrections; currently
the 
${\cal O}(\alpha_s^3)$ expression is
used.\cite{chetyrkin97,colangelo97}
Also the condensates up to $d=6$ are included in the OPE. 
The results  of this set of determinations are collected in 
Table~\ref{table:ms}.\cite{old_ms} 
The main uncertainty in the determination of $m_s$ is due
to the procedure of reconstructing the spectral function 
$\rho(s)$.\cite{gupta}

One could also use 
the correlator of the divergences of the strangeness changing axial current, 
proportional to ${m_s}+{m_u}$. The corresponding sum rule
has been investigated,\cite{jamin95,dominguez99} 
but the result for $m_s$ has a larger uncertainty caused by  
an insufficient information on  $K \pi \pi$ resonances
needed to  reconstruct  the  hadronic spectral function.

\begin{table}[ht]
\begin{center}
\begin{tabular}{| c | c | c |}
\hline
${\overline m_s}(\mu=1$ GeV) (MeV)  & Ref. & comments\\ \hline 
$171 \pm 15 $& CDPS95~\cite{CDPS95} & ${\cal O}(\alpha_s^2)$ +resonances\\
$189 \pm 32 $& JM95~\cite{jamin95}  &  ``  \\
$203.5 \pm 20$& CPS97~\cite{chetyrkin97}&${\cal O}(\alpha_s^3)$ +resonances\\
             &                           &     \\
$140\pm 20$  & CDNP97~\cite{colangelo97} &${\cal O}(\alpha_s^3)$ + Omn\`es   \\
$160\pm 30$ & J98~\cite{Jamin:1998sa}    &    `` \\
             &                           &       \\
\hline
$159\pm11$ & M99~\cite{maltman99} & FESR + Omn\`es  \\ 
\hline
$155 \pm 25$&DPS99~\cite{dominguez99}& axial current div. \\ 
\hline 
$200\pm40\pm30$ & CKP98~\cite{CKP98} & $\tau$ decays\\
$164\pm33$ & PP99~\cite{PP99} &                   `` \\
$176\pm37\pm13$ & KKP00~\cite{KKP00} &            `` \\
$158.6\pm18.7\pm16.3\pm13.3$ & KM00 \cite{KM00} & `` \\
\hline \hline
$184\pm26$&   E97~\cite{eicker97}&  (unquenched) lattice QCD\\
$124 \pm 21$& G97~\cite{gough97}&  (quenched) lattice QCD\\
$139\pm9$&   A99~\cite{aoki99}& `` \\
$138\pm5$&   G99~\cite{gockeler99}& `` \\
$146\pm12$&  BGLM00~\cite{BGLM00}& `` \\
$128\pm5$&   G00~\cite{garden00}& `` \\
$144\pm5$&   A00~\cite{alikan00}& `` \\
$117\pm9$&   A00~\cite{alikan00}& (unquenched) lattice QCD \\

\hline
\end{tabular}
\caption{QCD sum rule results for ${\overline m_s}(\mu=1$ GeV)
compared to recent lattice QCD determinations (renormalized to $\mu=1$ GeV).
\label{table:ms}}
\end{center}
\end{table}
In Table~\ref{table:ms} we also collect the outcome of
the analysis of strange and nonstrange hadronic $\tau$ decays.
In this case, $m_s$ is determined retaining in the OPE the
SU(3)-flavor breaking contributions,
which depend on the strange quark mass. 
The spread of  the results is  mainly caused by differences 
in the treatment of  
the higher-order perturbative terms.
The sum rule results can be compared with recent
lattice QCD  determinations of $m_s$, also given in Table~\ref{table:ms},
both in  quenched approximation and for dynamical fermions.

The compilation of results presented above clearly demonstrates
the unique ability of QCD sum rules
to analytically determine the light-quark masses, employing various
correlation functions and systematically including,
order by order, the contributions of the perturbative series and of the
nonperturbative condensates. The main limitation
is caused by an insufficient knowledge of the hadronic
spectral function in the scalar and pseudoscalar channels.
In principle, this difficulty can be solved  when
detailed experimental data become available.

\subsection{Heavy quark masses: $m_c$, $m_b$}
The study of the charmonium system was probably one of the first applications
of the QCD sum rule  method.\cite{SVZ79} 
The $c$ quark mass can be determined, if 
one considers the two-point correlation function of two 
${\bar c}\gamma_\mu c$ currents (discussed in Sec.~2), and uses for  
the hadronic spectral density
the  experimental data on the $e^+ e^-$ cross section into 
charm-anticharm states, 
in particular the precise information on the masses and the electronic
widths of the  $J^P=1^-$ charmonium levels ($J/\psi$, $\psi^\prime$, \dots).  
So far, the sum rule was analyzed including the perturbative QCD 
two-loop corrections.  The nonperturbative contribution mainly depends on the 
gluon condensate $\langle\alpha_s G^2\rangle$, while the contributions of other
vacuum condensates are believed to be small numerically.

\begin{table}[h]
\begin{center}
\begin{tabular}{| c | c | c | c |}
\hline 
$m_c$ (GeV) & ${\overline m_c}(\overline m_c)$ (GeV) &Ref.  &comments\\ \hline
$1.46 \pm 0.05 $ & &DGP94~\cite{paver94}  &Borel        \\
$1.42 \pm 0.03 $ &$1.23^{+0.02}_{-0.04}\pm 0.03$ &N94~\cite{narison94} &Borel, moments\\ \hline \hline 
&$1.525\pm0.040\pm0.125$&APE98~\cite{ape98}& lattice QCD\\
&$1.33\pm0.08          $&FNAL98~\cite{fnal98}& ``\\
&$1.20\pm0.04\pm0.11\pm0.2$&NRQCD99~\cite{nrqcd99}& lattice NRQCD\\
\hline 
\end{tabular}
\caption{QCD sum rule determinations of the $c$ quark mass and  recent
lattice QCD results.
\label{table:mc}}
\end{center}
\end{table}

In  Table~\ref{table:mc} we present two results for the charmed quark
pole mass $m_c$,\cite{old_mc} together with the recent 
lattice QCD determinations. 
The difference between the two analyses is that in the 
first one~\cite{paver94}  the
pole mass is directly computed, while in the second one~\cite{narison94} 
the $\overline{MS}$
running quark mass $\overline m_c$ is determined, and
then related to the pole mass. The two results are in agreement with each other, 
and with the estimate~\cite{SVZ79} 
$m_c(p^2=-m_c^2)=1.26$ GeV (having about $\pm 0.1$ GeV uncertainty) 
obtained in the original SVZ analysis. 
It would be important to update the sum rule determination of the charm
quark mass, including the $O(\alpha_s^2)$ perturbative 
correction~\cite{chetyrkin97b} and 
the three- and four-gluon condensate contributions.\cite{Rad3G}

In recent years, there has been an impressive progress in the determination 
of the $b$-quark mass, from the analysis of the two-point correlation
function of $\bar b\gamma_\mu b $ currents. In this analysis,
the hadronic spectral function is obtained taking into account six 
${\Upsilon}(nS)$ resonances, their masses and leptonic  decay constants 
being precisely measured in $e^+ e^-$ annihilation.
The current activity aims at  working with the highest 
possible moments of sum rules, where  NRQCD
is a good approximation. In this framework one is able to perform the Coulomb 
resummation,  taking into account relativistic and radiative corrections
order by order. Notice that the gluon condensate contributions are 
negligible in the heavy quarkonium system. 

In Table~\ref{table:mb} we present a 
summary of recent  determinations,\cite{old_mb}
together with an average of the lattice QCD results.\cite{hashimoto00}
From this Table we see that the preferred interval for the sum rule results is  
$m_b\simeq 4.8 \pm 0.1 $ GeV.

A detailed discussion of NRQCD and  its application to  
the $b$ quark mass problem is beyond our task, and the interest reader is addressed to 
more specialized reviews.\cite{heavymass}  
We only notice that there exists
an alternative approach, based on taking the first few moments of the
standard SVZ sum rules. In this case, the $b$-quark mass 
is determined in a purely relativistic way. The drawback is the
sensitivity to the tail of the hadronic spectral function corresponding
to the open beauty production at energies larger than the resonance 
masses. This spectral function can be handled by employing the duality approximation,
combining it with the  experimentally measured inclusive $e^+e^- \to \bar{b}b$ cross section.  
This kind of  analysis was employed in the early 
papers~\cite{RRY85} and still deserves attention, having in mind
the possibility of including  the recently calculated  
$O(\alpha_s^2)$  correction~\cite{chetyrkin97b}  which will improve 
the accuracy of the perturbative part.
\begin{table}[h]
\begin{center}
\begin{tabular}{| c | c |  c | c |}
\hline $m_b$ (GeV) & $\overline{m_b}(\overline{ m_b})$ (GeV)& Ref.  &
Method\\ \hline $4.72 \pm 0.05$ & & DP92~\cite{dominguez92} & 
standard SVZ
\\ $4.62 \pm 0.02$ & & N94~\cite{narison94} & `` \\ $4.827
\pm 0.007$& & V95~\cite{voloshin95} & NRQCD  \\ 
$4.84 \pm 0.08$ &$4.19 \pm 0.06 $&JP99~\cite{pich99} & ``\\ 
&$4.20 \pm
0.10 $&MY99~\cite{MY99} & `` \\ $4.88 \pm 0.10$ &$4.20 \pm
0.06 $&H99~\cite{H99} & `` \\ $4.80 \pm 0.06$ & &PP98~\cite{PP99a} &
`` \\ &$4.25 \pm 0.08 $&BS99~\cite{BS99} & `` \\
\hline \hline
&$4.26\pm0.11$&H00~\cite{hashimoto00}& lattice QCD (average)\\
\hline
\end{tabular}
\caption{The $b$ quark mass from SVZ and NRQCD sum rules and the lattice QCD average.
\label{table:mb}}
\end{center}
\end{table}

The task of an accurate evaluation of the $b$ quark mass is 
very important. As we shall see in the following, a precise determination of $m_b$ will
critically reduce the uncertainty in various sum rules for $B$ mesons.

\subsection{Heavy-light mesons}

The decay constants $f_{D_{(s)}}$ and $f_{B_{(s)}}$ of the heavy-light
mesons $D_{(s)}$, $B_{(s)}$ are defined by the matrix element of the
heavy flavored axial-vector current
\begin{equation}
<0|\bar q \gamma_\mu \gamma_5 Q |H(p)> = i f_H p_\mu\,,
\label{eq:hq}
\end{equation}
where $Q=c,b$, $H=D_{(s)}, B_{(s)} $ and $q=u,d (s)$, in the same 
normalization as $f_\pi$ defined in Eq.~(\ref{pionconst}).  
An accurate  calculation of these decay constants 
is very important  for heavy flavor
phenomenology, and QCD sum rules  were among
the first analytical methods to predict 
 $f_B$ and $f_D$.\cite{6auth,RRY85,oldfb} 
In these sum rules, perturbative ${\cal O}(\alpha_s)$ 
corrections and $d\leq 6$ condensates were taken in account.

We would like to emphasize two aspects. Firstly, the result for $f_B$ 
is very sensitive to the value of the $b-$quark
pole mass. The latter is extracted from the analysis of the 
$\Upsilon$ system discussed in the previous subsection.
Lowering the $b-$quark mass by $100$ MeV increases $f_B$ by
30 to 40 MeV. The heavy
quark mass dependence is less significant in the determination of $f_D$.
Secondly, the resummation of leading logarithmic
contributions in the sum rule for $f_B$,  
obtained in the framework of HQET, revealed that $\alpha_s$  
has to be taken at a low energy scale, $\mu \simeq 1$ GeV, rather then at
$\mu\simeq m_b$.\cite{broadhurst,neubert92} In full QCD with  finite 
$b-$quark mass, the optimal position of the scale, which has to be
determined from higher orders of QCD perturbation theory, remains an
unsolved problem.  Current sum rule calculations use a somewhat
intermediate scale, equal to the Borel parameter: $\mu^2 = M^2 \simeq 3-5$
GeV$^2$.  The shift from $m_b$ to lower scales produces another
significant increase of $f_B$. In the case of $f_D$ this effect is
again milder.

In Table~\ref{table:fb}, we collect several results and compare them with
the recent lattice QCD determinations.  Notice that the account of the
abovementioned effects has produced larger values for $f_B$ than in
the original calculations.
\begin{table}[h]
\begin{center}
\begin{tabular}{| c | c | c |  }
\hline $f_D$ (MeV) & $f_B$ (MeV) & Ref.  \\ \hline & & \\ $189 \pm 49$
& $158 \pm 25$& DP87~\cite{dominguez87a} \\ $173 \pm 22$ & $168 \pm
18$& D93~\cite{dominguez93} \\
             &              & \\
$200 \pm 20$ &$180 \pm 30$& KRWY99~\cite{KRWY2,KRWWY}\\
             &             &                     \\
\hline\hline
             &             &                      \\
$195\pm10^{+22}_{-10}$&$161\pm16^{+24}_{-13}$ & UKQCD99~\cite{UKQCD99} \\
             &             & \\
$216\pm11^{+5}_{-4}$&$173\pm13^{+34}_{-2}$    & APE00~\cite{APE00fb} \\
             &             &  \\
$220\pm3^{+2}_{-24}$&$218\pm5^{+5}_{-41}$      & UKQCD00~\cite{UKQCD00fb} \\
                    &                         & \\
\hline
\end{tabular}
\caption{$D$ and $B$ meson leptonic decay constants $f_D$ and $f_B$
from QCD sum rules and lattice QCD.
\label{table:fb}  }
\end{center}
\end{table}

The most recent sum rule results presented 
in Table~\ref{table:fb} correspond to the following intervals:
\begin{equation}
f_D = 180 \pm 30 \; {\rm MeV}\,,~~ 
f_B=  170 \pm 30 \; {\rm MeV} \,,
\end{equation}
where the rather large uncertainties suggest that there is still a room for an
improvement, as we shall argue below.

The  ratios $f_{D_s}/f_D$  and $f_{B_s}/f_B$  are calculated 
by including ${\cal O}(m_s)$, SU(3)-flavor symmetry breaking corrections
in the sum rules. In these ratios, the dependence on the heavy quark mass, as well as the
effects of the radiative ${\cal O}(\alpha_s)$ corrections, 
are less significant. 
A compilation of various  determinations is presented in 
Table~\ref{table:fbs}; again, one can
try to summarize the results as~\cite{babarbook}
\begin{equation}
f_{D_s}/f_D=1.19\pm 0.08\,,~~  f_{Bs}/f_B=1.16\pm 0.09 \,.
\end{equation}
\begin{table}[h]
\begin{center}
\begin{tabular}{| c | c | c | }
\hline
$f_{D_s}/f_D$  & $f_{Bs}/f_B$  &    Ref.        \\ \hline 
               &               &                \\
$1.21\pm0.06$& $1.22\pm0.02$& D93~\cite{dominguez93} \\
             & $1.09\pm0.03$& BCNP94~\cite{blasi94}     \\
$1.15\pm0.04$& $1.16\pm0.04$& N94~\cite{narison94a}  \\
$1.17\pm0.03\pm0.03$& $1.20\pm0.04\pm0.03$& HL96~\cite{huang96}  \\
                    &                     & \\
\hline\hline
                    &                     & \\
$1.15\pm0.04^{+0.02}_{-0.03}$&$1.16\pm0.06^{+0.02}_{-0.03}$&UKQCD99~\cite{UKQCD99}\\
                    &                     & \\
$1.11\pm0.01^{+0.1}_{-0}$&$1.14\pm0.02^{+0.}_{-0.1}$&APE00~\cite{APE00fb} \\
                    &                     & \\
$1.09\pm0.01^{+0.05}_{-0.02}$&$1.11\pm0.01^{+0.05}_{-0.03}$&UKQCD00~\cite{UKQCD00fb} \\
                         &                          &\\
\hline
\end{tabular}
\caption{The ratios $f_{D_s}/f_D$ and $f_{Bs}/f_B$: QCD sum rules
(the four upper lines) versus lattice QCD predictions (the three lower lines). 
\label{table:fbs}}
\end{center}
\end{table}
The sum rule predictions for $f_{D_s}$  and $f_D$ have to be compared with 
the results of the experimental measurements:
$f_{D_s}= 280 \pm 19 \pm 28 \pm34 $ MeV
and $f_D =300^{+180+80}_{-150-40}$ MeV, respectively.\cite{pdg00} 
The data are still too uncertain to challenge theory.

The decay constants of the  vector mesons $D^*$ and $B^*$, defined by the
matrix elements:
\be
\langle 0 | \bar q \gamma_\mu Q |H^* \rangle = 
m_{H^*} f_{H^*} \epsilon_\mu^{(H^*)}
\ee
($Q=c,b$, $H^*=D^*, B^*$) have also been updated:\cite{KRWY2}
\begin{equation}
f_{D^*}= 270 \pm 35~\mbox{MeV}\,, ~~f_{B^*} =195 \pm 35~\mbox{MeV}\,.
\end{equation}
The corresponding matrix elements for the orbitally excited 
heavy mesons have been determined, both for finite heavy 
quark mass~\cite{CNOP91} and in HQET as discussed below in Sec.~3.9.

Is it possible to  improve 
the determination of the heavy meson  decay constants?
The analysis of the heavy-light quark systems does not allow
an accurate, independent determination of the heavy quark masses,
the latter should be obtained from the heavy quarkonium channels. 
However, once the accuracy of the
$b$ and $c$ quark masses is increased, this  immediately
reduces the uncertainty of the heavy meson decay constants.
The issue of $\alpha_s$ corrections and of  their proper scale and 
resummation requires  further investigations, at least in the 
case of $f_B$. In this case,   
the ${\cal O}(\alpha_s)$  corrections in the sum rule seem to be unusually large, and
one may speculate that anomalously large soft parts of perturbative diagrams 
should be somehow separated. Furthermore, 
the contribution of the $d=7$ $\langle G G \bar q q \rangle$ condensate 
in the same sum rule is proportional to the heavy quark mass 
and could be sizeable, hence, it should be taken into account. 
Finally, improvement of the duality approximation 
for the hadronic spectral density is possible when 
enough experimental information on radially excited $B$ and $D$ 
resonances will be available.

\subsection{$B_c$ meson}
The studies of the charmonium 
system, in particular the determination of the masses and decay rates
of $J/\psi$, $\eta_c$ and $\chi_c$, were among the first applications  
of the SVZ method  and  provided the first
physical predictions of this approach. It is worth reminding
the successful prediction of the mass of $\eta_c$.\cite{SVZ79}
The analysis was then  extended to the bottomonium system 
($\Upsilon$,$\eta_b$,$\chi_b$,...). With this rich experience in analyzing
heavy quarkonia,  QCD sum
rules can be used to investigate the 
$B_c$ meson, the   state with open beauty and charm
observed recently by the CDF Collaboration.\cite{cdf98}
From the point of view of quark-gluon interactions,
$B_c$ is intermediate between the $\bar c c$ and
$\bar b b$ systems, and  it shares with the two heavy quarkonia
common dynamical properties.
For example, it is possible to consider the heavy quark and antiquark as 
nonrelativistic particles, and describe the bound state,
adding then the relativistic corrections.
On the other hand,  $B_c$, being the lightest hadron with
open beauty and charm, decays weakly. Therefore, it  provides
us with a rather unique possibility of investigating weak decay form factors 
in a quarkonium system.
 
In the framework of QCD sum rules, $B_c$ is investigated using the
interpolating current $j_5=i \; \bar c \gamma_5 b$. 
The analysis of the two-point correlator is very similar to that
discussed in Sec.~2, including the gluon condensate and the 
$O(\alpha_s$) correction, and summing
up the Coulomb part of this correction in the nonrelativistic approximation. 
Various calculations of the $B_c$ leptonic constant yield results
in the range $f_{B_c}=300 - 420$ MeV.\cite{chabab,paverbc}
Clearly, the accuracy of this determination can still be improved. 
This would be an important outcome,
since the purely leptonic mode $B_c \to \ell \bar \nu$ can be used
 to access the CKM matrix element $V_{cb}$.
Concerning the semileptonic $B_c$  decays, such as
 $B_c \to J/\psi (\eta_c) \ell \nu$ and $B_c \to B^{(*)}_s \ell \nu$,
they have been investigated
by three-point sum rules,\cite{paverbc,kiselev00}
following the method we shall illustrate below. 
In particular, in this framework it is  possible~\cite{kiselev00} 
to derive the relations
among the form factors determined by the heavy-quark spin symmetry, 
i.e. exploiting the decoupling of the spin of the heavy quarks
in the infinite quark mass limit.
The numerical determination  of the form factors is still 
hampered by the absence  of  ${\cal O}(\alpha_s)$ 
corrections, which are currently estimated  in  the
nonrelativistic Coulomb approximation at the zero-recoil 
point.\cite{kiselev00}
Abundant  $B_c$ production is expected at  
hadron colliders, and a careful  experimental investigation of this
system will be possible in the near future.\cite{Balllhc} $B_c$ will represent
an interesting testing ground for the  QCD sum rule approach, and 
therefore the refinement of the theoretical predictions concerning
this system should be in the working plans of the sum rule practitioners.  

\subsection{Ioffe currents and sum rules for baryons} 
QCD sum rules for  baryons suggested by Ioffe~\cite{ioffe81}
provide an important demonstration of the universality of the method,
generalizing it  from the quark-antiquark states to  the three-quark states.
To construct the correlation functions~\cite{ioffe81,oldbaryons}
one needs a baryon current, that is, a composite operator having the same
quantum numbers as a given baryon. For the proton, 
several possibilities were studied:
\begin{equation}
J^N(x)= \epsilon_{abc} (u^{a T }(x) {\cal C} \gamma_\mu u^b(x) )
 \gamma_5 \gamma^\mu d^c(x) \label{jN}
\end{equation}
or
\begin{equation}
J^{\prime N}(x)= \epsilon_{abc} (u^{a T }(x) {\cal C} \sigma_{\mu \nu} u^b(x) )
 \gamma_5 \sigma^{\mu \nu}  d^c(x) \label{jNp} \;\;\; ,
\end{equation}
where $a,b,c$ are color indices and $\cal C$ is
the charge conjugation matrix.
Other possible quark currents involve derivatives.
Some criteria  have to be adopted to single out the optimal 
interpolating current for a particular baryon. 
The first criterion is to choose a current with a
minimal number of derivatives, in order to deal with low-dimensional
spectral densities and, consequently, to minimize the contribution
of the excited states. Furthermore, the currents
should maximize
the projection onto the considered baryon state. This can be done,
for example, by considering 
linear combinations of the interpolating currents,
with the coefficient suitably
chosen in order to maximize the overlap. Finally, it is possible 
 to choose currents in such a way that the
two-point functions are dominated 
by the perturbative contribution, with the 
condensate terms producing a hierarchical set of corrections.
This last requirement suggests to use  the current (\ref{jN}),
instead of (\ref{jNp}), to interpolate the proton. 
Studying  the two-point correlation function
\begin{equation}
\Pi(q) = i \int e^{i q \cdot x} 
\langle 0 | T \{J^N(x) \bar J^N(0)\} | 0 \rangle = 
\Pi_1(q^2) + \not\!q \; \Pi_2(q^2)
\end{equation}
and neglecting the contribution of the continuum and of higher dimensional
condensates,  an
astonishingly  simple  expression for the nucleon mass can be 
obtained:\cite{ioffe81}
\begin{equation}
m_N^3 \simeq - 2(2 \pi)^2 \langle \bar q q \rangle(\mu=1 {\rm GeV})\,, 
\label{nucleon_mass}
\end{equation}
in agreement with experiment. 

With a little effort, one  defines the interpolating currents for
the $L=0$ baryonic octet:
\begin{eqnarray}
J^\Sigma(x)\!\!\!\!\! &=&\!\!\!\!\!\epsilon_{abc} (u^{a T }(x) {\cal C} \gamma_\mu u^b(x) )
 \gamma_5 \gamma^\mu s^c(x)\,, \nonumber \\
J^\Xi(x)\!\!\!\!\! &=&\! \!\!\!\!- \epsilon_{abc} (s^{a T }(x) {\cal C} \gamma_\mu s^b(x) )
 \gamma_5 \gamma^\mu u^c(x) \,,\\
J^\Lambda(x)\!\! \!\!\!&=& \!\!\!\!\!\sqrt{2 \over 3}\epsilon_{abc} \Big[
(u^{a T }(x) {\cal C} \gamma_\mu s^b(x) ) \gamma_5 \gamma^\mu d^c(x) - 
(d^{a T }(x) {\cal C} \gamma_\mu s^b(x) ) \gamma_5 \gamma^\mu u^c(x) \Big],
\nonumber
\end{eqnarray}
and for the $L=0$ decuplet:
\begin{eqnarray}
J^\Delta_\mu(x)\!\!\!&=& \!\!\!\epsilon_{abc} (u^{a T }(x) {\cal C} \gamma_\mu u^b(x) )
 u^c(x)\,, \nonumber \\
J^{\Sigma^*}_\mu (x)\!\!\! &=&\!\!\!\sqrt{1\over 3} \epsilon_{abc} 
\Big[2(u^{a T}(x) {\cal C} \gamma_\mu s^b(x)) u^c(x) + 
(u^{a T }(x) {\cal C} \gamma_\mu u^b(x)) s^c(x) \Big],\nonumber \\
J^{\Xi^*}_\mu(x)\!\!\! &=&\!\!\! \sqrt{1\over 3} \epsilon_{abc} 
\Big[2(s^{a T }(x) {\cal C} \gamma_\mu u^b(x)) s^c(x) + 
(s^{a T}(x) {\cal C} \gamma_\mu s^b(x)) u^c(x) \Big],\nonumber \\
J^\Omega_\mu(x)\!\!\! &=& \!\!\!\epsilon_{abc} 
(s^{a T }(x) {\cal C} \gamma_\mu s^b(x) ) s^c(x)\, .
\end{eqnarray}

The ${\cal O}(\alpha_s)$ radiative corrections to the two-point
functions of baryon currents
are known for three massless quarks,\cite{pivovarov91} and have been 
worked out also for the case
of baryons containing one heavy and two massless 
quarks.\cite{pivovarov99} This opens up interesting perspectives for 
more precise determinations of the baryon properties, both in the light quark
and  in the heavy quark sector. Let us remind the reader that
the analysis of the two-point correlator of baryonic currents was of
prime importance for determining some characteristics of the QCD vacuum. 
As a matter of fact, from the
fit of the masses of baryons belonging to the $L=0$ octet and 
decuplet, it was possible to  determine the values of 
the mixed quark-gluon condensate (\ref{qqbarG2})
and of the $SU(3)$-flavor breaking parameter 
$\gamma = {{\langle \bar s s \rangle} / {\langle \bar q q \rangle}} -1 
\simeq -0.2$.\cite{RRY85}

The extension  to the case of baryons containing
charm and beauty  quarks is also straightforward.
In the heavy quark sector, open problems concern 
the spectra of baryons containing more than one heavy ($c$ and $b$) quark. 
These baryons will be experimentally observed and investigated 
at hadron colliders. Two-point correlation functions of heavy-baryon
currents allow one to predict the spectrum of these states.\cite{kiselev99} 
For an overview of sum rule  applications to the baryonic problems, 
and for an analysis of the features of various three-quark interpolating
currents one can consult the available reviews.\cite{cohen}

An important technique for the studies of the baryon dynamics 
is {\em the method of external fields},\cite{external} 
in which two-point correlators in the 
external static fields (such as magnetic field) are introduced. Originally, 
many static properties of nucleons, such as magnetic moments, 
or nucleon matrix elements of axial-vector currents and other operators
have been successfully calculated.
The limits of this review do not allow us to discuss this technique and its
applications in more detail. A thorough presentation can be found, e.g., in the lecture 
by Ioffe.\cite{revIoffe} To demonstrate the variety of problems 
that can be solved by employing the external field method or the 
closely related approach of the soft-pion field, let us 
mention some recent applications.  In the light baryon sector,  
attention has been recently paid to 
the properties of negative parity baryons, such as $N^*(1535)$.
In particular, the experimental analysis has shown a suppression
of the strong  $N^* N \pi$ coupling and an enhancement of the $N^* N \eta$
coupling. Interestingly, estimates of these couplings 
using the two-point correlators in the 
external (soft) pion field  are in agreement with this observation.\cite{jido98}
Other strong pion-baryon couplings have been analyzed by the same
method.\cite{birse96}

\subsection{Three-point correlation functions: 
form factors and decay amplitudes}
The method of QCD sum rules can be generalized in order to calculate
the hadronic matrix elements of electromagnetic and weak
transitions. In this case one starts from  three-point vacuum 
correlation functions and uses double dispersion relations.
This approach has
 been extensively used, 
both for light and heavy hadrons. The applications include
the pion electromagnetic form factor,\cite{rad82} 
radiative charmonium decays 
such  as  $J/\psi \to \eta_c \gamma$,\cite{beilin85}
$D$ and $B$ semileptonic and 
flavor-changing neutral current (FCNC) 
transitions~\cite{AElKogan}$^{\!-}$\cite{raredecays2}
 and, more recently, the radiative decays
$\phi \to (\eta, \eta^\prime) \gamma$.\cite{fdf00} 

In order to discuss the advantages and the difficulties of three-point
sum rules,  let us outline, as an example,
the calculation~\cite{rad82}
of the pion electromagnetic form factor
defined by the matrix element:
\be
\langle \pi(p^\prime)| j^{em}_\mu | \pi(p)\rangle =
F_\pi(q^2) (p+p^\prime)_\mu \,,  
\label{pionformf}
\ee
where $q=p^\prime-p$ and
$j_\mu^{em}$ is the electromagnetic current 
\be
j_\mu^{em}=e_u \bar u \gamma_\mu u + e_d \bar d \gamma_\mu d \,.
\label{emcurrent}
\ee
The starting point is the correlator of $j_\mu^{em}$ with two 
pion-interpolating currents (\ref{eq:axialcurrent}): 
\bea
T_{\mu \nu \lambda}(p,p^\prime)&=& (i)^2 \int d^4x \; d^4y \;  
e^{i (p^\prime \cdot x - p \cdot y)}
\langle 0|T\{ j_{\mu}^{(\pi) \dagger}(x) j_\lambda^{em}(0) 
j_{\nu}^{(\pi)}(y)\}|0\rangle \nonumber \\
&=& p^\prime_\mu p_\nu (p+p^\prime)_\lambda T(p^2,p^{\prime 2}, q^2) 
+ \dots \,, 
\label{eq:3point}
\eea
where the momenta $p$, $p^\prime$ and $q$
flow through the axial-vector and the electromagnetic currents, 
respectively. In Eq.~(\ref{eq:3point}) we have only shown the relevant
kinematical structure, the others being  denoted by the ellipses.

Inserting in Eq.~(\ref{eq:3point})
two complete sets of hadronic states with the 
quantum numbers of the pion one obtains the dispersion relation:
\bea
T(p^2,p^{\prime 2},{q^2}) & = & 
{f_{\pi}^2   F_\pi(q^2)\over 
(m_\pi^2-p^2)(m_\pi^2-p^{\prime 2})} +
\int_{R_{12}} ds \; ds^\prime {\rho^h(s, s^\prime) \over (s-p^2)
(s^\prime - p^{\prime 2})} \nonumber \\
&+&P_1(p^2) \int_{R_2} ds^\prime 
{\rho_2(s^\prime) \over s^\prime - p^{\prime 2}} 
+P_2(p^{\prime 2}) \int_{R_1} ds {\rho_1(s) \over s -p^2}\,,
\label{eq:3pt}
\eea
where the double  dispersion integral receives contributions
from the excited and continuum states located in the region $R_{12}$
of the $(s, s^\prime)$ plane.
The terms containing the polynomials  $P_1$ and $P_2$ 
arise from subtractions in the dispersion relation, and, therefore, two
independent Borel transformations in $p^2$ and $p^{\prime 2}$ are needed to
get rid of them.
The same  transformations 
enhance the double pole term with respect to the 
double integral in Eq.~(\ref{eq:3pt}).

On the other hand, the amplitude  (\ref{eq:3point}) can be
computed by a short-distance expansion,
in terms of perturbative and condensate contributions:
\be
T(p^2,p^{\prime2},q^2)= 
\sum_d C^d(p^2,p^{\prime 2}, q^2,\mu) \langle O_d(\mu)\rangle
\,. \label{eq:3ptOPE}
\ee
The expansion is valid
for large spacelike external momenta: 
$|p^2|, |p^{\prime 2}| \gg \Lambda^2_{QCD}$;
the squared momentum transfer  $Q^2=-q^2$ is also kept large
in order to stay far away 
from the hadronic thresholds in the $q^2$-channel.
%
\begin{figure}[ht]
\vspace{-1.9cm}
\centerline{\hspace*{1cm}\epsfig{file=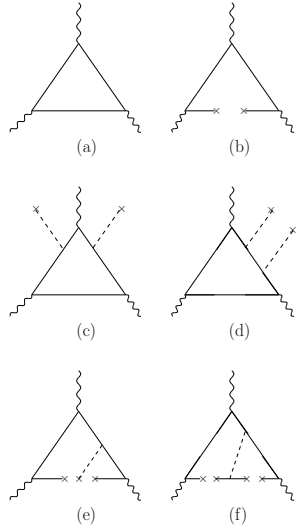,height=4.5in}}
\vspace{-1cm}
\caption{Contributions to the three-point correlator 
(\ref{eq:3ptOPE}): ${\cal O}(\alpha_s=0)$
perturbative term (a)
and  some nonperturbative corrections (b$-$f).
\label{fig:3point}}
\end{figure}
%
%
In Fig.~\ref{fig:3point} we depict the perturbative diagram
 and several contributions of the condensates.
In the chiral limit, the contributions of the quark  and 
quark-gluon condensates  vanish.

To proceed further, one has  to match the expression (\ref{eq:3ptOPE})
with the hadronic representation (\ref{eq:3pt}).
Invoking quark-hadron duality, one approximates the
double dispersion integral in the region $R_{12}$ with the perturbative
one. After that, the double Borel transformation in $p^2, p^{\prime 2}$ is
performed, 
introducing two corresponding  parameters  $M, M^{\prime}$, so
that the sum rule for $F_\pi(Q^2)$ reads:
\bea
f^2_\pi F_\pi(Q^2) &=& {1 \over \pi^2} 
\int_{\tilde R_{12}} ds \; ds^\prime \;
{\rm Im \;} C^0 (s, s^\prime, Q^2) 
e^{-{s \over M^2}-{s^\prime \over M^{\prime 2}}} \nonumber \\
&+&\sum_{d=3}^n C^d (M^2, M^{\prime 2}, Q^2, \mu) 
\langle O_d(\mu)\rangle \;\;\;.
\label{eq:3ptc}
\eea
The domain ${\tilde R_{12}(s^\pi_0)}$, characterized by the
threshold $s^\pi_0$,  is determined by the
duality approximation and by the
procedure of subtracting the contributions of excited and 
continuum states. More specifically,
taking $M^{\prime }=M$ and considering the contribution of
condensates up to $d=6$, yields the following sum rule:\cite{rad82}
\bea
F_\pi(Q^2)&=&{4 \over f_\pi^2} \Big[  \int_{\tilde R_{12}} ds \; ds^\prime 
\rho_0 (s, s^\prime, Q^2) e^{-{s + s^\prime \over M^{2}}} \nonumber \\
&+&{\alpha_s \over 48 \pi M^2} 
\langle G_{\mu \nu}^a G_{\mu \nu}^a \rangle + 
{52 \pi\over 81 M^4} \alpha_s \langle \bar \psi \psi \rangle^2 
(1 + {2 Q^2 \over 13 M^2}) \Big]\,,
\label{eq:3ptioffe}
\eea
where 
\be
\rho_0 (s, s^\prime, Q^2)={3 Q^4 \over 16 \pi^2} {1 \over \lambda^{7/2}}
\big[3 \lambda(\sigma+Q^2)(\sigma+2 Q^2) - \lambda^2 - 5 Q^2 (\sigma + Q^2)^3 
\big] \,,
\ee
$\lambda=(s+s^\prime+Q^2)^2- 4 s s^\prime$
and $\sigma=s+s^\prime$.

The numerical analysis of (\ref{eq:3ptioffe}) uses
the values of  condensates given in Sec.~2 and
the threshold  $s^\pi_0$ inferred from the two-point sum rule (\ref{SVZpi}).
For intermediate values of the momentum transfer, $Q^2=1\div 3$  GeV$^2$
the appropriate range of Borel parameter is
$0.7 < M^2 < 1.7$ GeV$^2$,
where the sensitivity to the duality approximation is low 
and where there is a hierarchy of power corrections, the
criteria adopted in the analysis of two-point
SVZ sum rules in Sec.~2. For larger values of $Q^2$ 
the contributions of the higher-dimensional condensates
containing terms $\sim Q^2/M^2$
overwhelm the contributions of the low-dimensional terms.
 Therefore, the three-point sum rule
(\ref{eq:3ptioffe}) cannot
be used to predict the large $Q^2$ behavior of  $F_\pi(Q^2)$. The reason
of the failure can be traced back to the truncated 
local condensate approximation, which is too crude to reproduce the 
physical mechanisms governing the exclusive $\gamma^* \pi \to \pi$ 
transition.\footnote{ As one remedy solving this problem
it was suggested to use (model-dependent) 
nonlocal condensates.\cite{BakRad}} We shall return to this point in Sec.~4.
On the other hand, there is no doubt that the sum rule 
(\ref{eq:3ptioffe}) is reliable 
in the region $Q^2=1\div 3$  GeV$^2$, where also
the numerical result  $Q^2 F_\pi(Q^2) \simeq 0.3$ GeV$^2$
is in agreement with the  experimental data
(the latter are presented in Fig.~\ref{fig:piformf}).

The calculation of the heavy meson form factors follows the
same strategy as  outlined above. Let us begin with some 
useful definitions. The matrix elements
governing  the weak transitions of $D$ meson  to a pseudoscalar
$P=K,\pi$ and vector $V=K^*,\rho$ final state are:
\be
\langle P(p^\prime)| \bar q \gamma_\mu c| D(p)\rangle =
f^+_{DP}(q^2) (p+p^\prime)_\mu + f^-_{DP}(q^2) (p-p^\prime)_\mu\,,
\label{heavylight}
\ee
and
\bea
&&
\!\!\!\!\!\langle~V(p^\prime, \lambda)|\bar q \gamma_\mu(1-\gamma_5)c|D(p)~\rangle = 
{2 V^{DV}(q^2) \over m_D + m_V} \epsilon_\mu^{\alpha \beta \gamma} 
\epsilon^{(V)*}_\alpha p_\beta p^\prime_\gamma 
\nonumber \\ 
-&&\!\!\!\!\! i (m_D + m_V) A_1^{DV}(q^2) \epsilon_\mu^{(V)*} 
+ {i A_2^{DV}(q^2) \over m_D + m_V} (\epsilon^{(V)*} \cdot p)(p+p^\prime)_\mu 
\nonumber \\ 
+&&\!\!\!\!\!i { 2 m_V \over q^2} [A_3^{DV}(q^2)-A_0^{DV}(q^2)]  
(\epsilon^{(V)*} \cdot p)(p-p^\prime)_\mu\,, 
\eea
where $\epsilon^{(V)*}_\mu$ is the polarization vector of $V$, 
\be
A_3^{DV}(q^2)= {m_D + m_V \over 2 m_V} A_1^{DV}(q^2) -  
{m_D - m_V \over 2 m_V} A_2^{DV}(q^2)
\ee
and $A_3^{DV}(0)=A_0^{DV}(0)$. 

Each of the form factors introduced above can be studied by
introducing an appropriate three-point correlation function.
For example, the form factor  $f^+_{DK}$ can be computed from
\be
T_{\mu \lambda}(p,p^\prime)= \!(i)^2\!\! \int\!\! d^4x  d^4y  
e^{i (p^\prime \cdot x - p \cdot y)}
\langle 0|T\{j_{\lambda}^{(K)}(x) 
\bar s(0) \gamma_\mu c(0)j_5^{(D)}(y)\} |0\rangle ,
\label{eq:3pta}
\ee
where $D$ and $K$ mesons are 
interpolated by the corresponding currents $j_5^{(D)}=i\; \bar c \gamma_5 u$
and $j_{\lambda}^{(K)}=\bar u  \gamma_\lambda \gamma_5 s$, respectively.
In this case, due to the presence of a large scale $m_c$, 
the accessible region of the squared momentum transfer $q^2$ includes also
small positive values: $q^2 \ll m_c^2$. 
There is another important difference with respect to the sum rule
for the pion form factor: In the OPE for the correlator 
(\ref{eq:3pta}), the most important nonperturbative effects 
are due to the  quark and quark-gluon condensates, and are
described by the diagrams in Fig.~\ref{fig:3point}b,e. 
The contributions of the gluon condensate and of the
four-quark condensates are negligible.
The subsequent  steps include the use of quark-hadron duality and the
 double Borel transformation in $p^2, p^{\prime 2}$. 
The Borel parameters $M, M^\prime $ are now kept different, since they
correspond to heavy and light mass scales, respectively. 
The sum rule has the form:
\bea
{m_D^2 f_D f_K \over m_c} f^+_{DK}(q^2) 
e^{-{m_D^2 \over M^2}-{m_K^2 \over M^{\prime 2}}}\!\!\!\! &=& \!\!\!\!{1 \over \pi^2} 
\int_{\tilde R_{12}} ds \; ds^\prime 
{\rm Im \;} C^0_{DK} (s, s^\prime, q^2, \mu) 
e^{-{s \over M^2}-{s^\prime \over M^{\prime 2}}} \nonumber \\
&+&\!\!\!\!\sum_{d=3}^n C^d_{DK} (M^2, M^{\prime 2}, q^2, \mu) 
\langle O_d(\mu)\rangle\,.
\label{eq:3ptb}
\eea
The explicit expressions for the Wilson coefficients 
$C_{DK}^d$ can be found in 
the literature.\cite{dsemilep}$^-$\cite{bsemilep}  
The threshold parameters  $s^D_0$ and $s^K_0$ determining  the domain
${\tilde R_{12}(s^D_0,s^K_0)}$
can be inferred, together with $f_D$ and $f_K$,  from the
study  of the corresponding two-point correlation functions.
\begin{table}[h]
\begin{center}
\begin{tabular}{|c | c | c | c | c | c | }
\hline
           &&$f^+_{DP}(0)$&$A_1^{DV}(0)$&$A_2^{DV}(0)$& $V^{DV}(0)$ \\ 
\hline 
           &&       &          &           &      \\       
$D \to K  $&    & $0.60\pm0.15$&   &           &       \\
           &exp.\cite{ryd}& $0.76\pm0.03$&   &           &       \\
$D \to K^*$&    &              &$0.50\pm0.15$&$0.60\pm0.15$&$1.1\pm0.25$  \\
           &exp.\cite{e791}  & &$0.58\pm0.03$&$0.41\pm0.06$&$1.06\pm0.09$ \\
           &&              &            &              & \\
$D \to \pi$&    &$0.50\pm0.15$&   &           &         \\
$D \to \rho$&   &                 &$0.5\pm0.2$&$0.4\pm0.1$&$1.0\pm0.2$ \\
            &&             &            &              & \\
\hline
\end{tabular}
\caption{Form factors of the weak $D \to P,V$ transitions at $q^2=0$.
The experimental numbers are obtained assuming the nearest pole dominance
for the form factors.
\label{table:DK}}
\end{center}
\end{table}

In Table~\ref{table:DK} we present a set of results obtained  
at $q^2=0$.\cite{dsemilep,dsemilep2,bsemilep}
The $q^2$ dependence of the form factors can be predicted,
although the procedure is non-trivial
since one should stay far from the thresholds appearing
in the double dispersion relation. 
The $q^2$ dependence of $A_1$ and $A_2$ turns out to be
quite mild; on the contrary, the dependence of $f^+$ and $V$ is compatible 
with the nearest pole dominance.
The predictions, within their uncertainties, 
are in agreement with experiment.
Note that the form factor $f^+_{DK}$ is important for an 
independent determination of the CKM parameter $|V_{cs}|$ from
the semileptonic $D\to K l\nu_l$ decays.\cite{pdg00}  

In order to predict the form factors  of $B$ transitions 
to light mesons $(\pi, K, \rho)$, one has to replace $c \to b$ and
$D \to B$ in the above sum rules yielding  results for 
semileptonic~\cite{bsemilep,col95,aliev99} and FCNC  $B$ 
transitions.\cite{raredecays1,raredecays2}
The method can 
be easily generalized to different processes and final states,
 including orbital excitations, and has
provided interesting predictions.
An example is the determination~\cite{Dominguez:1988wa,raredecays1,raredecays2}
of the form factors  relevant for the FCNC 
$B\to K^* \gamma$ and $B\to K^* \ell^+ \ell^-$ decays:
\bea
\langle~K^*(p^\prime, \lambda)|\bar s \sigma_{\mu \nu} q^\nu b_R|B(p)~\rangle 
=  i \epsilon_{\mu \nu \alpha \beta} \epsilon^{*\nu} p^\alpha p^{\prime \beta} T_1(q^2)
\nonumber \\
\!+ {1 \over 2} [e^*_\mu (m_B^2-m^2_{K^*}) - \epsilon^* \cdot q ( p+p^\prime)_\mu] T_2(q^2)
\nonumber \\
\!
+ {\epsilon^* \cdot q \over 2} \Big( q_\mu - {q^2 \over m_B^2 - m_K^{*2}}
(p + p^\prime)_\mu \Big ) T_3(q^2)\,,
\eea
where $b_R=1/2(1 + \gamma_5)b$, 
$q=p-p^\prime$ .
 The result~\cite{raredecays1,raredecays2}
$T_1(0) = 0.35 \pm 0.05$ allowed to predict the ratio
$\Gamma(B \to K^* \gamma)/ \Gamma(b \to s \gamma)=0.17 \pm 0.05$, which 
agrees  with the experimental measurements.\footnote{Recent results for the 
exclusive $B \to K^* \gamma$ transitions are:
${\cal B}(B^0 \to K^{*0} \gamma)= (4.55^{+0.7}_{-0.68}\pm 0.34) \cdot10^{-5}$
and
${\cal B}(B^+ \to K^{*+} \gamma)= (3.76^{+0.89}_{-0.83}\pm 0.28) \cdot 
10^{-5}$.\cite{cleoBKg}
For the inclusive decay, the most recent measurements give:
${\cal B}(b \to s \gamma)= 
(3.15\pm0.35\pm0.36\pm0.26)\cdot 10^{-4}$,\cite{cleobsg} and
${\cal B}(b \to s \gamma)= 
(3.11\pm0.35\pm0.80\pm0.72)\cdot 10^{-4}$.\cite{alephbsg}}

However, 
a close inspection of the general structure of three-point sum rules
for heavy-light transitions
reveals a difficulty  which is manifest in the parametric 
$m_b$ dependence of the various terms of the short-distance expansion. 
An example is the form factor  $A_1$ in  the $B \to \rho$ matrix element.
In the limit of large $m_b$, one observes that the coefficients of the quark
and  quark-gluon condensates grow  with $m_b$
faster than the coefficient of the perturbative contribution.
This is another manifestation of the difficulty in approximating
the OPE expansion by the first few terms, 
revealed in the sum rule for the  pion form factor at large 
$Q^2$. The physical origin of this difficulty can be traced back
to the mechanisms of producing the light state in the heavy meson
decay, for large heavy quark mass.\cite{BB97}
Of course, this problem could be irrelevant
for the actual value of the $b$-quark mass, and for particular
processes and final states: an example is the transition $B \to \pi$.
However,  in order to avoid the problem {\it ab initio}, 
an operator expansion on the light-cone can
be used  to  describe 
heavy-light transitions, as explained in the next Section.

Concerning $b \to c$ transitions,
$B$ decays to charmed states ($B \to D^{(*)}, D^{**}$)
were investigated in early papers~\cite{CNOP91,baier90}
for finite  $c$ and $b$  quark masses. 
Since that time,  the common attention has 
shifted towards  approaches incorporating the heavy quark 
flavor and spin symmetry,
with the development of  appropriate sum rules in HQET (see Sect. 3.9).
Our opinion is
that three-point sum rules
for $b\to c$ transitions and finite quark masses represent a viable
approach complementary to HQET, which allows one to 
control the accuracy of the heavy quark limit.\cite{ballbc}
Actually, in the finite-mass sum rules the important contributions 
of hard-gluon exchanges in the diagram in Fig.~\ref{fig:3point}a 
are  not yet available. With the current progress of methods for 
computing many-loop diagrams, it should become possible to calculate
these two-loop three-point diagrams with different quark masses.

\subsection{Hadron structure functions}

A standard application of QCD to the hadronic structure functions 
is to study their logarithmic dependence on the momentum transfer 
$Q^2$ using the perturbative evolution equations. The initial conditions
for these equations are usually parametrized 
and fitted to the experimental data at some intermediate value of $Q^2$. 
A direct analytical calculation of structure functions 
remains a challenging task. It is therefore very important  
that QCD sum rules are in a position to solve this problem,
albeit approximately and within a limited range of the Bjorken
variable. 
The method was suggested by Ioffe~\cite{Ioffe85} and was originally
applied to the nucleon structure functions, 
to obtain, in particular, the valence $u$ and $d$ quark-parton 
distributions in the nucleon at intermediate $Q^2$. 
The idea is to consider the four-point correlator 
\be
T_{\mu \nu}^\pm = -i \int d^4x \; d^4y \; d^4z \; 
e^{i q \cdot x} e^{i p \cdot (z-y)}
\langle 0|T\{J^N(y) j_\mu^\mp(x) j_\nu^\pm(0) {\bar J^N}(z)\}|0\rangle,
\label{eq:4point}
\ee
$j_\mu^-=\bar d \gamma_\mu(1-\gamma_5)u$ and
$j_\mu^+=\bar u \gamma_\mu(1-\gamma_5)d$ being 
the weak quark currents,  and $J^N$ 
the nucleon interpolating current (\ref{jN}).
%
\begin{figure}[ht]
\begin{tabular}{c c}
\psfig{figure=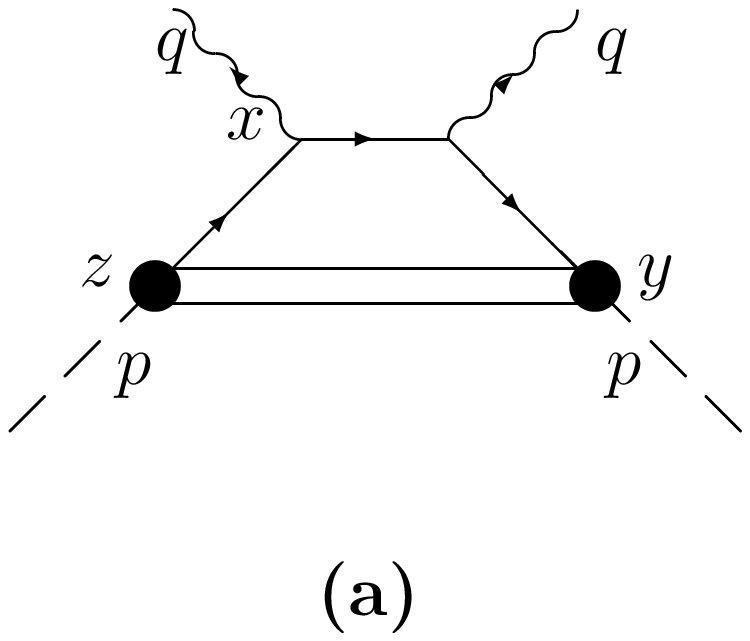,height=4.cm}&
\psfig{figure=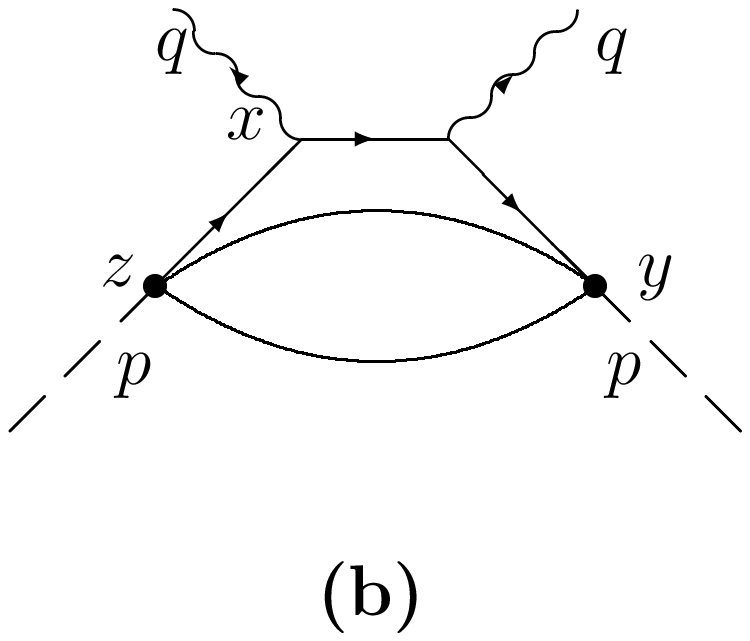,height=4.cm}
\end{tabular}
\caption{Four-point correlator in Eq.~(\ref{eq:4point}) (a) 
and its perturbative part (b).
\label{fig:4point}}
\end{figure}
Inserting  complete sets of states with the nucleon
quantum numbers and picking up the ground state,
one encounters the  forward scattering amplitude: 
\be     
T_{\mu\nu}^N=-i \int d^4x \; e^{i q \cdot x}\langle N |
T\{j_\mu^\mp(x) j_\nu^\pm(0)\}| N \rangle\,,
\label{ref:forward}
\ee 
which is directly related to the nucleon structure function $F_2(x_{Bj},Q^2)$ 
measured in  deep inelastic neutrino-nucleon scattering.
The key observation~\cite{Ioffe85} is 
that the imaginary part of the correlator (\ref{eq:4point}) 
\be
{\rm Im} \; T_{\mu\nu}={1 \over 2 i} \big[T_{\mu\nu}(p^2,q^2,s+i\epsilon)-
T_{\mu\nu}(p^2,q^2,s-i\epsilon)\big] 
\label{eq:im4point}
\ee
($s=(p+q)^2$) can be expanded at short distances
in the range of large Euclidean momenta $p$ and $q$ :  
$Q^2=-q^2\simeq 10$ GeV$^2$ and $|p^2|\simeq1$ GeV$^2 \ll Q^2$. 
For  example, considering
the lowest order perturbative 
diagram in Fig.~\ref{fig:4point}b and taking the imaginary part of this
diagram in $s$,  one notices that the quark
virtualities  in the $t$ channel 
are of the order of  $p^2 x_{Bj}$, with $x_{Bj}={Q^2 / 2 p \cdot q}$ being 
the Bjorken variable. Therefore, the quarks are far off-shell if $x_{Bj}$ 
is not too small. 

The OPE for the 
correlation function (\ref{eq:4point}) is obtained~\cite{nucleonsf} in a 
standard way in terms of perturbative and  condensate contributions. 
In this calculation 
 one  retains only the 
leading powers in the ratio ${|p^2|/Q^2}$ 
accounting for the leading twist in the structure functions
(parton distributions).
The dispersion relation is taken in  one variable
$p^2$, which is the virtuality of the baryon current. 
One then performs a Borel transformation in
$p^2$ and separates the contribution of the lowest-lying nucleon
state from the contributions of the other poles and 
of the continuum. This is a rather sensitive point of the 
procedure, because, even after the Borel 
transformation, there is still a contamination of the ground-state term  by 
"parasitic" contributions in the dispersion relation.\footnote{Recently, in calculating 
the valence quark distribution in the pion, this problem was 
solved~\cite{IoffeAO} by considering a more complicated kinematical 
configuration of the 4-point correlator, with two different virtualities of 
interpolating currents. In this case double dispersion relations and 
double Borel transformation are applicable, eliminating 
all ``parasitic'' terms and improving the accuracy of the sum rule. }
The range of the Bjorken variable, which is accessible to the method,
does not include small $x_{Bj}\simeq 0$, where OPE breaks down.
The region  $x_{Bj} \simeq 1$ is also out of the reach of this method, 
simply because it is the resonance region in the $s$ channel.

With all such caveats, the valence quark distributions in the nucleon
have been computed: $u_v(x_{Bj})$ in the region $0.2<x_{Bj}<0.7$ and 
$d_v(x_{Bj})$ in the region $0.3<x_{Bj}<0.5$.\cite{nucleonsf} 
The accuracy of the calculation
is at the level of 50\%;
within this accuracy the result is in agreement 
with the parton distributions extracted from experimental 
data. The mere fact that an analytical expression containing quark and 
gluon condensates reproduces  rather subtle dynamical 
features of the nucleon is impressive and promising.

Estimates of the polarized structure functions $g_1(x_{Bj})$ and $g_2(x_{Bj})$ 
obtained by the same method~\cite{polarsf} can be presented
as average values in the (rather narrow) accessible ranges of the variable
$x_{Bj}$:\cite{revIoffe}
\be
\overline{g_1(x_{Bj})} (0.5 <x_{Bj}<0.7)= 0.05 \pm 50\%\,
\label{eq:g1}
\ee
\be
\overline{g_2(x_{Bj})} (0.5 <x_{Bj}<0.8)= -0.05 \pm 50\% \, .
\label{eq:g2}
\ee
These estimates are in a  reasonable agreement with 
the data in the same intervals of the Bjorken variable:
\be
\overline{g_1(x_{Bj})}(0.5 <x_{Bj}<0.7)= 0.08 \pm 0.02\;,\mbox{\cite{spindata1}} 
\label{eq:g1exp}
\ee
\be
\overline{g_1(x_{Bj})}(0.4 <x_{Bj}<0.7)= 0.08 \pm 0.02 \pm 0.01\;,
\mbox{\cite{spindata2}} 
\label{eq:g1exp2}
\ee
and 
\be
\overline{g_2(x_{Bj})} (0.5 <x_{Bj}<0.8)= -0.037 \pm 0.020 \pm 0.003
\%\;\;.\mbox{\cite{spindata3}}
\label{eq:g2exp}
\ee

Having an approximate but reliable method to calculate the 
structure functions at intermediate $x_{Bj}$, it is 
interesting to apply it to a structure function which 
is completely unknown and not even directly measurable in deep 
inelastic scattering. This is the case of the chirality-violating 
structure function $h_1(x_{Bj})$   which characterizes 
the transverse spin dynamics in the nucleon.\cite{JaffeJi}
The dominant, $u$-quark contribution to the proton structure function
 $h_1$ has been estimated~\cite{Ioffehx} employing 
an appropriate four-point correlator. The prediction, in the region $0.3 <x_{Bj}<0.5$,
is:
\be
\overline{h_1(x_{Bj})}(0.3 <x_{Bj}<0.5)= 0.5 \pm 50\%\,. 
\label{eq:h1}
\ee

The photon structure function $F_{2\gamma}(x_{Bj},Q^2)$ can also be 
obtained~\cite{photonsf} starting from the correlator
of four electromagnetic currents with two different virtualities
$q^2$ and $p^2$. The imaginary part of this correlator, which determines 
the structure function of the photon with a virtuality $p^2$, is calculated in a form 
of a condensate expansion, and in the approximation of small $p^2/Q^2$.
In order to obtain the structure function of the real photon,
a dispersion relation in  $p^2$ is used, 
expressing the virtual photon structure function 
in terms of the integral over hadronic states and employing the quark-hadron duality. 
The analyticity of the dispersion relation in $p^2$ permits the extrapolation
to $p^2=0$.  This method 
allows one to separate the hard and soft (hadronic) parts in the photon structure function. 
The hard part corresponds to the pointlike quark-photon interaction, whereas the  
soft (hadronic) part receives contributions from the 
interactions of the photon with quark-antiquark states at large distances. The 
duality threshold $s_0^\rho$ serves as an effective boundary of this separation.
The result for the structure function $F_{2\gamma}(x_{Bj},Q^2)$, predicted 
in the range $0.2<x_{Bj}<0.7$ and at 
$Q^2 \sim$ a few GeV$^2$~~\cite{photonsf}  agrees with the experimental data. 
This agreement is impressive, because all inputs
in this calculation are fixed by two-point sum rules. 
Furthermore, in this approach it is possible to calculate the 
twist 4 corrections to $F_{2\gamma}(x_{Bj},Q^2)$,\cite{photonsftw4} and to 
find  the gluon distribution in the photon.\cite{photonsfglue}

The four-point correlator with two heavy-light quark currents 
$j_5=\bar{Q}\gamma_5 q$ 
and two electromagnetic heavy quark currents $j_\mu=\bar{Q}\gamma_\mu Q$ ( $Q=c$ or $b$) interpolates 
a deep inelastic scattering on a heavy meson ($D$ or $B$). The advantage in 
this case is that the correlator has a well-defined forward scattering limit,
since the $\bar{Q}Q$  states in the $t$ channel are far from $t=0$. Calculating this
correlator in terms of the perturbative part plus condensate expansion 
and performing the double Borel transformation, it is possible 
to calculate a few first moments of the heavy quark parton distribution 
in $D$,$B$ mesons,\cite{heavyparton} obtaining, in a certain approximation, 
also the fragmentation functions of heavy quarks into heavy mesons.  
The calculation was done for finite quark masses, it would be interesting to repeat it in HQET.

Finally, as a new direction in studying the structure functions,
which is, however, out of the scope of our main presentation,
one should mention the analysis of parton distributions
in terms of a coordinate-space variable called the Ioffe time.
These distributions have been obtained for proton and photon, 
employing the four-point correlation functions and sum rule 
technique.\cite{ioffetime}

\subsection{Matrix elements of effective operators}
The description of neutral meson oscillations involves a set of 
important hadronic parameters. The first one is
$B_K(\mu)$, defined by the matrix element 
\begin{equation}
\langle \bar K^0 | (\bar s \gamma^\mu (1-\gamma_5) d)
(\bar s \gamma_\mu (1-\gamma_5) d) |K^0 \rangle =
2 (1+ {1\over N_c}) (f_K m_K)^2 B_{K}(\mu)  \label{defbk}
\end{equation}
and representing the deviation  of the
$K^0-\bar K^0$ mixing amplitude
from the vacuum saturation approximation. A similar definition holds for 
$B_{d,s}$, and  the matrix elements analogous to
(\ref{defbk}) are expressed in terms of  $B_{B_d,s}(\mu)$.  The leading
dependence on the renormalization scale $\mu$ is:
\begin{equation}
B_K(\mu)=\hat B_K\left(\alpha_s(\mu)\right)^{2 \over 9} \;\;\; , \;\;\;  
B_{B_d,s}(\mu)=\hat B_{B_d,s} \left(\alpha_s(\mu)\right)^{6 \over 23},
\label{b_bk}
\end{equation}
$\hat B_i$ being renormalization invariant quantities.

There are two different methods to determine $\hat B_K$. 
The first one is based on 
the analysis of the two-point correlator of the
$\Delta S=2$ four-quark operator
$O_{\Delta S=2}=
(\bar s \gamma^\mu (1-\gamma_5) d)(\bar s \gamma_\mu (1-\gamma_5) d)$,
with the hadronic spectral function receiving a  
contribution from the $K K$ intermediate state. The second  method consists 
in the calculation of the three-point function of 
$O_{\Delta S=2}$ and of the interpolating currents for $K^0$ and $\bar K^0$.
The determination of $\hat B_K$ from the two-point sum rule
yields $\hat B_K=0.55\pm0.09$,\cite{narison95}
while three-point  sum rules give 
$\hat B_K=0.4-0.9$.\cite{BK_threep}

Similar two- and three-point sum rules involving the
relevant operator $O_{\Delta B=2}=
(\bar b \gamma^\mu (1-\gamma_5) d)(\bar b \gamma^\mu (1-\gamma_5) d)$
are employed to obtain the parameter $\hat B_{B_d}$.
The results are compatible with the vacuum saturation approximation,
namely  $\hat B_{B_d}=1.00 \pm 0.15$.\cite{pivovarov94}
All these calculations suffer from the uncertainties related to
the neglect of nonfactorizable $\alpha_s$ corrections. 
Notice that an accurate knowledge of the ratio
$\displaystyle r= {f_{B_d}^2 \hat B_{B_d}/ f_{B_s}^2 \hat B_{B_s}}$
is nowadays indispensable for 
the analysis of the CKM unitarity triangle using information
on $B_d$ and $B_s$ oscillations. In this ratio, several
uncertainties (from  $b$-quark mass, radiative corrections,
thresholds, etc.) should cancel out. Therefore, a direct extraction of 
the parameter $r$ from the ratios of sum rules deserves to be worked out.

\subsection{QCD sum rules in HQET}

QCD sum rules are frequently applied in the framework
of the Heavy Quark Effective Theory (HQET). One considers   
the correlation functions of quark
currents,  where the heavy quarks are represented by
their effective fields  $h_v(x)$, $v$ being the heavy quark 
four-velocity.\footnote{A more detailed discussion of HQET 
with the relation of the effective
fields $h_v(x)$ to the heavy quark fields $Q(x)$ can be found in the
chapter by De Fazio in this book.} 
An example is the calculation of the parameter $\hat F$
related to the $B$ meson leptonic decay constant $f_B$ by the
equation:
\be
f_B= \hat C(m_b) \hat F \left [ 1 - {A\over m_b} + {\cal O}\left( {1 \over m_b^2}
\right) \right ]\,,
\label{eq:fbinf}
\ee
where the coefficient $\hat C(m_b)$  can be computed in perturbation
theory. To determine $\hat F$, the  SVZ method can be applied 
to the two-point correlation function:
\be
\Pi(\omega) = i \int d^4 x \; e^{i k \cdot x} 
\langle 0| T\{j_M^\dagger(x) j_M(0)\}|0\rangle \label{eq:twophqet}
\ee
where $\omega= 2 v \cdot k$ and 
$j_M(x)= \bar h_v(x) i \gamma_5 q(x)$ is the interpolating current of 
the pseudoscalar heavy-light mesons in HQET. Due to the heavy quark spin symmetry,
 $\hat F$ can also be computed
from the two-point correlation function of the vector currents
$j_V(x)= \bar h_v(x) (\gamma_\mu-v_\mu) q(x)$ interpolating
heavy-light $1^-$ mesons.
The procedure starts from  writing down a dispersion relation
for  (\ref{eq:twophqet}) in the variable $\omega$:
\be
\Pi(\omega) = { \hat F^2 \over 2 \bar \Lambda - \omega } +
\int_{s_h}^\infty d\nu  {\rho^h(\nu) \over \nu - \omega}  + 
{\rm subtractions} ,
\label{eq:drhqet}
\ee 
isolating the ground state contribution from the integral
over the excited states and  the continuum.
The parameter $\bar \Lambda=m_B-m_b$  represents the binding energy
of the light degrees of freedom in the heavy meson. 
The dispersion relation (\ref{eq:drhqet}) 
is then matched with the QCD expression, obtained for negative $\omega$ in terms
of a perturbative term and condensate contributions:
\be
\Pi(\omega) = \Pi_{pert}(\omega) + 
\sum_d C_d {\langle O_d \rangle \over (-\omega)^d} \;\;\;. 
\label{eq:opehqet}
\ee 
Finally, the Borel transformation is performed and, 
invoking quark-hadron  duality, 
the contribution of higher states and of the continuum are approximated by the
perturbative contribution above the threshold $s_h$. 

Another example is the calculation of the Isgur-Wise function
$\xi(y)$.  
At the leading order in the $1/m_{b,c}$ expansion, this function 
parametrizes the semileptonic $B \to D^{(*)}$ matrix elements.
The form factor $\xi$ can be obtained from the sum rule for the
three-point correlation function:\cite{neubert92,radyushkin91,BBG93}
\be
\Xi(\omega, \omega^\prime, y) = i \int d^4 x \;d^4 x^\prime \; 
\exp^{i (k \cdot x - k^\prime \cdot x^\prime)} 
\langle 0| T\{j_{M^\prime}^\dagger(x) J_\mu(0) j_M(x^\prime) \}|0\rangle,\, 
\label{eq:threephqet}
\ee
$y=v \cdot v^\prime$ being the product of the initial and final meson
four-velocities and  $J_\mu= \bar h_{v^\prime} \gamma_\mu  h_v$.
Similar analyses can be applied to the form factors $\tau_{1/2}(y)$,   
$\tau_{3/2}(y)$, etc, of  $B$ transitions
to orbitally excited charm mesons.

There are several advantages in applying QCD  sum rules in HQET.
For example, the relatively simple form of the Feynman rules in the
effective theory allows to compute  
two-loop radiative corrections,  not only for two-point correlators,
such as (\ref{eq:twophqet}),  but also for three-point ones similar to
(\ref{eq:threephqet}).\cite{BBG93,neubertas,defazio98}
As we  mentioned in Sec.3.6, such corrections
are not yet available for finite quark masses in the three-point functions.
Furthermore, a systematic renormalization group
improvement in the current correlators can be performed; this is
important, for example, in determining the choice of the scale 
of $\alpha_s$ corrections in the calculation of the HQET parameters.

Let us briefly mention a few  numerical results obtained in this framework.
For the parameters $\hat F$ and $A$ defined in Eq.~(\ref{eq:fbinf})
the sum rule 
predictions~\cite{broadhurst,neubert92}$^,\,$\cite{es92}$^-$\cite{neubertrev} 
are, typically,
\bea
\hat F= 0.40 \pm 0.06~{\rm GeV}^{3\over 2}\,,
~~A=0.9 \pm 0.2~{\rm GeV}\,. 
\label{eq:fhat}
\eea
The parameter $\bar \Lambda= 570 \pm 70~\mbox{MeV}$~\cite{neubertrev}
is consistent with the determination
of $m_b$ from the bottomonium system.\footnote{For a more
detailed discussion of the definition of $\bar{\Lambda}$ and its relation
to the $b$ quark mass see the chapter by Uraltsev in this book.}
The analogous quantities $\bar \Lambda^+$ and  $\bar \Lambda^\prime$
have been computed for the $0^+,1^+$ and $1^+,2^+$ heavy-light
doublets, respectively, with the results 
$\bar \Lambda^+\simeq \bar \Lambda^\prime =0.9 - 1.0$ 
GeV.\cite{defazio00}
Combining this result with the calculation of $\bar \Lambda$ 
one can infer a prediction
of the mass of P-wave $\bar q Q$ states which is consistent 
with the experimental data.

The mass splitting between the lowest-lying $1^-$ and $0^-$ states is
related to the  chromomagnetic interaction parameter $\lambda_2$;
the  result~\cite{neubert92,BB94,neubertvirial}
$m^2_{(1^-)} - m^2_{(0^-)} = 0.46 \pm 0.14$ GeV$^2$  fits well to 
the experimental measurement.

More involved is the determination of the HQET parameter $\lambda_1$,
related to the kinetic energy of the $b-$quark in the $B$ meson, where the
results still have a large uncertainty.\cite{BB94,neubertvirial}
The role of 
nondiagonal contributions (i.e. matrix elements between different 
radial excitations) to the double dispersion relation has to be better
understood.\cite{BSU97}

The Isgur-Wise form factors $\xi(y)$ and 
$\tau_{1 \over 2}(y)$
have been computed at the next-to leading order in
$\alpha_s$;\cite{neubertas,defazio98} 
the  functions
$\tau_{3 \over 2}(y)$ and 
$\tau_{5 \over 2}(y)$ are known at the leading order in the strong coupling
constant,\cite{defazio00}
together with the form factors
of subleading terms in the inverse heavy quark mass 
expansion.\cite{BBG93,neubertsublead}

The analysis has been
extended to baryons containing one heavy quark, with the determination
of the binding energy $\bar \Lambda_{\Lambda_b}$,
the kinetic energy of the heavy quark in the 
heavy baryon and four-quark matrix elements on 
the $\Lambda_b$,\cite{dominguez96} 
the Isgur-Wise  form factor governing the semileptonic
 $\Lambda_b \to \Lambda_c$ transition,\cite{dai96}
and   the matrix
elements of decays such as  $\Lambda_b \to p \ell \nu$ and
$\Lambda_b \to \Lambda \ell^+ \ell^-$.\cite{baryonsl} 

It would be interesting to enlarge 
the field of applications of HQET  sum rules by analyzing 
also four-point correlators with heavy-light currents in order to 
determine the heavy and light quark distributions in the heavy hadrons.

\section{Light-Cone Sum Rules} 

\subsection{The basics of the method}

The method of light-cone sum rules (LCSR)~\cite{BBK89}$^-$\cite{CZ90} 
is a fruitful hybrid of the SVZ technique and the theory of 
hard exclusive processes.\cite{BL79}$^-$\cite{CZ80}
The basic idea is to expand the products of currents near the 
light-cone. This procedure involves  a partial resummation of local operators 
and avoids certain irregularities of the truncated OPE 
in the three-point sum rules. In recent years, the LCSR  
approach proved very useful in  calculating various hadronic transition 
form factors. Within this approach, one is able to take into account both 
hard scattering  and soft (end-point)~\cite{soft} contributions. 
 
\begin{figure}[t]
\vspace{-1.2cm}
\hspace{3cm}
\psfig{figure=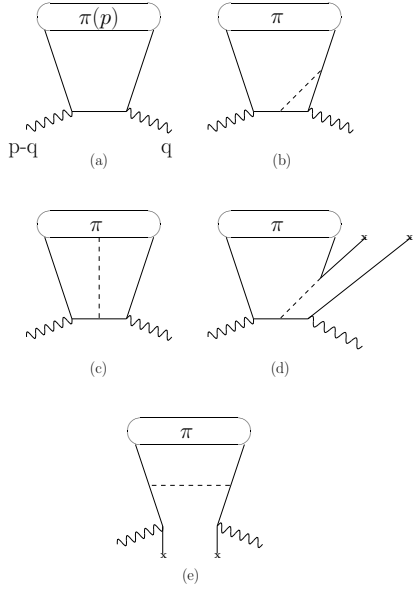,height=4.0in}
\vspace{-0.5cm}
\caption{ Light-cone expansion 
of the correlation function (\ref{ampl}).}
\label{fig:pionwf}
\end{figure}

While SVZ sum rules employ 
vacuum-to-vacuum correlation functions,  the starting object  of LCSR   
is different. One considers a correlation 
function which is a T-product
of two quark currents sandwiched between vacuum and 
an on-shell state.$\;$\cite{Craigie} 
The latter can be  a light-quark hadron 
(pion, kaon, $\rho$, $K^*$, nucleon) or a photon. 
A physical example is provided by the process $e^+e^- \to \pi^0 e^+e^-$.
The hadronic part of this reaction is a fusion of two virtual photons 
into a single $\pi^0$  via quark e.m. currents.
The corresponding amplitude, shown
schematically  in Fig.~\ref{fig:pionwf}a,  has the structure 
of a typical LCSR correlation function:
\bea
 F_{\mu\nu}(p,q) &=&
i\int \! d^4x \,e^{-iq\cdot x}\langle \pi^0 (p)\mid
T\{ j^{em}_\mu(x) j^{em}_\nu(0)\}\mid 0\rangle
\nonumber
\\
&=&\epsilon_{\mu\nu\alpha\beta}p^\alpha q^\beta F(Q^2,(p-q)^2)\,,
\label{ampl}
\eea
where $p$ is the pion momentum, $q$ and $(p-q)$ are the photon momenta, 
$Q^2=-q^2$, $j^{em}_\mu$ 
is the quark electromagnetic current (\ref{emcurrent})
and $F$ is the invariant amplitude encoding the dynamics of the process. 
The chiral limit is adopted and $p^2=m_\pi^2 =0$. 

To derive the LCSR, one has to calculate the 
correlation function  (\ref{ampl}) in QCD, in the region 
of large $Q^2$ and $|(p-q)^2|$ and to use
dispersion relation to match the result of this calculation with
hadronic matrix elements. 
Let us explain this procedure in more detail. Using unitarity  in the channel of 
the current $j_\nu^{em}$ with the momentum $p-q$,
i.e., inserting the complete set of hadronic states,  
one derives a dispersion relation 
for (\ref{ampl}) in the variable $(p-q)^2$ keeping the second variable $Q^2$ fixed: 
\bea
F_{\mu\nu}(p,q) 
&=&2\frac{\langle \pi^0(p)\!\mid j^{em}_\mu\mid \!\rho^0(p-q)\rangle 
\langle \rho^0(p-q)\!\mid j_\nu^{em} \mid \! 0 \rangle}{ m_{\rho}^2-(p-q)^2}
\nonumber\\
&+& \frac{1}{\pi}\!
\int\limits_{s_0^h}^\infty \!ds~ \!
\frac{Im F_{\mu\nu}(Q^2,s)}{s-(p-q)^2}\,.
\label{disprhopi}
\eea
In the above, the ground-state contribution of the $\rho$ meson 
contains the hadronic matrix element determining 
the  $\gamma^* \rho \to \pi $ transition form factor
multiplied by the $\rho$ meson decay constant: 
$\langle \rho^0(p-q)\!\mid\! j_\nu^{em}\! \mid \! 0 \rangle \!= \!
(f_\rho/\sqrt{2})m_\rho\epsilon_\nu ^{(\rho)*}$. The dispersion integral 
includes the contributions of excited and continuum states at $s>s_0^h$. 
The coefficient 2 takes into account the contribution of $\omega$ meson 
which is approximately equal to that of $\rho$. 
Calculating  the amplitude $F_{\mu\nu}$ in QCD, one then applies
the standard SVZ technique: the Borel transformation in the variable
$(p-q)^2$ and quark-hadron duality. The resulting sum rule 
allows to obtain the  
$\gamma^*\rho \to \pi  $ form factor and to determine its dependence 
on the momentum transfer $Q^2$. 

The correlation function 
(\ref{ampl}) can  be calculated by expanding the $T$ product 
of quark currents near the light-cone $x^2 = 0$. This expansion 
is different from the local OPE used before and, as we shall 
see below, incorporates summation of infinite series of local operators. 

Before proceeding with the light-cone expansion, we first have  
to convince ourselves that at sufficiently large 
$Q^2=-q^2$ and $|(p-q)^2|$ 
the dominant part of the integrand in the correlation function
(\ref{ampl}) stems from the region near the light-cone $x^2 =0$.
Importantly, the pion momentum $p$ does not need to vanish. 
Hence, the invariant variable 
$\nu= q\cdot p= (q^2-(p-q)^2)/2$ 
can also be large:
\be
|\nu| \sim |(p-q)^2| \sim Q^2 \gg \Lambda_{QCD}^2\,. 
\label{region}
\ee
It is convenient to define the ratio $\xi$ 
\be
\xi= 2\nu /Q^2\,,
\label{ksi}
\ee 
so that the region (\ref{region}) corresponds 
to finite values of this ratio, $\xi\sim 1$.    
Consider a reference frame where the pion three-momentum $\vec{p}$ is 
finite but small as compared to the photon virtualities: 
$|\vec{p}\,| \sim \mu$, $|p_0| \sim \mu$ and $\mu^2 \ll Q^2,\nu$.
In this frame, $q_0 \sim Q^2\xi/(4\mu)+ O(\mu)$ and 
the argument $q\cdot x$  of the exponential function  
in Eq.~(\ref{ampl}) can be approximated as:
$$
q\cdot x 
= q_0x_0-q_3x_3\simeq \frac{Q^2\xi}{4\mu}x_0-
\left(\sqrt{\frac{Q^4\xi^2}{16\mu^2}+Q^2} \right)x_3 
\simeq \frac{Q^2\xi}{4\mu}(x_0-x_3) - 
\frac{2\mu}{\xi}x_3\,.
$$
In order to avoid strong oscillations of the integrand, one has 
to demand $x_0 -x_3 \sim 4\mu/(Q^2\xi)$, and simultaneously 
$x_3\sim\xi/(2\mu)$. These two conditions yield 
$$
x_0^2 \simeq (x_3+4\mu/Q^2\xi)^2 
\simeq x_3^2 +4/Q^2 + O(\mu^2/Q^4)\,,
$$
and, hence, $x^2 \sim 1/Q^2 \to 0$ in the region (\ref{region}).
At the same time, there is no short-distance dominance, 
because $x_0 \sim x_3 \sim \xi/(2\mu) \gg 1/\sqrt{Q^2}$,
indicating that an expansion in local operators around $x=0$ 
is not applicable.\footnote{ One may still use the local OPE 
 for the soft pion, $p=0$ ($\xi=0$). 
In this case, (\ref{ampl}) reduces to a one-variable amplitude
similar to (\ref{pimunu}) and 
the short-distance dominance follows from the arguments presented in 
Section 2.1.}

To proceed, let us calculate the leading-order contribution
to the light-cone OPE of the correlator (\ref{ampl})
corresponding to the diagram in Fig.~\ref{fig:pionwf}a.
For simplicity we consider only the $u$-quark
part of the currents without the electromagnetic charge factor. 
Contracting the $u$ quark fields in (\ref{ampl}), using
the propagator of the free massless quark 
\be
iS_0(x,0) = \langle 0 \mid T \{u(x) \bar{u}(0)\} \mid 0 \rangle= 
\frac{i\not\!x}{2\pi^2x^4}\,, 
\ee
and transforming $\gamma_\mu\gamma_\alpha\gamma_\nu \to
-i\epsilon_{\mu\alpha\nu\rho} \gamma^\rho\gamma_5$ + ... ~, 
we obtain
\be
F_{\mu\nu}(p,q)= -i\epsilon_{\mu\nu\alpha\rho}
\int d^4x \frac{x^\alpha}{\pi^2x^4} e^{-i q \cdot x}
\langle \pi^0(p) \mid  \bar{u}(x)\gamma^{\rho} \gamma_5 u(0)
\mid 0 \rangle\,.
\label{ampl2}
\ee
To investigate the structure of the nonlocal quark-antiquark 
operator in Eq.~(\ref{ampl2}), let us 
expand it in local operators around $x=0$:
\be
\bar{u}(x)\gamma_\rho\gamma_5u(0)=\sum_r\frac{1}{r!}
\bar{u}(0)(\stackrel{\leftarrow}{D}\cdot x)^r\gamma_\rho\gamma_5 d(0)~.
\label{expan4}
\ee
The matrix elements of these operators have the following general 
decomposition:
\bea
\langle\pi^0(p) |\bar{u}\stackrel{\leftarrow}{D}_{\alpha_1}
\stackrel{\leftarrow}{D}_{\alpha_2}...
\stackrel{\leftarrow}{D}_{\alpha_r}\gamma_\rho\gamma_5 u |0\rangle        
=\!(-i)^r  p_{\alpha_1} p_{\alpha_2}...p_{\alpha_r}p_\rho M_r 
\nonumber
\\
\!+ (-i)^r g_{\alpha_1\alpha_2} p_{\alpha_3}...p_{\alpha_r} 
p_\rho M_r' +...~.
\label{local4}
\eea
In the above, the first term  is totally symmetric and traceless  
(at $p^2=0$) and contains only 
4-vectors. There are  other terms containing one or more 
$g_{\alpha_i\alpha_k}$, one of them  
displayed explicitly. Substituting the decomposition (\ref{expan4})
in (\ref{ampl2}), integrating over 
$x$ and using the definitions (\ref{local4}) and (\ref{ksi})
one obtains 
\be
F(Q^2,(p-q)^2) = 
\frac1{Q^2}\sum_{r=0}^{\infty}\xi^rM_r
+ \frac{4}{Q^4}\sum_{r=2}^{\infty}
\frac {\xi ^{r-2} }{r(r-1)}M_r'+ ...~.
\label{expanM}
\ee
Since the variable $\xi\sim 1$ 
in a generic exclusive kinematics with $p \neq 0$, 
all terms should be kept in each series in this expression.
We have explicitly confirmed the qualitative conclusion
made above: the expansion of $F$ in local operators
cannot be truncated at any finite order. One has to take into
account and to sum up  an infinite series  of  matrix elements 
$M_r$ , $M'_r$,... of local 
operators.  On the other hand,
there is a distinct hierarchy on the r.h.s. of (\ref{expanM}). 
The second term containing $M'_r$ and further similar
terms are suppressed by powers of a small parameter $1/Q^2$
as compared with the first term containing $M_r$. 
A closer investigation reveals that the difference between
the local operators entering the first
and the second term in (\ref{expanM}) is in their {\em twist}.    
Twist is defined as the difference between
the dimension and the spin of a traceless
and totally symmetric local operator. 
The lowest twist of the operators entering (\ref{local4}) is equal to two, simply 
because the operator without derivatives has dimension 3 and 
Lorentz spin 1. Furthermore, after taking the matrix elements, 
the twist 2 components of the operators contribute only to the 
first, symmetric and traceless term   
of Eq.~(\ref{local4}), containing $M_r$.   
After multiplying 
both parts of Eq.~(\ref{local4}) by $g_{\alpha_1\alpha_2}$ 
it becomes clear that the matrix elements 
$M'_r$ receive their contributions  from the twist 4 operators,
the lowest-dimension operator being
$\bar{u}(\stackrel{\leftarrow}{D})^2\gamma_\rho\gamma_5 u$. 
We conclude that  one has to treat the nonlocal operator in 
Eq.~(\ref{ampl2}) by expanding it near the light-cone $x^2=0$ in components 
corresponding to different twists.

In the leading order of this expansion, at $x^2=0$ (and $p^2=0$),  
the matrix element in Eq.~(\ref{ampl2})  has the following 
parametrization: 
\be
\langle \pi^{0}(p)|\bar{u}(x)\gamma_\mu\gamma_5u(0)|0\rangle_{x^2=0}=
-ip_\mu\frac{f_\pi}{\sqrt{2}}\int_0^1du\,e^{iu p\cdot x}
\varphi_\pi (u,\mu)\,,
\label{pionwf4}
\ee
where the function $\varphi_\pi(u,\mu)$ is the
 pion {\em light-cone distribution amplitude} of twist 2, 
normalized to unity: $\int_0^1\varphi_\pi(u,\mu) du =1$.\footnote{ The complete 
definition includes the path-ordered factor necessary for gauge invariance:
$Pexp \{ig_s \int ^1_0d\alpha~ x_\mu 
A^{a\mu} (\alpha x)\lambda^a/2\}$, which 
is unity in the light-cone gauge, $x_\mu A^{a\mu}=0$ employed here.}
The normalization scale $\mu$ 
emerges due to the logarithmic dependence on $x^2$ and
reflects the light-cone separation between the quark and antiquark fields 
in the operator.
At $x=0$, Eq.~(\ref{pionwf4}) is simply reduced to the matrix element 
(\ref{pionconst}) defining the pion decay constant. 
Furthermore, expanding both sides
of Eq.~(\ref{pionwf4}) and comparing the l.h.s. with the
expansions (\ref{expan4}) and (\ref{local4}) we find 
that the moments of $\varphi_\pi(u)$  are related to 
the matrix elements of local twist-2 operators:
\be
M_r= -i\frac{f_\pi}{\sqrt{2}}
\int_0^1du ~u^r\varphi_\pi(u,\mu).
\label{local}
\ee 
The function $\varphi_\pi(u)$, multiplied by $f_\pi$,
is a  universal nonperturbative object encoding the long-distance dynamics of the pion.
Together with the corresponding higher-twist distribution amplitudes,  $\varphi_\pi(u)$
plays a similar role as the vacuum condensates play in SVZ sum rules.

Substituting the definition (\ref{pionwf4}) in 
Eq.~(\ref{ampl2}), integrating over $x$, restoring the electromagnetic charge factor 
and adding the $d$-quark part  we obtain the correlation function
in the twist 2 approximation:
\be
F^{(tw2)}(Q^2,(p-q)^2)= 
\frac{\sqrt{2}f_\pi}{3}\int\limits_0^1 
\frac{du~\varphi_\pi(u,\mu)}{\bar{u}Q^2 -u(p-q)^2}\,,
\label{leading}
\ee
where $\bar{u} =1-u$. Note that this representation has the  
form of a convolution
\be
F^{(tw2)}(Q^2,(p-q)^2) = \frac{\sqrt{2}f_\pi}{3}\int\limits^1_0 
du\, \varphi_\pi (u,\mu) T(Q^2,(p-q)^2,u,\mu)
\label{convol}
\ee
of the hard scattering amplitude $T$ with the distribution amplitude
$\varphi_\pi$. The scale $\mu$ plays the role of the factorization 
scale which separates the contributions at $x^2<1/\mu^2$ entering the hard
scattering amplitude from the long-distance 
effects at $x^2>1/\mu^2$ parametrized by the distribution amplitude. 
At zeroth order in $\alpha_s$, the hard amplitude  
\be
T^{(0)}(Q^2,(p-q)^2,u)=
\frac{1}{\bar{u}Q^2-u(p-q)^2}
\label{T}
\ee
is $\mu$-independent.  

At this point one has to emphasize that 
the definition (\ref{pionwf4}) 
is actually almost ten years older than
the method of LCSR. It was introduced~\cite{BL79}$^-$\cite{CZ80} 
in perturbative QCD studies of exclusive processes with large
momentum transfer. In this approach, $\varphi_\pi(u)$ 
describes the distribution in the fraction of the longitudinal pion momentum 
carried by a valence quark in the infinite-momentum frame. 
The convolution (\ref{convol}) determining the asymptotic limit of the 
hard exclusive $\gamma^*\gamma^* \to \pi^0$ amplitude 
has also been obtained at that time.~\cite{BL79} 
The $O(\alpha_s)$ correction to $T$ has been calculated~\cite{rad} 
from the diagrams
similar to the one shown in Fig.~\ref{fig:pionwf}b.

We are now in a position to obtain a  sum rule from the 
dispersion relation (\ref{disprhopi}) matching it with the result of the 
light-cone expansion. Defining the matrix element
\be
\langle \pi^0(p) \mid j_\mu^{em} \mid \rho^0(p-q) \rangle
= F^{\rho\pi}(Q^2)m_\rho^{-1}\epsilon_{\mu\nu\alpha\beta}
\epsilon ^{(\rho)\nu} q^\alpha p^\beta \,,
\ee
in terms of the transition form factor $F^{\rho\pi}(Q^2)$,
we obtain, to leading twist 2 accuracy:
\bea
\frac{\sqrt{2}f_\rho F^{\rho\pi}(Q^2)}{m_\rho^2-(p-q)^2} +
\int\limits_{s_0^h}^\infty \!ds~ \!
\frac{\frac{1}{\pi}\mbox{Im}\, F(Q^2,s)}{s-(p-q)^2}
=
\frac{\sqrt{2}f_\pi}{3}\int\limits_0^1 
\frac{du~\varphi_\pi(u)}{\bar{u}Q^2 -u(p-q)^2}\,.
\label{dispers1}
\eea
Furthermore, representing Eq.~(\ref{leading}) in a form of  the dispersion integral
\be
F^{(tw2)}(Q^2,(p-q)^2)=
\frac{1}{\pi}\int\limits_{0}^\infty ds~ 
\frac{\mbox{Im}\,F^{(tw2)}(Q^2,s)}{s-(p-q)^2}\,  
\label{disp}
\ee
with
\be
\frac{1}{\pi}\mbox{Im}\,F^{(tw2)}(Q^2,s)= \frac{\sqrt{2}f_\pi}{3}\int\limits^1_0 
du \varphi_\pi (u)\delta(\bar{u}Q^2-us)\,,
\ee
we obtain the duality approximation for the contribution of excited 
and continuum states:
\bea 
\int\limits_{s_0^h}^\infty \!ds~ \!
\frac{\frac{1}{\pi}\mbox{Im}\, F(Q^2,s)}{s-(p-q)^2}= 
\!\int\limits_{s_0^\rho}^\infty \!ds~ 
\frac{\frac{1}{\pi}\mbox{Im}\,F^{(tw2)}(Q^2,s)}{s-(p-q)^2} 
=\!\frac{\sqrt{2}f_\pi}{3} \!\!\!\int\limits_{0}^{u_0^\rho}\!\! 
\frac{du~\varphi_\pi(u)}{\bar{u}Q^2 -u(p-q)^2}, 
\label{contin}
\eea
where $u_0^\rho=Q^2/(s_0^\rho+Q^2)$. The duality threshold 
parameter $s_0^\rho$ can be taken from  the SVZ sum rule (\ref{SVZrho}).
Using Eq.~(\ref{contin}), one can simply subtract the integral 
on the l.h.s. of Eq.~(\ref{dispers1}) from the r.h.s. 
Performing the Borel transformation, 
we finally obtain the LCSR for the form factor of the $\gamma^* \rho\to \pi$ 
transition:\cite{AK99} 
\bea
F^{\rho\pi}(Q^2)& = &\frac{f_\pi}{3f_\rho} 
\int^1_{u_0^\rho}\frac{du}u \varphi_\pi (u,\mu)
\exp\left(-\frac{\bar{u}Q^2}{uM^2} +\frac{m_\rho^2}{M^2}\right)\,.
\label{rhopi}
\eea
The light-cone distribution amplitude $\varphi_\pi$, which is 
the necessary input for this sum rule, will be discussed below.

To improve the accuracy of Eq.~(\ref{rhopi}), 
one has to include into the sum rule not only $O(\alpha_s)$ perturbative 
QCD corrections to the hard amplitude, but also higher twist 
effects. Physically, the latter take into account 
both the transverse momentum of the quark-antiquark state and 
the contributions of higher Fock states in the pion wave function.
There are several sources of  higher twist corrections
in the light-cone expansion. First, one encounters the twist 4 
contributions expanding the matrix element (\ref{pionwf4}) at $x^2=0$ beyond the 
leading order:
\bea
\langle\pi(p)|\bar{u}(x)\gamma_\mu\gamma_5u(0)|0\rangle =
-ip_\mu f_\pi\int_0^1du\,e^{iup \cdot x}
\left(\varphi_\pi (u)+x^2g_1(u)\right)
\nonumber \\
+
f_\pi\left( x_\mu -\frac{x^2p_\mu}{p \cdot x}\right)\int_0^1
du\,e^{iu p\cdot x}g_2(u)\,,
\label{phitw4}
\eea
where $g_{1,2}(u)$ are the light-cone distibution amplitudes 
of twist 4.
Furthermore, the expansion of the quark propagator near the light-cone 
yields the matrix elements of  the quark-antiquark-gluon
operators corresponding to three-particle distribution amplitudes 
(diagram in Fig.~\ref{fig:pionwf}c). 
There are also four-quark contributions stemming from the propagator
expansion, some of them shown in Fig.~\ref{fig:pionwf}d. 
These effects, together with other 
four-quark diagrams (Fig.~\ref{fig:pionwf}e) give rise 
(in the twist 6 approximation) to  contributions which can 
be factorized into a product of the quark condensate 
and a two-particle distribution amplitude. 
This effect can be potentially important at intermediate 
$Q^2$ and has to be investigated case by case. 
Nonfactorizable four-quark contributions and all 
other higher twist terms suppressed by high powers of $Q^2$
can safely be neglected.\footnote{Note that twist 3
contributions to the correlation function (\ref{ampl})  
are proportional to $m_\pi^2$ and vanish 
in the chiral limit.}

Replacing the currents $j_{\mu}^{em}$ in the correlation function 
(\ref{ampl}) by quark currents  with different 
spin-parity and flavor, one is able to interpolate other
hadronic transitions involving the pion. This replacement  
will only change the hard scattering
amplitude. One and the same set of distribution amplitudes will enter the 
light-cone expansion
of the new correlation function.
The idea of combining perturbatively calculable, process-dependent 
coefficients with a universal long-distance input works here in complete analogy 
with SVZ sum rules.

\subsection{Light-cone distribution amplitudes}

As we have seen in the case of $\varphi_\pi(u)$, distribution amplitudes  
are defined through 
the vacuum-hadron matrix elements  of nonlocal operators
composed of a certain number of quark and gluon fields
taken at light-like separations. These operators emerge 
in the light-cone OPE 
of the T-product of currents.
The relevant technique including the quark propagator expansion near the light-cone
and the extraction of various twist components 
is quite general.\cite{Balitsky83,BB89}  
It shares many common features  with the technique
used nowadays to study  forward and nonforward deep-inelastic amplitudes.
The distribution amplitude
can be expanded employing 
the conformal symmetry of massless QCD.
The conformal spin (partial wave) decomposition 
allows to represent each distribution amplitude as  a sum of 
certain orthogonal polynomials in the variable $u$. The coefficients of these polynomials
are multiplicatively renormalizable, and have  growing anomalous dimensions,
so that, at sufficiently large normalization scale $\mu$, only the first  few 
terms in this expansion are relevant. The part of the distribution amplitude, 
which does not receive logarithmic renormalization is called {\em asymptotic}.
The discussion of many important aspects of this analysis and of several 
interesting results obtained in recent years, is beyond the scope of this review, 
and we refer the reader to the literature.\cite{BF90,BallBraun} Here, we only
present the most important conformal expansion of the leading twist 2
distribution amplitude:
\be
\varphi_\pi(u,\mu) = 6u\bar{u} \left[1+ \sum_{n=2,4,..} 
a_{n}(\mu)C_{n}^{3/2}(u-\bar{u})\right]\,, 
\label{gegenb}
\ee
where $C_{n}^{3/2}$ are the Gegenbauer polynomials (for a  derivation, 
see, e.g., Ref.~167). 
The coefficients $a_{n}$ are multiplicatively renormalizable:
\be 
a_{n}(\mu)=a_{n}(\mu_0)\left( 
\frac{\alpha_s(\mu)}{\alpha_s(\mu_0)}\right)^{\gamma_n/\beta_0}~,
\label{anom}
\ee 
and
\be
\gamma_{n}=C_F\left[-3 -\frac{2}{(n+1)(n+2)}+4\left(\sum_{k=1}^{n+1}
\frac1k\right)\right]
\label{gamman}
\ee
are the anomalous dimensions.\cite{CZrep}
At $\mu \rightarrow \infty $, $a_{n}(\mu)$ vanish,
and  the limit $a_{n}=0$ 
corresponds to the asymptotic distribution amplitude
\be
\varphi_\pi^{(as)}(u)=6u\bar{u}~.
\ee

The values of the nonasymptotic coefficients 
$a_{n}$ at a certain intermediate scale $\mu_0$ 
can be estimated from  two-point sum rules~\cite{CZrep} for the moments
$\int u^n \varphi_\pi(u,\mu) du$ at low $n$. This method
is attractive because it employs nonperturbative information encoded in quark and gluon 
condensates. However, in practice the two-point sum rule determination  
of $a_n$ is not very accurate.   
One can also determine or, at least, restrict the nonasymptotic coefficients 
from the light-cone sum rules for measured hadronic quantities.
Some examples will be discussed below. 
Regarding other methods, lattice QCD has an almost unexplored potential to 
calculate distribution amplitudes.\cite{latticepion} 
Furthermore,    
$\varphi_\pi$ was evaluated in the instanton vacuum model, at a low normalization 
scale.\cite{piinst} Interestingly, this calculation has produced 
a distribution close to the asymptotic one.

Twist 3 and 4 distribution amplitudes for the pion 
have also been worked out.\cite{BF90,Ballhandbook}  
Use of the QCD equations of motion~\cite{BF90,Gorsky} yields
rigorous relations between three- and two-body distribution amplitudes
of the same twist, considerably reducing the number of independent
parameters.

To apply LCSR to hadronic matrix elements involving 
other pseudoscalar mesons ($K,\eta $), 
one has to assess the SU(3)-violation effects in the light-cone 
expansion. There are several sources of such effects.
The virtual $s$-quark propagator creates twist 3 contributions 
proportional to $m_K^2$, which differ from the analogous $O(m_\pi^2)$ effects
in the case of $u,d$ quarks; 
the ratio of nonperturbative parameters 
$f_K/f_\pi $ is larger than unity by about 20 \% ; finally, 
the asymmetry between strange and nonstrange quark momentum distributions  
in the kaon has to be taken into 
account by introducing nonzero odd coefficients $a_1,a_3,...$ in the Gegenbauer expansion
(\ref{gegenb}) for $\varphi_K(u)$. 

The light-cone analysis for the vector mesons
($\rho,\omega,K^*,\phi$) has been worked out, including
the update of the twist 2,\cite{Ball96}  and a 
detailed study of twist 3 and 4 distribution amplitudes,\cite{BallBraun}
taking into account the finite meson-mass effects. 
The distribution amplitudes for other than 
pseudoscalar and vector mesons still need to be determined.
It is already possible to apply LCSR to
hadronic matrix elements involving nucleons, since the nucleon 
higher twist distribution amplitudes
have been classified and studied.\cite{Braunetal} Finally,  
studies of the two-pion light-cone distribution amplitudes~\cite{twopion}
extend the field of applications of 
the LCSR method to the exclusive processes with  two pions.\cite{2gamma2pi} 

To complete our survey, let us 
mention the photon light-cone distribution amplitudes. They are  
important for many applications of LCSR to exclusive processes, where the photon is emitted at 
large distances. The photon distribution amplitudes were used, e.g.,
in the early study of the weak radiative transition 
$\Sigma \to p \gamma$.\cite{BBK89} In particular, the leading twist-2 
distribution amplitude of the photon has the following definition:
\be
\langle \gamma |\bar{u} (x) \sigma_{\alpha\beta} u(0)|0 \rangle=
e_u\langle\bar{u}u\rangle\int _0^1 du \varphi_\gamma(u)F_{\alpha\beta}(uq,x)\,, 
\ee
where  $F_{\alpha\beta}(q,x)=(\epsilon_\beta q_\alpha -\epsilon_\alpha 
q_\beta)e^{iq\cdot x}$ is 
the photon field strength tensor, and the asymptotic distribution amplitude
is $\varphi_\gamma(u)=6\chi u(1-u)$. 
The nonperturbative parameter $\chi\simeq-4~\mbox{GeV}^2$ (at $\mu=1$ GeV) 
normalizing this distribution 
in units of the quark condensate density,  can be determined from two-point sum 
rules with experimentally known hadronic parts.
The  update of this parameter and of  higher twist photon distribution amplitudes 
is desirable.

\subsection{LCSR for the pion form factor}

An example of the application of LCSR  
is the calculation of the pion electromagnetic form factor defined 
in Eq.~(\ref{pionformf}). The original sum rule~\cite{BH94} 
was recently improved~\cite{BKM99} by calculating 
the $O(\alpha_s)$ perturbative  contribution of twist 2 and  
the factorizable twist 6 corrections. The starting object is the 
vacuum-pion correlation function 
similar to (\ref{ampl}), but containing a pion interpolating current $j_\nu^{(\pi)}$ 
instead of   $j_\nu^{em}$. The resulting LCSR,
at zeroth order in $\alpha_s$ and  in the twist 2 approximation,
reads:\cite{BH94}
\begin{equation}
F_{\pi}(Q^2)\! =\!\! \int\limits_{u_0^\pi}^1 \!du\, \varphi_{\pi}(u,\mu_u) 
\exp \left( - \frac{\bar uQ^2}{u M^2} \right)
\!\stackrel{Q^2\to\infty}{\longrightarrow}
\frac{\varphi_{\pi}'(0,M^2)}{Q^4} 
\!\!\int\limits_0^{s_0^\pi}\!\! ds\, s\, e^{-s/{M^2}}, 
\label{SR1}
\end{equation} 
where $\varphi_{\pi}'(0)=-\varphi_{\pi}'(1)$, and  
$u_0^\pi= Q^2/(s_0^\pi+Q^2)$, $s_0^\pi$ is the 
duality threshold in the pion channel, taken from Eq.~(\ref{SVZpi}). 
The factorization scale $\mu_u^2= \bar{u}Q^2+uM^2 $  
corresponds to the average quark virtuality in the correlation function. 
This sum rule has a regular behavior at large  $Q^2$: 
the  $1/Q^4$ dependence of Eq.~(\ref{SR1}) at $Q^2\to \infty$   
corresponds to the soft end-point mechanism, provided that 
the integration region shrinks to the point $u=1$.

At $O(\alpha_s)$, one recovers the  leading $\sim 1/Q^2$ asymptotic 
behavior  corresponding to the hard scattering mechanism. 
Including this contribution in the LCSR and retaining the 
first two terms of the sum rule expansion in powers 
of $1/Q^2$ one obtains:\cite{BKM99}
\bea 
F_{\pi} (Q^2)\! =\! 
\frac{2\alpha_s}{3\pi Q^2}\!\! \int\limits_0^{s_0^\pi}\!\!ds\,  e^{-s/{M^2}}\! \!\!
\int\limits_0^1\! du\, \frac{\varphi_{\pi}(u)}{\bar u}
+\varphi_{\pi}'(0) \!\!\int\limits_0^{s_0} \!\frac{ds\, s\, e^{-s/{M^2}}}{Q^4}
+O\!\left(\frac{\alpha_s}{Q^4}\right)\!.
\label{pert}
\eea
The $O(1/Q^2)$ term in (\ref{pert})
coincides with the well known expression for the asymptotics 
of the pion form factor:\cite{BL79}$^-$\cite{CZ80} 
\begin{equation}
 F_\pi(Q^2) = \frac{8\pi\alpha_s f_\pi^2}{9Q^2}
  \left|\int\limits_0^1\!\! du \frac{\varphi_{\pi}(u)}{\bar u}\right|^2\,,
\label{asympt}
\end{equation}  
obtained by the convolution of two twist-2 distribution amplitudes
$\varphi_\pi(u)$ of the initial and final pion 
with  the  $O(\alpha_s)$ quark hard-scattering kernel. 
This coincidence can be easily traced 
provided that the SVZ sum rule (\ref{SVZpi}) for $f_\pi$ 
yields  in the leading order $\int_0^{s_0}\!ds\,e^{-s/M^2}= 4\pi^2f_\pi^2$,
and that $\int_0^1 \!\!du\varphi^{\rm as}_{\pi}(u)/\bar{u} =3$.

In addition to the twist 4 terms,\cite{BH94} the factorizable 
twist 6 contributions determined by 
the quark condensate density, have been calculated and turned out to be 
small.\cite{BKM99} 
\begin{figure}[t]
\vspace{-0.2cm}
\hspace{2cm}
\psfig{figure=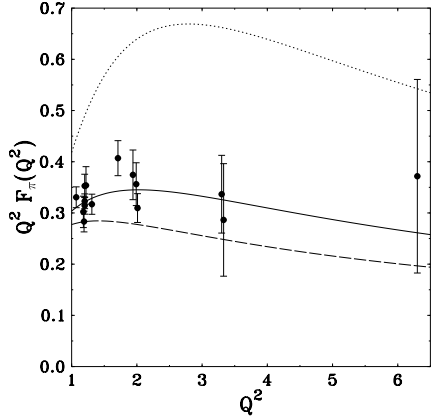,height=2.7in}
\vspace{-0.5cm}
\caption{The light-cone sum rule predictions for the pion electromagnetic 
form factor 
using asymptotic  
(dashed), CZ  (dotted) distribution amplitude 
and the fit to the data (solid). $Q^2$ is in GeV$^2$.}
\label{fig:piformf}
\end{figure}
Adding twist 4,6 terms to the twist 2 (leading and $O(\alpha_s)$) parts 
yields the LCSR prediction for $F_\pi(Q^2)$  shown in 
Fig.~\ref{fig:piformf}~\cite{BKM99} for 
the pion distribution amplitude (\ref{gegenb}) 
with two choices: $a_2=0$ (asymptotic) and $a_2( 1\,\mbox{GeV})= 2/3 $ 
(the CZ distribution~\cite{CZrep}). For the twist 4 distribution amplitude, 
their asymptotic expressions are taken, which provides a sufficient accuracy. 
The fit of the LCSR  to the experimental data~\cite{exppiformf} 
(with only $a_2\neq 0) $ yields 
\be
a_2( 1 \,\mbox{GeV} ) = 0.12 \pm 0.07
^{+0.05}_{-0.07}\,, 
\ee
where the first (second) uncertainty is 
experimental (theoretical). Within errors, this determination is compatible with 
the asymptotic distribution amplitude.

Similar to the pion elastic form factor $F_\pi$ , the 
LCSR for other transition form factors involving the pion 
have been obtained, including $\gamma^* \rho \to \pi $ 
(chosen as a study case in Sec. 4.1),\cite{AK99} and  
$\gamma^* \pi \to a_1 $.\cite{Belyaev}  
Transition form factors of the pion to other light mesons are still unexplored
within the LCSR method.

\begin{figure}[t]
\hspace{2cm}
\psfig{figure=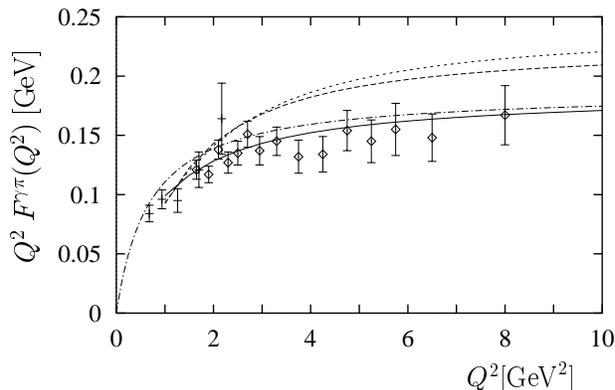,height=2.1in}
\vspace{-0.5cm}
\caption{ Form factor of the $\gamma^*\gamma \to \pi^0$ transition 
calculated with the asymptotic 
(solid), CZ ($a_2(\mu_0)=2/3$; long dashed) and 
BF($a_2(\mu_0)=2/3$, $a_4(\mu_0)=0.4$; short-dashed) 
distribution amplitude of the pion (where $\mu_0=0.5$ GeV)
in comparison with the experimental 
data points. The dash-dotted line corresponds to the  
Brodsky-Lepage interpolation formula.
}
\vspace{-0.5cm}
\label{fig:pigamma}
\end{figure}

Finally, LCSR contributed to 
the study of the $\gamma^*\gamma \to  \pi$ transition.
The corresponding form factor is simply equal to the amplitude 
Eq.~(\ref{ampl}) at zero virtuality of one of the photons: 
$F^{\gamma\pi}(Q^2)\equiv F(Q^2,0)$. 
Since the real photon is a  long-distance  object, 
the form factor  $F^{\gamma\pi}(Q^2)$ 
contains nonperturbative contributions which are beyond the light-cone
expansion of two electromagnetic quark currents. 
The leading $1/Q^2$ asymptotics of $F^{\gamma\pi}(Q^2)$ 
is well known~\cite{BL79} and given by Eq.~(\ref{leading})
at $(p-q)^2\to 0$. The calculation of the contributions 
suppressed by powers of $1/Q^2$  is a nontrivial task, at least 
when using three-point QCD  sum rules and short-distance OPE.\cite{RadRus}
Within LCSR approach, one possibility not yet exploited is to employ the photon 
distribution amplitudes. Another way to estimate the $\gamma^*\gamma \to  \pi$
form factor is to use the hadronic dispersion relation (\ref{dispers1})
where the resonance term is determined by the LCSR (\ref{rhopi})
and the integral over higher states is estimated using duality.\cite{AK99}   
One can analytically continue  this relation 
to $(p-q)^2 \to 0$ provided that it does not contain  
subtraction terms. The result 
shown in Fig.~\ref{fig:pigamma}~\cite{AK99} 
again has a better agreement with the experimental data~\cite{CELLO,CLEOgammapi}
in the case of the asymptotic pion distribution amplitude. 
The $O(\alpha_s)$ correction~\cite{rad,SY}  decreases the leading order (twist 2,4) result 
by  15-20\% leaving some room for nonasymptotic coefficients. The fit to the
data yields  $a_2=0.12 \pm 0.03$ at $\mu=2.4$ GeV (if all other coefficients are neglected).\cite{SY} 

Summarizing, the LCSR studies indicate  that the twist 2  
distribution amplitude of the pion   
is close to its asymptotic shape, already at intermediate scales.   
More precise data on various form factors involving pion are needed to increase 
the accuracy of this determination.

\subsection{Strong Couplings}

The correlation function (\ref{ampl}) can also 
be used to calculate the strong 
$\rho \omega \pi$ coupling, defined by the hadronic 
matrix element
\be
\langle \pi^0(p) \omega(-q) | \rho^0(p-q) \rangle = \epsilon^{\mu\nu\alpha\beta}
\epsilon_\mu^{(\omega)}\epsilon_\nu^{(\rho)} p_\alpha 
q_\beta\,g_{\omega\rho\pi}\,.
\label{strong}
\ee
The idea is to employ the analyticity of the amplitude $F(q^2,(p-q)^2)$ 
in two independent invariant variables $q^2$ and $(p-q)^2$,
the squares of momenta flowing through the two channels of electromagnetic currents.  
Inserting in these channels the complete sets of hadronic states 
with the quantum numbers of $\omega$ and $\rho$, one obtains 
a double dispersion relation:
$$
F_{\mu\nu}(p,q)=\!\!\frac{ 
\langle \pi^0(p) \omega(-q)|\rho^0(p-q) \rangle  
\langle 0 |j_\mu^{em}|\omega(-q) \rangle
\langle \rho^0(p-q) |j^{em}_\nu|0 \rangle }{(m_\omega^2-q^2)(m_\rho^2-
(p-q)^2)} + \{\omega \leftrightarrow \rho \}
$$
\be
+\int\limits_{R_{12}} ds_1 ds_2\frac{\rho^h_{\mu\nu}(s_1,s_2)}{(s_2-q^2)(s_1-(p-q)^2)} + 
...\,, 
\label{ddisp}
\ee 
where $\rho^h_{\mu\nu}$ is the double spectral density of the excited and 
continuum states and $R_{12}$ is the region occupied by these states 
in the $(s_1,s_2)$ -plane. 
Eq.~(\ref{ddisp}) is very similar to the double
dispersion relation (\ref{eq:3pt}) used in Sec.~3 for the three-point 
correlation function. The terms arising from 
subtractions are denoted by ellipses. The ground-state resonance 
contribution includes two (approximately equal) combinations
of $\rho$ and $\omega$. Substituting the definition (\ref{strong}) in the 
above relation and expressing the matrix elements of the electromagnetic currents 
through the decay constants $f_{\rho,\omega}$ 
yields the dispersion relation for the invariant amplitude:
\be
F(q^2,(p-q)^2)= \frac{f_\rho f_\omega g_{\omega\rho\pi}}{(m_\omega^2-q^2)(m_\rho^2-
(p-q)^2)} + ...~.
\label{ddispomega}
\ee
To obtain the sum rule, one has 
to match Eq.~(\ref{ddispomega}) to the light-cone expansion
(\ref{leading}), in the region of large $|q^2|$ and $|(p-q)^2|$.
The double dispersion integral 
over continuum and excited states can be replaced by its dual counterpart, 
obtained by integrating the double imaginary part of $F(q^2,(p-q)^2)$ over a certain region
in the $(s_1,s_2)$-plane. The lower boundary is $s_{1,2}=0$ and the upper 
boundary is characterized by a single threshold parameter
$s_0^\rho$ (again assuming that $s_0^\omega \simeq s_0^\rho$). The shape of this region 
is not very important because
the double imaginary part of the amplitude $Im_{s_1}Im_{s_2}F(s_1,s_2)$ 
obtained from (\ref{leading}) is concentrated 
on the diagonal  $s_1=s_2$. A detailed discussion of this procedure
is available in the literature.\cite{BBKR,KR}

Furthermore, one applies the Borel transformations with respect to 
the variables $q^2$ and $(p-q)^2$, introducing two independent
Borel parameters $M_1$ and $M_2$, respectively. 
The following formula is useful for this derivation:  
\bea
{\cal B}_{M_1^2}{\cal B}_{M_2^2}\frac{(l-1)!}{(-\bar{u}q^2 -u(p-q)^2)^l}=
(M^2)^{2-l}\delta(u-u_0)\,,
\eea
where $M^2=M_1^2M_2^2/(M_1^2+M_2^2)$ and $u_0=M_1^2/(M_1^2+M_2^2)$.
It is natural to take $M_1^2=M_2^2=2M^2$, i.e. $u_0=1/2$, having in mind  
almost equal mass scales in the $\rho$ and $\omega$ channels. 
After Borel transformation, all subtraction terms
vanish and the sum rule converts into a simple relation: 
\be
g_{\omega\rho\pi} 
=\frac{\sqrt{2}f_\pi}{f_\rho^2m_\rho^2}e^{m_\rho^2/M^2}M^2 (1-e^{-s_0^\rho/M^2})
\varphi_\pi(1/2) + O(\alpha_s)+\mbox{higher twists}\,,
\label{couplsr}
\ee
where we used  $f_\omega\simeq f_\rho/3$, and 
the subtracted exponent is the result of the quark-hadron duality approximation. 
The twist 4,6 terms not shown explicitly have also been calculated,\cite{BF89} 
whereas the $O(\alpha_s)$ correction is not yet 
available. The interesting point is that the sum rule for the strong coupling
is determined by the value of the light-cone distribution amplitude at the middle point 
$u=1/2$.
This value  is particularly  sensitive to the nonasymptotic coefficients $a_{2n}$.
Therefore, the sum rule (\ref{couplsr}) together with analogous relations 
for other measured strong couplings (like the LCSR for the 
$\pi NN$ coupling~\cite{BF89}) is  
useful for constraining the nonasymptotic parts of distribution amplitude.
For the asymptotic $\varphi_\pi$ the numerical result 
$g_{\omega\rho\pi}= 12.5\pm 2~\mbox{GeV}^{-1}$ 
obtained  from Eq.~(\ref{couplsr}) agrees with the experimental 
value $g_{\omega\rho\pi}= 14 \pm 2~\mbox{GeV}^{-1}$.\cite{pdg00}

The method of external fields, briefly mentioned in Sec.~3.5,
also allows to calculate $g_{\omega\rho\pi}$.\cite{omegarhopi}
In fact, this method corresponds to the soft pion limit of the 
light-cone expansion.\cite{BBKR} In this limit one can apply a short-distance 
expansion in terms of local operators with increasing dimensions. From that one may argue 
that sum rules in the external field
can still be used for determinations of the strong couplings if the pion is soft,
say, $p_\pi\sim 100$ MeV. On the other hand, the external field technique 
is not applicable in cases when the pion has a relatively large momentum,
like, e.g., in the $\rho \to \pi \pi$ decay, whereas LCSR work for any
pion momentum. 
Moreover, in the one-variable dispersion relation used
in the external field  approach, the contributions containing 
transitions between the ground and excited states remain
unsuppressed after Borel transformation, thus, reducing the 
accuracy of the sum rule.

\subsection{Heavy-to-light form factors and couplings} 

During the last years, LCSR have been extensively used 
to predict the form factors of various transitions of heavy  
($B$,$D$) to light ($\pi$, $K$, $\rho$, $K^*$,...) hadrons. Reviews 
describing these studies are available.\cite{KR,Brauntalk1}
Here we only briefly outline the main points, and present some 
recent results.

Let us consider, as an example, the calculation of the form factor $f_{B\pi}^+(q^2)$   
(the definition is similar to Eq.~(\ref{heavylight})). 
The underlying correlation function in this case is
\bea
\label{corr}
F_\mu(p,q) & = & i \int dx e^{i q \cdot x} \langle \pi(p) |
T \left\{ \bar{u}(x) \gamma_\mu b(x), m_b \bar{b}(0) i\gamma_5 d(0)
\right\} | 0 \rangle
\nonumber
\\
&&
= F(q^2,(p+q)^2) p_\mu + \tilde{F}(q^2,(p+q)^2) q_\mu\,,
\eea
where the weak current $\bar{u} \gamma_\mu b$ is combined with the current 
$\bar{b} i\gamma_5 d$ interpolating the $B$ meson. 
The  leading order in $\alpha_s$ 
corresponds to the diagram
in Fig.~\ref{fig:pionwf}a with the virtual $b$ quark.
In the twist 2 approximation, one obtains, for the relevant 
invariant amplitude,
\be
F(q^2,(p+q)^2)=m_bf_\pi\int_0^1\frac{du 
~\varphi_\pi(u,\mu_b) }{m_b^2-\bar{u}q^2-u(p+q)^2} ~.
\label{Fzeroth4}
\ee
Comparison of this expression with 
the result (\ref{leading}) for the correlator  of the light-quark currents
nicely demonstrates the universality 
of the method: both expressions are determined by one and the same distribution 
amplitude $\varphi_\pi$. The hard  
scattering amplitudes are different, the one in Eq.~(\ref{Fzeroth4}) 
being generated by the heavy quark propagator. 
Also the factorization scale of $\varphi_\pi$  in Eq.~(\ref{Fzeroth4}) 
should be adjusted to the characteristic virtualities of the heavy quark.

The LCSR is obtained by matching the light-cone expansion of $F$ with 
the dispersion relation in the $B$ channel. The result reads:\cite{CZ90,BKR}
\bea
f_{B\pi}^+(q^2)& = &
\frac{1}{2m_B^2f_B} e^{m_B^2/M^2}m_b^2 f_{\pi}\int\limits_{u_0^B}^1\frac{du}{u} 
exp\left(-\frac{m_b^2-q^2(1-u)}{uM^2} \right)\varphi_\pi(u,\mu_b)
\nonumber
\\
&&
+ ~O(\alpha_s) +\mbox{higher twists}\,,
\label{lcsrBpi}
\eea
where $u_0^B=(m_b^2-q^2)/(s_0^B -q^2)$, $s_0^B$ being the duality-threshold
parameter in the $B$ channel. This sum rule is valid at $q^2$, sufficiently lower 
than $m_b^2$, in order to stay far away from the hadronic states in the channel of 
the weak current. In contrast to the case of the LCSR for the pion form factor, 
the twist 3 contributions, not shown explicitly in Eq.~(\ref{lcsrBpi}), 
are important.

Importantly, LCSR (\ref{lcsrBpi}) has a regular behavior in the 
$m_b\to \infty$ limit. If one employs the scaling relations for mass parameters and 
decay constants: $ m_B =m_b+\bar{\Lambda}$, $s_0^B =m_b^2 +2m_b\omega_0$, $M^2=2m_b\tau$,
$f_B =\hat{f}_B/\sqrt{m_B}$, where $\bar{\Lambda}$, $\omega_0$, $\tau$ and $ \hat{f}_B$ 
are independent of $m_b$, the LCSR can be expanded in the heavy mass. 
The higher-twist corrections either scale with the same power of 
$m_b$ as the leading-twist term, or they are suppressed by extra powers of $m_b$. 
Furthermore, the asymptotic scaling behavior sharply differs at small and large
momentum transfers: at $q^2=0$ one has $f^+(0) \sim m_b^{-3/2}$,\cite{CZ90} and at 
$p^2=m_b^2-2m_b\chi$, where $\chi$ does not scale with $m_b$, 
$f^+(p^2) \sim m_b^{1/2}$.\cite{scalar} Hence, 
the large $p^2$ behavior is consistent with the asymptotic
dependence predicted in HQET for the region of small pion momentum.\cite{Isgur}  
Another important feature is that, similar to case of the pion 
form factor, the heavy-to-light form factors calculated from LCSR 
receive contributions from both soft and hard mechanisms of the 
exclusive transition.\cite{Brauntalk}
\begin{figure}[t]
\hspace{4cm}
\psfig{figure=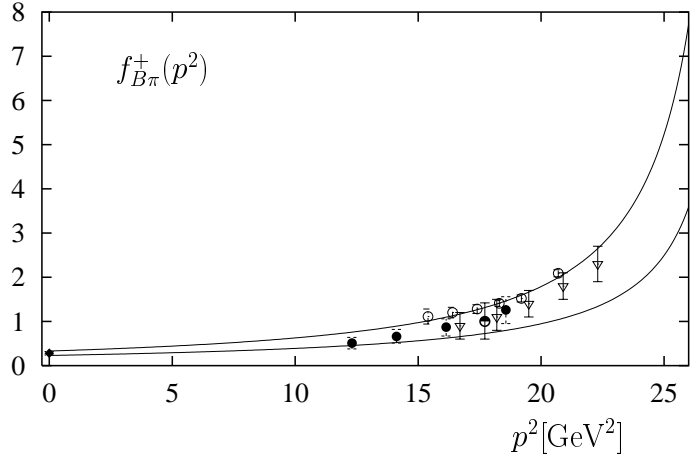,height=2.5in}
\caption{The LCSR predictions for the $B\to \pi$ 
transition form factor in comparison to  lattice results.
The solid curves indicate the size of the LCSR uncertainties.}
\label{fig:Bpi}
\end{figure}

At the current accuracy, Eq.~(\ref{lcsrBpi}) includes the 
$O(\alpha_s)$ correction to the twist 2 term~\cite{KRWY,Bagan} and all twist 
3 and 4 quark-antiquark and quark-antiquark-gluon contributions.\cite{BKR,BBKR} 
Concerning the numerical analysis of this LCSR, one has to mention that  
the sensitivity of $f^+(q^2)$  to $m_b$ is considerably reduced if $f_B$ and $s_0^B$ 
are taken from the two-point sum rule considered in Sec.~3.3. 
There is also a spectacular cancelation between $O(\alpha_s)$ corrections
to these two sum rules.
Furthermore, the sensitivity to the light-cone distribution amplitudes is also not high
because in the sum rule (\ref{lcsrBpi}) these 
normalized distributions are convoluted with relatively smooth coefficient functions 
over a wide region of $u$ starting from $\sim$ 0.5 to 1.
In particular, the result obtained from the sum rule (\ref{lcsrBpi}) is less 
sensitive to the values of nonasymptotic coefficients in the 
distribution amplitude $\varphi_\pi$, than the LCSR result for the pion form factor. 
A typical overall  uncertainty of heavy-to light form factors 
obtained by this method amounts to 15-20 \%, if one employs
constraints on nonasymptotic coefficients $a_{2,4,..}$ provided by the 
comparison of the LCSR for the pion form factors ($F_\pi$, $F^{\gamma\pi}$) 
and for the strong couplings with experimental data. To somewhat reduce this uncertainty,
one needs better determination of these coefficients. It is also important 
to calculate the $O(\alpha_s)$ correction to the twist 3 contribution in Eq.~(\ref{lcsrBpi}).

\begin{figure}[p]
\vspace{-1.2cm}
\hspace{2cm}
\psfig{figure=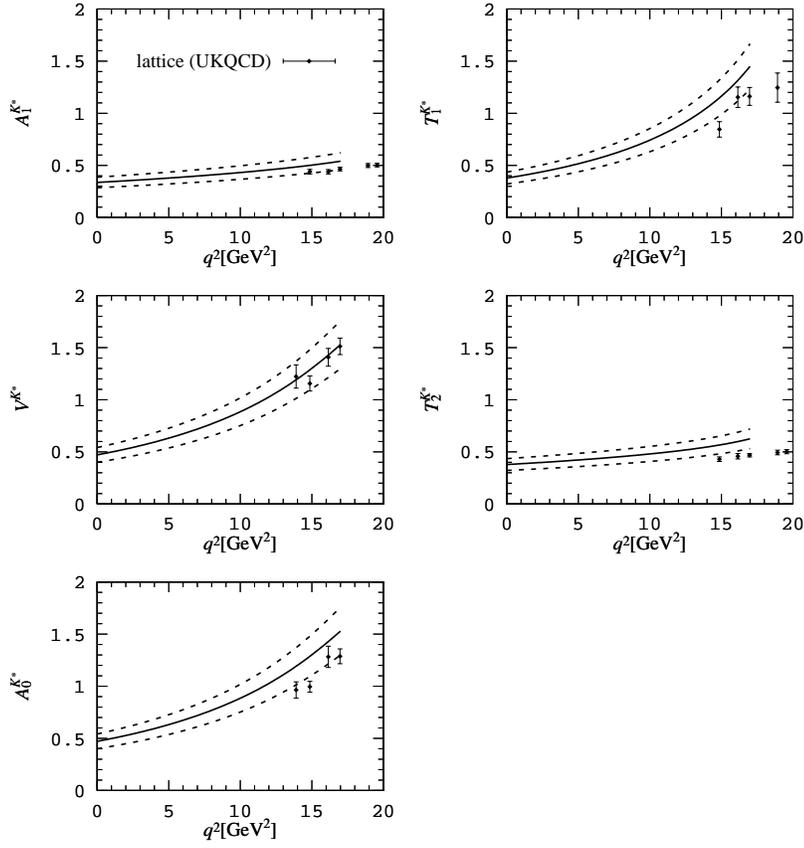,height=4.5in}
\caption{~LCSR predictions for the $B\to K^*$ 
transition form factors. The dashed lines indicate theoretical 
uncertainties, the points represent lattice QCD
results.} 
\vspace{-0.5cm}
\label{fig:Brho}
\end{figure}

The same correlation function (\ref{corr}) can be used to estimate 
the $B^*B\pi$ coupling, defined as 
$\langle \bar{B}^{*0}(q)\pi^-(p)\mid B^-(p+q)\rangle =
-g_{B^*B\pi}(p \cdot\epsilon^{(B^*)}) $, 
from the double dispersion relation in $B$ and $B^*$ 
channels, as explained in Sec.~4.4. The corresponding LCSR reads:\cite{BBKR}
\bea
f_{B^*}g_{B^*B\pi}&=&\!\!\!\!\frac{1}{m_B^2m_{B^*}f_B f_{B^*}}
e^{\frac{m_{B}^2+m_{B^*}^2}{2M^2}}
m_b^2 f_\pi M^2\left(e^{-\frac{m_b^2}{M^2}} - 
e^{-\frac{s_0^B}{M^2}}\right)\varphi_\pi(1/2,\mu_b)
\nonumber
\\
& & \!\!\!\!
+~O(\alpha_s) +\mbox{higher twists} \,.
\label{srcoupl}
\eea
Here again the NLO accuracy in twist 2 has recently been achieved.\cite{KRWY2}
In Fig.~\ref{fig:Bpi}~\cite{KRWWY} 
the form factor $f^+_{B \pi}(q^2)$, 
obtained  using Eqs.~(\ref{lcsrBpi}) and (\ref{srcoupl}),
is displayed in comparison with the recent lattice results (see references 
in the original paper~\cite{KRWWY}).
Other applications of LCSR  include the $B\to K$,\cite{BKR,Ball98}
$B\to \rho,K^*,\phi $ form  factors of 
weak and FCNC  transitions.\cite{BB97,ABS,BB98} Fig.~\ref{fig:Brho}~\cite{BB98} shows,
for example,  the recent predictions on the form factors of $B\to K^*$ transition.
It is easy to convert the sum rules for $B$ mesons to the corresponding sum rules
for $D$ mesons by replacing $b$ with $c$, and $B^{(*)}$  with $D^{(*)}$
in Eqs.~(\ref{lcsrBpi}) and (\ref{srcoupl}). The $D\to 
\pi,K$ form factors (the latter being quite sensitive to the SU(3)-flavor
violation) calculated within this method~\cite{dsemilep2,BBKR,KRWWY} are in a good agreement 
with the experiment and with the lattice QCD.   
\begin{table}[h]
\begin{center}
\begin{tabular}{| c | c| c | c | c | }
\hline
$\hat{g}$&$g_{D^* D \pi}$&$g_{B^* B \pi}$&method &Ref.  \\ \hline 
   &               &               &       &      \\
0.22 $\pm$ 0.06& $10.5\pm 3.0$  &$22\pm7$&LCSR (NLO) &KRWY99~\cite{KRWY2}\\
\hline\hline
  &               &               &       &       \\
   &           &$32\pm6 $&SP (LO) &  O89~\cite{ovc89}\\
$0.39\pm 0.16$ &$9\pm1$  &$20\pm4$&''  `&CNDDFG95~\cite{CNDDFG95}\\ 
& $11\pm2$  &$28\pm 6 $&''  &BBKR~\cite{BBKR}\\
\hline\hline
    &               &              &       &       \\
$0.27^{+0.04+0.05}_{-0.02-0.02}$&& &$\chi$PT+data&  S98~\cite{stewart98}\\  
   &               &               &       &       \\ \hline \hline
    &               &               &       &       \\
$0.42\pm0.04\pm0.08$& &  & lattice QCD&UKQCD98~\cite{dedivitiis98}\\  
   &               &             &         &       \\
\hline
\end{tabular}
\caption{Strong $B^* B \pi$ and $D^* D \pi$ couplings from LCSR 
and from the sum rules in the soft pion limit (SP)
compared with some recent results of other methods. The scale-independent 
coupling is defined as $\hat{g}=f_\pi g_{H^*H\pi}/2m_H$, $H=B,D$. 
\label{table:strongc}}
\end{center}
\end{table}
In Table~\ref{table:strongc} 
the sum rule predictions for the strong couplings 
$g_{D^* D \pi}$ and $g_{B^*B\pi}$ are compared with 
the result obtained on the lattice, and with the 
recent prediction of $\chi$PT
constrained by experimental data. A compilation of results of many other approaches
is also available.\cite{BBKR}  LCSR have been applied  to obtain 
the strong couplings $g_{D^* D \rho}$ and  $g_{B^* B \rho}$,\cite{aliev96} 
as well as the matrix elements for the radiative transitions between 
heavy mesons ($B^* \to B \gamma$)~\cite{aliev97} and heavy baryons.\cite{dai99} 
The strong couplings of negative and  positive parity heavy mesons and pions
have also been determined, both in the heavy quark mass limit and
for finite $c$ and $b$ quark masses, with  predictions
on the widths of $0^+$, $1^+$ and $2^+$ charmed and beauty 
mesons.\cite{defazio95,defazio98a} LCSR with  photon distribution amplitudes
have been employed to estimate the long-distance contributions 
to  $B\to \rho \gamma$ weak radiative decays.\cite{Brhogamma} 

Finally, it is important to mention the use of QCD sum rules
for the analysis of exclusive nonleptonic decays
of heavy mesons. Nowadays, these decays attract a lot of attention, being  
one of the main objects for studying the mechanism of CP-violation.
The form factors
and decay constants calculated from LCSR and SVZ sum rules 
can immediately be used for estimates of the nonleptonic 
decay amplitudes in the factorization approximation, in particular, for assessing
the violation of the $SU(3)$-flavor (or $U$-spin) symmetry  in various $B$-decays. 
More subtle effects, such as violation of factorization, annihilation 
and penguins can, in principle, also be treated within the sum 
rule framework.  First attempts to analyze nonfactorizable 
effects in $B$ and $D$ decays and related long distance effects 
in $B\to K^* \gamma$ decays employed short-distance 
OPE.\cite{KR,blokShifman,KRSW} The  LCSR have only been partially used 
in order to obtain the form factors of effective operators emerging in 
the short-distance expansion of  correlation functions.\cite{nonfact} 
It would be extremely  interesting to continue these studies 
trying to create a framework where the light-cone OPE is consistently
employed at all stages of calculation.

\section{Summary}

In this review we have tried to present a concise and updated 
picture of QCD sum rules and their applications. This was not an easy 
task, because after more
than twenty years of development, the manifold of works using this method 
resembles a large tree with many branches penetrating into different fields 
of QCD and hadron phenomenology. 
Some of the branches are quite distant with respect to  each other, and 
in order to find the relevant papers one has to search 
not only in the hep-ph, but also  in the
hep-th and nucl-th electronic archives.
Nevertheless, the procedure used in all these applications 
is essentially one and the same, formulated in the original work:\cite{SVZ79}
A) construct a correlation function,  B) calculate it at 
some Euclidean scale using the operator product
expansion and C)  match it with the hadronic dispersion integral. 
Above, in Sec.2,  we have described this procedure step by step,
deriving the SVZ sum rule for the decay constant $f_\rho$ as a study case. 

The applications of QCD sum rules considered in this review cover only 
a part of the work done in this field. 
Following our particular line of discussion, we could have missed
some important references. 
The aim of our presentation was 
not just to demonstrate how the problems of hadron phenomenology 
are treated within this method, but to help the reader to
assess the current status  of QCD sum rules in general.
Let us try to formulate  the main conclusions of our discussion:

$-$ QCD sum rules, together with experimental data 
on hadronic spectral densities can successfully be used to determine 
quark masses and universal nonperturbative 
parameters, such as vacuum condensates;

$-$Within the sum rule framework, operating with a 
handful of inputs, one is able to reproduce many hadronic 
observables, such as decay constants, form factors and 
parton distributions, in a reasonable agreement 
with the experimental data;

$-$The accuracy of the method is essentially limited. 
Nevertheless, since the correlation 
functions are field-theoretic objects defined in QCD 
and related to hadrons via rigorous dispersion relations,
the theoretical uncertainties can be traced and  estimated;

$-$Today QCD sum rules are not limited by 
the ``classical'' SVZ  approach
based on  two-point correlators. 
As we have seen,  combining the sum rule technique
with the light-cone expansion, it is possible to develop the  
LCSR method which avoids certain problems of the local 
condensate expansion and adequately describes 
QCD mechanisms of exclusive hadronic transitions.

Concerning the open problems, the major one is 
to reduce theoretical uncertainties in the inputs 
and in the procedure.  Here QCD sum rules alone are not sufficient.
Interaction with other nonperturbative methods 
is very important, e.g.  lattice QCD
determination of condensates or light-cone distributions.
Some new experimental results of hadron physics
could be very helpful. As we have seen,
accurate measurements of the form factors of light hadrons
will allow us to fix the nonasymptotic coefficients 
in the light-cone distribution amplitudes. Furthermore, a better knowledge 
of excited hadronic resonances with different quantum numbers 
could reduce systematic uncertainties 
introduced by the quark-hadron duality 
approximation. One may even suggest dedicated experimental studies.
For example,  semileptonic decays of charmed mesons can provide 
important information about the resonances in the $K\pi$ and $K\pi\pi$ system
with different spin-parities. 

To outline the future perspectives of the method, let us give only one 
important example. 
In order to extract fundamental parameters of the electroweak theory
and to search for new physics employing  
the current and future data on exclusive $B$ and $D$ decays, 
accurate predictions on the  heavy meson decay
amplitudes are needed. Approximate QCD methods are the only 
tools we have at our disposal, and the analytical method of QCD sum rules has, 
undoubtfully, a large unexplored potential in this field.

\section*{Acknowledgments}
We are grateful to M. Shifman for offering us to write this review.  
Numerous discussions with our colleagues, with whom 
we collaborated on different applications 
of QCD sum rules during many years, are acknowledged.  
We thank  P. Ball and N. Paver for useful remarks on the manuscript. 
A.K. is grateful to the High Energy Group at the Niels Bohr Institute for
hospitality and support during the initial stage of this work.

\section*{References}


\end{document}